%% file: main.tex
\documentclass[10pt,journal,compsoc,twoside]{IEEEtran}
\usepackage{booktabs} 
\usepackage{balance}
\usepackage{algorithm,algpseudocode}
\usepackage{amsmath}
\usepackage{graphicx}
\usepackage{pdfpages}
\usepackage{comment}
\usepackage{balance}
\usepackage{soul}
\usepackage{amssymb}
\usepackage{pifont}
\newcommand{\Checkmark}{\ding{51}}
\newcommand{\xmark}{\ding{55}}

\usepackage[caption=false,font=footnotesize]{subfig}
\usepackage{fancyhdr}
\usepackage{mdwlist}
\usepackage{url}
\usepackage{color}
\usepackage{multirow}
\usepackage{amsmath}        
\graphicspath{{figures/}}

\makeatletter
\def\blfootnote{\xdef\@thefnmark{}\@footnotetext}
\makeatother

\usepackage{setspace}

\usepackage{soul}
\newcommand{\drop}[1]{\textcolor{red}{#1}}
\renewcommand{\drop}[1]{}

\ifCLASSOPTIONcompsoc
  \usepackage[nocompress]{cite}
\else
  \usepackage{cite}
\fi

\ifCLASSINFOpdf
\else
\fi

\begin{document}

\newcommand{\maintitle}{Opening the Doors to 
\textit{Dynamic} Camouflaging}
\newcommand{\subtitle}{Harnessing the Power of Polymorphic Devices}
\newcommand{\thetitle}{\maintitle: \subtitle}

\title{\thetitle}

\author{Nikhil~Rangarajan\textsuperscript{*},~\IEEEmembership{Member,~IEEE,}
Satwik~Patnaik\textsuperscript{*},~\IEEEmembership{Graduate~Student~Member,~IEEE,}
Johann~Knechtel,~\IEEEmembership{Member,~IEEE,}
Ramesh~Karri,~\IEEEmembership{Fellow,~IEEE,}\\
Ozgur~Sinanoglu,~\IEEEmembership{Senior~Member,~IEEE,} and~Shaloo~Rakheja,~\IEEEmembership{Member,~IEEE}

~\thanks{\textsuperscript{\textbf{*}}Nikhil~Rangarajan and Satwik~Patnaik contributed equally.}
\IEEEcompsocitemizethanks{\IEEEcompsocthanksitem Satwik~Patnaik and Ramesh~Karri are with the Department of Electrical and Computer Engineering, Tandon School of Engineering, New York University (NYU), Brooklyn, NY, 11201, USA.
\IEEEcompsocthanksitem Nikhil~Rangarajan, Johann~Knechtel and Ozgur~Sinanoglu are with the Division of Engineering, New York University Abu
Dhabi (NYU AD), Abu Dhabi, 129188, UAE.
\IEEEcompsocthanksitem Shaloo~Rakheja is with the Holonyak Micro and Nanotechnology Laboratory, University of Illinois at Urbana-Champaign (UIUC), Urbana, IL, 61801.
\IEEEcompsocthanksitem Corresponding authors: Nikhil~Rangarajan (nikhil.rangarajan@nyu.edu) and Satwik~Patnaik (sp4012@nyu.edu).}
}

\markboth{IEEE Transactions on Emerging Topics in Computing,~Vol.~X, No.~X, Month~202X}
{Rangarajan and Patnaik \MakeLowercase{\textit{et al.}}: Dynamic Camouflaging: Harnessing the Power of Polymorphic Devices}

\IEEEtitleabstractindextext{

\begin{abstract}
The era of widespread globalization has led to the emergence of hardware-centric security threats throughout the IC supply chain. 
Prior defenses like logic locking, layout camouflaging, and split manufacturing have been researched extensively to protect against intellectual property (IP) piracy at different stages.
In this work, we present \emph{dynamic camouflaging} as a new technique to thwart IP reverse engineering at all stages in the supply chain, viz., the foundry, the test facility, and the end-user. 
Toward this end, we exploit the multi-functionality, post-fabrication reconfigurability, and run-time polymorphism of spin-based devices, specifically the magneto-electric spin-orbit (MESO) device. 
Leveraging these unique properties, \emph{dynamic camouflaging} is shown to be resilient against state-of-the-art analytical SAT-based attacks and test-data mining attacks.
Such dynamic reconfigurability is not afforded in CMOS owing to fundamental differences in operation.
For such MESO-based camouflaging, we also anticipate massive savings in power, performance, and area over other spin-based camouflaging schemes, due to the energy-efficient electric-field driven reversal of the MESO device.
Based on thorough experimentation, we outline the  promises of \emph{dynamic camouflaging} in securing the supply chain end-to-end along with a case study, 
demonstrating the efficacy of \emph{dynamic camouflaging} in securing error-tolerant image processing IP.
\end{abstract}

\begin{IEEEkeywords}
Hardware security,
IP protection,
Layout camouflaging,
Dynamic camouflaging,
Post-fabrication reconfigurability,
Dynamic morphing,
Functional polymorphism,
Spin devices.
\end{IEEEkeywords}}

\maketitle

\renewcommand{\headrulewidth}{0.0pt}
\thispagestyle{fancy}
\lhead{}
\rhead{}
\chead{\copyright~2020 IEEE.
This is the author's version of the work. It is posted here for personal use.
Not for redistribution.	The definitive Version of Record is published in
IEEE TETC, DOI 10.1109/TETC.2020.2991134}
\cfoot{}

\IEEEdisplaynontitleabstractindextext
\IEEEpeerreviewmaketitle

\IEEEraisesectionheading
{\section{Introduction}
\label{sec:introduction}}

\IEEEPARstart{A}{s} the aggressive scaling of complementary metal-oxide-semiconductor (CMOS) technology nodes reaches physical device limits, traditional CMOS-based architectures face significant challenges ranging from saturated performance gains to increased power density and variability, coupled with concerns for reliability. 
The continual miniaturization of CMOS technology has not only complicated chip design but also necessitated advanced and expensive fabrication facilities.
Alternative technologies are being pursued extensively, which can augment CMOS in enabling higher memory and logic efficiency. 
These include several emerging devices such as silicon nanowire field-effect transistors (SiNW-FETs)~\cite{de2012polarity}, memristors~\cite{chua1971memristor},
negative capacitance FETs (NCFETs)~\cite{salahuddin2008use}, spin devices~\cite{manipatruni2019scalable}, etc.
Spintronic devices, in particular, have emerged as one of the top contenders for the post-CMOS era~\cite{manipatruni2019scalable}.

\begin{table}[ht]
\centering
\footnotesize
\caption{IP Protection Techniques Versus 
Untrusted Entities
in IC Supply Chain
(\Checkmark: Protection Offered, \xmark: No Protection Offered)}
\label{tab:protection_comparison_2}
\setlength{\tabcolsep}{1mm}
\renewcommand{\arraystretch}{1.6}
\begin{tabular}{*{4}{c}}
\hline
\textbf{Technique} & 
\textbf{FEOL/BEOL} & 
\textbf{Test Facility} & 
\textbf{End-User} 
\\ \hline
Logic Locking & 
\Checkmark/\Checkmark & \Checkmark~\,\,(\cite{yasin16_test}) & 
\Checkmark 
\\ \hline
Layout Camouflaging & \xmark/\xmark~\,\,(\Checkmark/\xmark~\cite{patnaik2020obfuscating}) & 
\xmark~\,\, & 
\Checkmark   
\\ \hline
Split Manufacturing & 
\Checkmark/\xmark~\,\, & 
\xmark & 
\xmark~\,\,(\Checkmark~\cite{patnaik2018best,patnaik2019modern}) 
\\ \hline
\textbf{Dynamic Camouflaging} & 
\Checkmark & 
\Checkmark & 
\Checkmark 
\\ \hline
\end{tabular}
\end{table}

\begin{figure*}[ht]
\centering
\includegraphics[width=\textwidth]{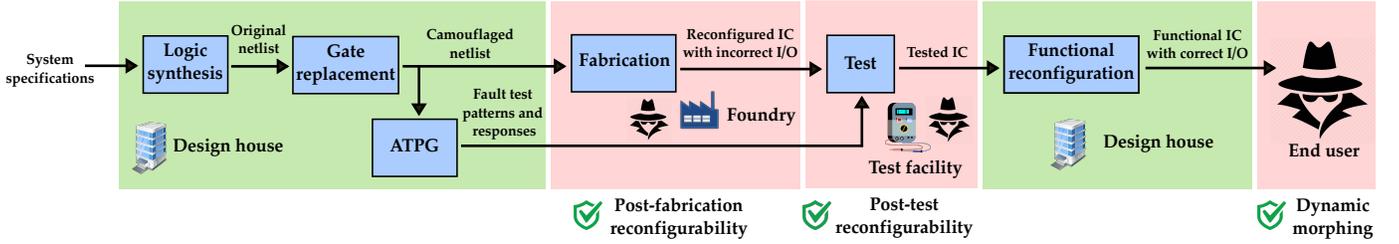}
\caption{Threat model for 
\textit{dynamic camouflaging}-based IP protection.
Green and red blocks represent the \textit{trusted} and \textit{untrusted} entities, respectively. 
The protection schemes---which are all flavors of dynamic camouflaging---employed for each of the untrusted entities are mentioned below the respective red blocks, and indicated by green shields.
Gate replacement, which can either be random or through some designer's chosen heuristic, involves the selective replacement of gates in the original netlist with polymorphic gates (magneto-electric spin-orbit (MESO) gates in this work).
After fabrication and testing, the design is sent back to the design house (or some trusted facility) for functional reconfiguration before being deployed in the open market.
ATPG stands for automatic test pattern generation.
}
\label{fig:threat_model}
\end{figure*}

The expeditious globalization of the electronics industry has resulted in the outsourcing of the integrated circuit (IC) supply chain. 
Such a distributed supply chain, which is often spread across geographically different locations, enables various attacks, ranging from piracy of intellectual property (IP) to illegal and unauthorized overproduction of ICs, and targeted insertion of malicious circuits known as hardware Trojans.
IP piracy, in particular, is quite multi-faceted and an attacker has different avenues to mount such an attack, ranging from an \textit{untrustworthy foundry}, an \textit{untrustworthy test facility}, to \textit{malicious end-users} (Fig.~\ref{fig:threat_model}).  
Estimates suggest loss to the tune of billions of dollars annually due to infringement of IP cores.
While malicious employees residing in an untrusted foundry or an end-user could pirate the design by reverse engineering (RE) and/or mounting Boolean satisfiability (SAT)-based attacks~\cite{subramanyan15,massad15} to decipher the chip IP, an adversary in the test facility can misuse test patterns to compromise the security of a chip~\cite{yasin16_test,yasin17_TIFS}.
Various design-for-trust schemes have been proposed in the literature (including few which have been demonstrated on silicon) over the past decade to counter IP piracy. 
Table~\ref{tab:protection_comparison_2} summarizes the protection offered by some of these techniques in the face of untrusted entities; they are discussed briefly next.

\subsection{An Overview of IP Protection Schemes}

\textbf{Logic locking} (LL) protects the underlying design IP by inserting dedicated locks, which are controlled by a secret key. 
A locked circuit contains additional inputs, which are referred to as \textit{key inputs}, and are driven by an on-chip \textit{tamper-proof memory} (TPM).
Most common locking mechanisms are realized by inserting additional logic (e.g., XOR/XNOR gates, AND/OR gates or look-up tables (LUTs)).
The locked IC is activated by a trusted facility or the design house after fabrication and testing (but before deploying in the open market), namely by loading the secret key onto the chip's dedicated TPM.
Examples include random logic locking (RLL), fault analysis-based locking (FLL)~\cite{rajendran15}, 
Anti-SAT~\cite{xie2016mitigating}, and stripped functionality logic locking (SFLL)~\cite{yasin17_CCS}.
Note that the overall security of LL hinges on the \textit{secure realization of TPMs}, which remain under active research and development~\cite{anceau17}.

\textbf{Layout camouflaging} (LC) obfuscates the layout implementation---and thereby attempts to obfuscate the
functionality---by using specialized camouflaged cells which aim to be indistinguishable across several functions. 
This can be achieved by (i)~using dummy contacts~\cite{rajendran13_camouflage}, 
(ii)~leveraging threshold voltage-dependent cells~\cite{erbagci16}, 
(iii)~incorporating AND-tree camouflaging~\cite{li16_camouflaging}, and (iv) obfuscating the interconnects~\cite{patnaik2020obfuscating}.
An important consideration for LC is that almost all prior works need to \textit{trust the foundry} for implementing their obfuscation mechanisms.

\textbf{Split manufacturing} (SM) entails the physical separation of the entire chip stack into front-end-of-line (FEOL) and back-end-of-line (BEOL) layers, across geographically distinct foundries.
Typically, the FEOL consists of transistors (device layer) and lower metal layers (M1--M3) which are fabricated by an advanced, off-shore \textit{untrustworthy} foundry, while the remaining metal layers are manufactured on top of the incomplete chip at a \textit{trustworthy}, in-house, low-end facility~\cite{rajendran2013split}.
This physical separation of the design IP avoids dissemination of the complete layout information to one untrustworthy foundry.
A multitude of techniques has been proposed in the recent literature to safeguard FEOL layouts for SM,
e.g.,~\cite{rajendran2013split,patnaik18_SM_ASPDAC,patnaik18_SM_DAC}.
However, it is essential to note that, SM can safeguard the design IP from untrusted foundries only, \textit{but not against untrusted end-users}.

To summarize, although IP protection techniques have been proposed to safeguard the supply chain 
against malicious entities, each of these solutions have some caveats.
Logic locking has the potential to protect the IC supply chain end-to-end but, in its current state, the resilience depends on a TPM to store the secret key. 

\subsection{Role of Emerging Devices in Securing Hardware}

Emerging devices are prime candidates for augmenting hardware security~\cite{bi16_JETC,alasad2017leveraging,patnaik18_GSHE_DATE,patnaik2019spin}. 
The controllable ambipolarity in
SiNW-FETs has been exploited to implement camouflaged 
layouts in~\cite{bi16_JETC}. 
Recent research in the field of emerging device-based security has explored the domain of spintronics~\cite{ghosh2016spintronics}.
Spin devices like the charge-spin logic
and magneto-electric spin-orbit logic (MESO)~\cite{manipatruni2019scalable} possess innate \textit{run-time polymorphism} and 
\textit{post-fabrication reconfigurability} capabilities, which are typically not afforded by CMOS and other emerging devices.
The additive nature of the input spin currents coupled with a magnetic tunnel junction (MTJ)-based differential voltage readout enables these spin devices to implement majority logic directly and exhibit polymorphic characteristics.
Recent works~\cite{patnaik18_GSHE_DATE,patnaik2019spin} on using emerging devices for LC have leveraged polymorphic logic for static camouflaging.
However, the \textit{true potential} of polymorphic devices lies in \textit{dynamic camouflaging}, 
which is unexplored yet---therefore, exploring dynamic camouflaging is the focus of this paper.

\subsection{Dynamic Camouflaging}
Dynamic camouflaging involves obfuscating and switching the device-level functionality \textit{post-fabrication}, as well as \textit{during run-time}, thereby hindering various attacks throughout the IC supply chain.
\ul{We study \textit{dynamic camouflaging} using polymorphic spin devices 
and establish \textit{security} and \textit{computational accuracy} as two entangled design variables, especially for error-tolerant applications such as image processing and 
machine learning.}
We focus on such scenarios as we believe they are meaningful, but we note that polymorphic gates can in
principle result in any arbitrary dynamic behavior. However, such behavior can be impractical,
as it would come along with an excessive loss of computational accuracy.
In other words, we study dynamic camouflaging based on run-time reconfiguration among functionally equivalent or approximately equivalent circuit structures
with the help of polymorphic gates, while maintaining the practicality of such circuits.
For such applications, dynamic camouflaging can thwart both exact~\cite{subramanyan15,massad15} and approximate SAT (\textit{AppSAT})
attacks~\cite{shamsi17}, as we show in this work.

In general, we discuss extensively about securing the supply chain 
end-to-end using spin-based devices, 
and circumventing the risks associated with 
untrusted foundries, test facilities, and end-users (Fig.~\ref{fig:threat_model} and Table~\ref{tab:protection_comparison_2}).

\subsection{Contributions}

The contributions of this work are summarized 
as follows:

\begin{enumerate}

\item We introduce the concept of \textit{dynamic camouflaging} leveraging the inherent functional polymorphism of spin devices. 
Toward this end, we demonstrate the promising security properties pertaining to the MESO device as a representative spin device. We choose the MESO device owing to its superior performance metrics and CMOS compatibility.

\item We propose a secure end-to-end solution to 
counter IP piracy across the distributed IC supply chain, encompassing an untrusted foundry, untrusted test facility, and an untrusted end-user.
This is the \textit{first} work in the context of LC to safeguard the supply chain end-to-end.
Extensive simulations demonstrate the superior resilience of our proposed scheme against state-of-the-art attacks. 

\item From the purview of an untrusted foundry, we show that advanced ``inside foundry'' attacks do not compromise our security claims, as we rely on the concept of \textit{post-fabrication reconfigurability}.

\item The idea of post-fabrication 
reconfigurability is also leveraged to demonstrate resilience against attackers in an 
untrusted testing facility.
By employing \textit{post-test configuration}, we
protect the design IP against test-data mining attacks like \textit{HackTest}~\cite{yasin17_TIFS}. 
We carry out detailed simulations on various test cases for static and dynamic camouflaging.

\item We extend the benefits of dynamic camouflaging, through \textit{dynamic morphing}, to protect also against untrusted
end-users, especially for error-tolerant applications such as image processing.
We show the implications of using approximate SAT-based attacks (\textit{AppSAT})~\cite{shamsi17} for the same.

\item Finally, we project the superior cost in terms of synthesis-level power, performance, and area (PPA) for full-chip camouflaging in contrast 
with other selected, spin-based camouflaging schemes.

\end{enumerate}

\section{Background and Motivation}
\label{sec:background}

Here, we discuss the recent advancements in LC 
along with demonstrated attacks, which have been tailored toward static camouflaging. 
Further, we report on some early studies directed toward the notion of dynamic camouflaging.

\subsection{Static Layout Camouflaging \& SAT-Based attacks}
\label{sec:static_camouflaging}

\begin{figure*}[ht]
\centering
\includegraphics[width=2\columnwidth]{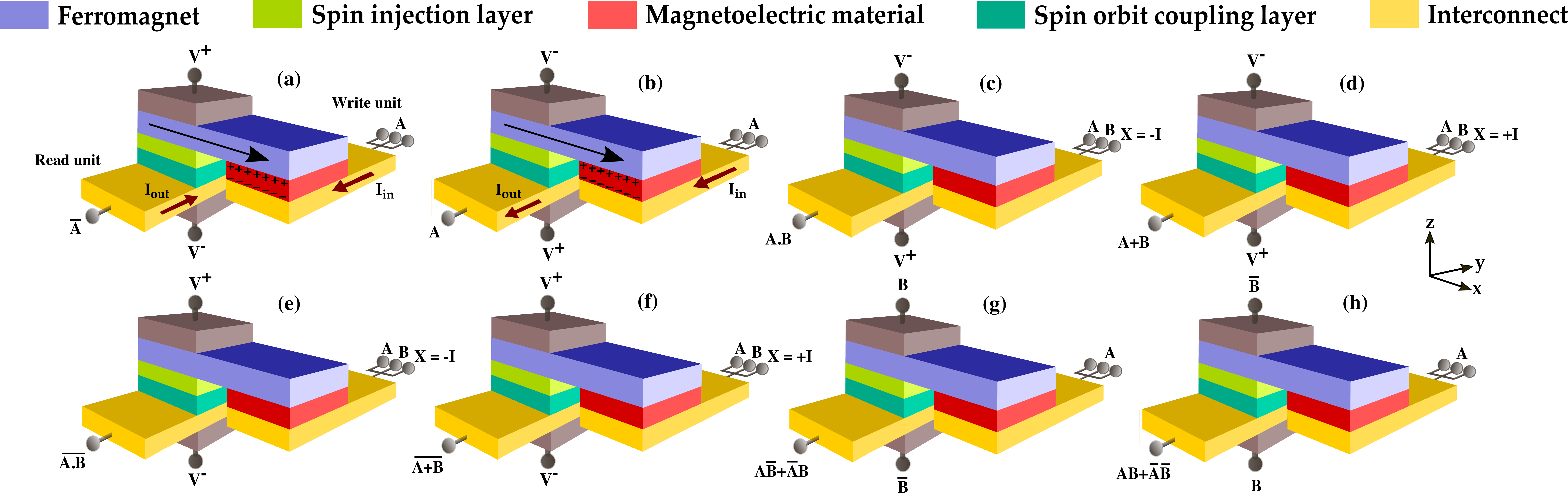}
\caption{(a-h) Implementation of INV, BUF, AND, OR, NAND, NOR, XOR, XNOR with a single MESO device, using different input configurations.
Signals $A$ and $B$ are logic inputs, and $X$ is a control input required for some functionalities.
Note that INV, BUF, XOR, and XNOR gates have dummy wires/contacts at their input terminals, to make them optically indistinguishable from other implementations.}
\label{fig:primitive}
\end{figure*}

Early research in the field of LC was aimed primarily toward (i)~selection of gates to be 
camouflaged, and (ii)~the design of camouflaged cells.
Most of the existing LC schemes have a high layout cost (in terms of PPA) 
and are therefore limited for practical implementation.
The ambiguous XOR-NAND-NOR camouflaged cell proposed
in the seminal work of ~\cite{rajendran13_camouflage} has a power overhead of 5.5$\times$, 
timing overhead of 1.6$\times$, and area overhead 
of 4$\times$, respectively, when compared to a 
conventional 2-input NAND gate.
Promising works such as the threshold-dependent, full-chip LC proposed in~\cite{erbagci16} induces overheads of 14\%, 82\%, and 150\% on PPA, respectively.
Therefore, existing LC schemes can be applied 
only selectively due to their significant impact 
on PPA budgets.
Such a constrained application of these 
techniques (e.g., camouflaging fixed set of gates) leads to a compromise in security,
which is discussed next.
\ul{It should also be noted that most existing camouflaging schemes
necessitate the use of a \textit{trusted} foundry and the camouflaging effected by them is \textit{static}.}

In 2015, Subramanyan \textit{et al.}~\cite{subramanyan15} and Massad \textit{et al.}~\cite{massad15} independently 
challenged the security guarantees offered by LL and LC, respectively.
The attack---commonly referred to as SAT-based attack in the literature---leverages Boolean satisfiability 
to compute so-called \textit{discriminating input patterns} (DIPs).
By definition, a DIP generates different outputs for the same input pattern across two (or more) different keys, which indicates that at least one of the keys is incorrect.
The attack then proceeds in a step-wise fashion
where different DIPs are evaluated until all wrong keys have been eliminated.
Inspired by the promise raised by the SAT-based attack, research groups focused on SAT-resilient camouflaging
schemes~\cite{li16_camouflaging} which force the attack to explore exponential numbers of DIPs.
Such SAT-resiliency is achieved by inserting so-called point functions for, e.g., AND-trees, OR-trees, which ultimately leads to very low output corruptibility.
High-corruptibility schemes like
FLL~\cite{rajendran15} are integrated for such SAT-resilient schemes to improve output corruptibility, thereby providing a two-layer defense.
Shamsi \textit{et al.}~\cite{shamsi17} formulated \textit{AppSAT}, which reduces such compound schemes to their low-corruptibility constituent by ``peeling off'' the high-corruptibility portion. 

\subsection{Toward Dynamic Camouflaging}
\label{sec:toward_dynamic_camo}

Dynamic camouflaging builds on the foundations of polymorphic computing which is
a subset of reconfigurable computing.
Reconfigurable computing using programmable devices (such as field-programmable gate arrays, FPGAs) 
typically fix the logic functionality of the chip 
\textit{before} run-time. 
In polymorphic computing, however, the devices are 
reconfigured in time and space \textit{during} run-time. 
Therefore, dynamic camouflaging involves dynamically obfuscating the circuit at the device/circuit level. 
\ul{Individual gates are configured correctly \textit{only after fabrication and testing}, and these gates can further switch between
different functionalities at run-time by application of certain control inputs---we refer to this approach as \textit{dynamic morphing}.}
Contrary to static camouflaging schemes like~\cite{rajendran13_camouflage,erbagci16,li16_camouflaging,patnaik2020obfuscating}, 
dynamic camouflaging requires polymorphic logic gates.

Prior work using programmable CMOS for IP protection 
leverage reconfigurable logic barriers~\cite{baumgarten2010preventing} 
and reconfigurable key gates~\cite{liu2014embedded}. 
These techniques \textit{do not} use 
functional polymorphism, but rather fix the logic functionality using select lines and/or key-bits. Although it is possible to 
implement functional polymorphism using CMOS-based reconfigurable units, such as LUTs within FPGAs, the overheads incurred by such schemes can be
high, as discussed further in Section~\ref{sec:PPA_cost_analysis}.

The notion of dynamic functional obfuscation was put forward by Koteshwara \textit{et al.}~\cite{koteshwara17}, where sequentially triggered counters are leveraged to provide security guarantees. This scheme requires additional circuitry to alter the key, which is potentially prone to removal attacks. 
Another study leverages \textit{hot-carrier injection} 
to program threshold voltage-based CMOS gates post-fabrication~\cite{akkaya2018secure}.
The authors also showed a proof-of-concept implementation by fabricating obfuscated adders 
in 65-nm bulk CMOS process.
However, they \textit{do not support run-time reconfiguration} and suffer from large PPA 
overheads.
For example, a camouflaged NAND gate incurs power overhead of 9.2$\times$, delay overhead of 6.6$\times$, and area overhead of 7.3$\times$,
    all with respect to a regular 2-input NAND gate.

Run-time polymorphism and, hence, 
dynamic camouflaging is challenging
to implement for CMOS at the device level,
owing to fundamental limits. 
Our scheme enables a radically different solution, wherein we use the unique properties of
spin devices to achieve truly polymorphic chips.\footnote{While we choose the MESO device as a representative example for our work,
the concepts presented in this work can be readily extended to any emerging device which exhibits qualities like functional polymorphism and post-fabrication functionality reconfiguration.}
This is especially useful for error-tolerant applications such as image processing.
We argue that dynamic camouflaging is also
particularly promising for
approximate computing applications, which trade-off computational accuracy for better energy-efficiency
(Sec.~\ref{sec:case_study}).

\section{Dynamic Camouflaging: Working Principle}
\label{sec:working_principle}

\subsection{The Magneto-Electric Spin-Orbit (MESO) Device: Construction and Operation}

The spin device considered in this study is the MESO device, whose operation is based on the phenomena of magneto-electric (ME) switching~\cite{lottermoser2004magnetic}
and inverse spin-orbit effects~\cite{dyakonov1971current}.
The schematic of the MESO device implementing different Boolean functions is shown in Fig.~\ref{fig:primitive}. 
The inputs/outputs are electric currents, and the logical information is encoded in the direction of the current flow. 
A detailed description can be found in~\cite{manipatruni2019scalable}. 

During the writing phase, an input electric
current flowing in the $\pm \hat{y}$ direction through the non-magnetic interconnect sets up
an electric field in the $\pm \hat{z}$ direction within the ME capacitor (red in Fig.~\ref{fig:primitive}). 
The resulting ME field switches the magnetization state of the ferromagnet (purple) along the $\pm \hat{x}$ direction. 
Information is written into the MESO device by transducing 
the input electric current
into the magnetization state of the device. 
Typical room-temperature multiferroics used for the ME capacitor include BiFeO$_3$ and LuFeO$_3$. 
The charge accumulation across an ME capacitor in response to an applied electric field is given as $Q_{\text{ME}}=A_{\text{ME}}(\epsilon_{\text{0}}\epsilon_{\text{mf}}E+P_{\text{mf}})$, where $A_{\text{ME}}$ is the cross-sectional area of the capacitor, $\epsilon_{\text{0}} = 8.85\times 10^{-12}$ F/m is the permittivity of free space, $\epsilon_{\text{mf}}$ is the relative dielectric permittivity of the ME, and $P_{\text{mf}}$ is the saturated ferroelectric polarization.
For the BiFeO$_3$ capacitor considered in~\cite{manipatruni2019scalable}, $A_{\text{ME}}=10^{-16}$ m$^2$, while $\epsilon_{\text{mf}}=54$. 
The electric field to be applied to the ME capacitor to switch it all-electrically is $E=E_{\text{mf}}B_{\text{c}}/B_{\text{mf}}$, where $E_{\text{mf}}= 1.8\times10^6$ V/m refers to the electric switching field, $B_{\text{mf}}=0.03$ T is the exchange bias at switching field, 
and $B_{\text{c}}=0.1$ T is the ME switching field.

After the writing process is complete, which takes $\sim 200$ ps~\cite{manipatruni2019scalable}, 
the supply voltages $V^+$ and $V^-$ are turned on to initiate the reading phase. 
In the reading phase, a spin-polarized current is 
injected into the spin-orbit coupling (SOC) layer, 
which converts the spin current into electric 
current at the output node ($I_{\text{out}}$), 
due to the inverse spin-Hall and inverse Rashba-Edelstein
effects~\cite{shen2014microscopic}. 
These topological effects result in the shifting of the Fermi surface of the high-SOC material 
in k-space. 
This shift causes a charge imbalance and hence a charge current in the Fermi surface, 
in a direction orthogonal to the injected spin density. 
The magnitude of the charge current transduced by 
the SOC layer as a result of the applied spin 
density is given by
\begin{equation}
j_{\text{c}} = \frac{\alpha_{\text{R}}\tau_{\text{s}}}{\hbar} j_{\text{s}}=\lambda_{\text{IREE}} \>j_{\text{s}},
\end{equation}
where $\alpha_{\text{R}}$ is the Rashba coefficient, $\tau_{\text{s}}$ is the spin relaxation time and $\lambda_{\text{IREE}}$ ($\sim1.4\times10^{-8}$m) is the inverse Rashba-Edelstein length of the SOC material~\cite{manipatruni2019scalable}. 

The direction of the output current is determined 
by the polarity of the supply voltages $V^+$/$V^-$ 
(+/- 100 mV) applied on the nanomagnet, and the final
magnetization state of the ferromagnet. 
For instance, when the ferromagnetic moment is along $+\hat{x}$ and the flow of the injected spin current is along $-\hat{z}$, with spin polarization along $+\hat{x}$, the direction of the charge current generated is along $+\hat{y}$ (Fig.~\ref{fig:primitive}a). 
However, when the ferromagnet is reversed to the $-\hat{x}$ direction, with the injected 
spin current direction unchanged 
but the spin polarization now along $-\hat{x}$, 
the output charge current reverses to $-\hat{y}$. 
The same reversal in the direction of output 
current can also be achieved by keeping the ferromagnetic moment constant and flipping 
the voltage polarities $V^{+}/V^{-}$. 

The total intrinsic switching time of the MESO device is a combination of the time taken to charge the multiferroic
capacitor, $\tau_{\text{ME}}$, and the ferroelectric polarization/magnetization reversal time, $\tau_{\text{mag}}$. 
These are given as $\tau_{\text{ME}}= 2Q_{\text{ME}}/ I_{\text{ISOC}}$ and
$\tau_{\text{mag}}= \pi/ \gamma B_c,$
where 
$ I_{\text{ISOC}}$ is the current produced by the spin-orbit
effect and $\gamma$ is the gyromagnetic ratio of the electron. 
Evaluating these switching times according to the parameters provided in the supplementary material of~\cite{manipatruni2019scalable} yields an
intrinsic switching time of $\sim$230 ps. The total switching time of the MESO device is then obtained as $\sim$258 ps, by adding the
interconnect delay of $2.9$ ps (quoted from the supplementary material of~\cite{manipatruni2019scalable}) and the extrinsic peripheral delay of
$\sim$25 ps which corresponds to
multiplexers (MUXes) simulated using \textit{Cadence Virtuoso} for the 15-nm CMOS node, considering the NCSU FreePDK15 FinFET library, for a supply voltage of 0.8V.

For a further, in-depth analysis about the switching and transduction processes in the MESO device, 
interested readers are kindly referred to~\cite{manipatruni2019scalable}. 
Finally, we note that the MESO device has sufficient gain, namely $\sim10$~\cite{manipatruni2019scalable}, to drive multiple fan-out stages.

\subsection{Polymorphic Gates}
\label{sec:polymorphic_gates}

By switching the polarity of the supply voltages, we can implement a buffer (BUF) or an inverter (INV) using the same device (Fig.~\ref{fig:primitive}(a,b)).
Further, we can implement complex gates such as majority logic, by leveraging the additive nature of the input signals. 
As shown in Fig.~\ref{fig:primitive} (c,d), $A$ and $B$ are the signal inputs and $X$ is the tie-breaking control input. 
The polarity of $X$ decides the functionality of the MESO gate. 
Here, for $X=-I$, it realizes an AND gate and for $X=+I$, it realizes an OR gate. 
To implement NAND and NOR gates, the polarities of the supply voltages are flipped 
(Fig.~\ref{fig:primitive} (e,f)). 
For XOR and XNOR gates, the tie-breaking input $X$ is eliminated, and one signal is provided at the input terminal. 
The other input signal is encoded in the voltage domain and applied directly at the $V^+$/$V^-$ terminals 
(Fig.~\ref{fig:primitive} (g,h)). 

\begin{figure}[ht]
\centering
\includegraphics[width=0.9\columnwidth]{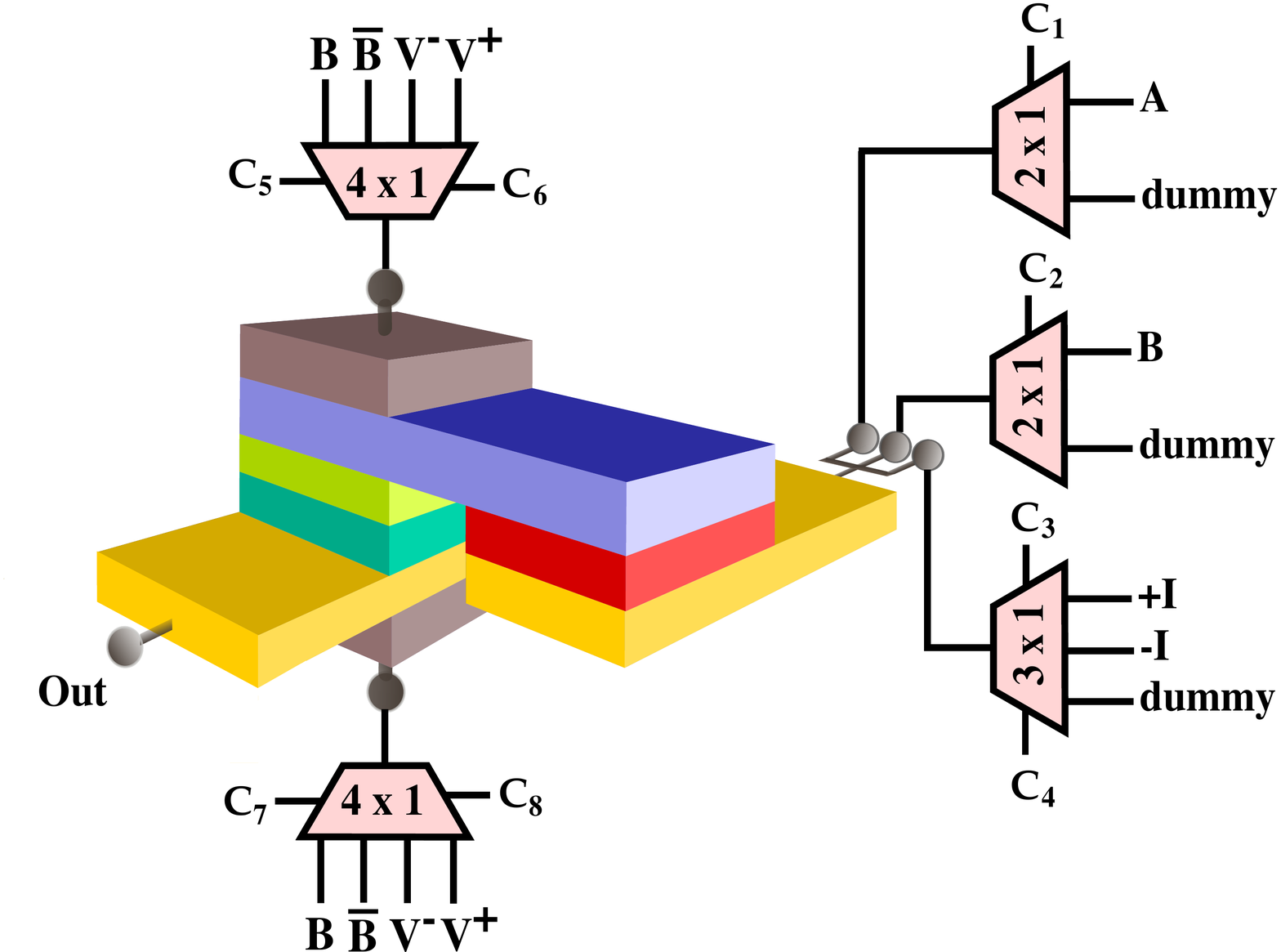}
\caption{A generic MESO gate with peripheral 
MUXes, which dictate the input and control 
signals through control bits $\text{C}_1$--$\text{C}_8$.
This generic structure implements any of the Boolean functionalities in Fig.~\ref{fig:primitive} (a-h) once the appropriate control bits are provided.}
\label{fig:MESO_peripherals}
\end{figure}

\begin{figure}[ht]
\centering
\includegraphics[width=0.9\columnwidth]{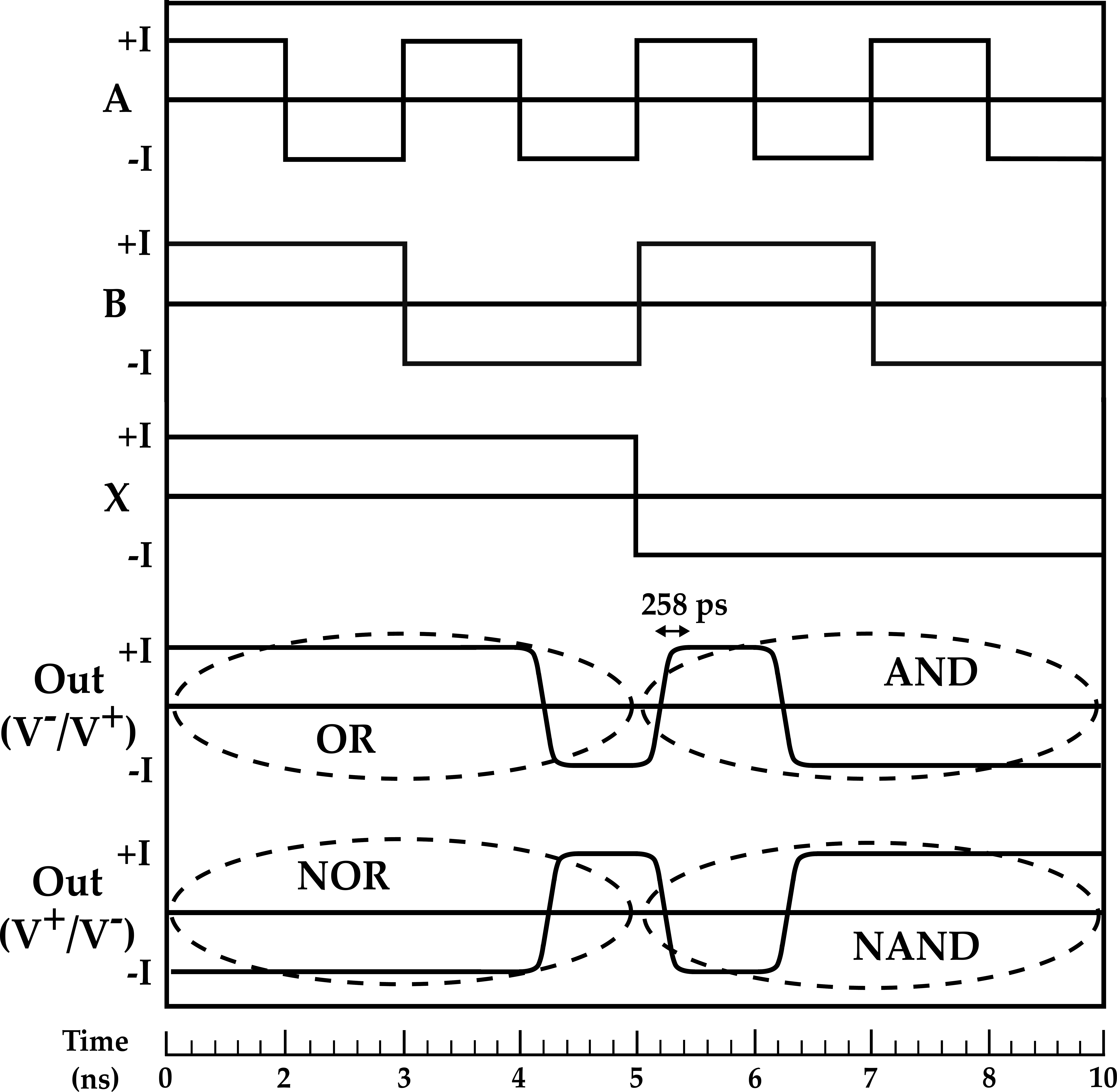}
\caption{Timing waveforms for MESO-based AND / OR / NAND / NOR 
gates from \textit{behavioral Verilog} models,
which represent the estimated overall delays of the MESO device along with their intended functionality.
The MESO primitive's function morphs based on the value of the 
voltage terminals and the control signal $X$. Toggling $X$ allows one to morph between OR $\leftrightarrow$ AND and NOR $\leftrightarrow$ NAND. 
Morphing between OR and NOR involves setting the 
top/bottom voltage terminals as $V^-$/$V^+$ or $V^+$/$V^-$; the converse is true for AND $\leftrightarrow$ NAND morphing.
Note that the morphing time is included in the total switching time of
$\sim$258 ps, in the form of the peripheral MUX delay.}
\label{fig:MESO_Timing}
\end{figure}

\begin{figure}[tb]
\centering
\includegraphics[width=0.8\columnwidth]{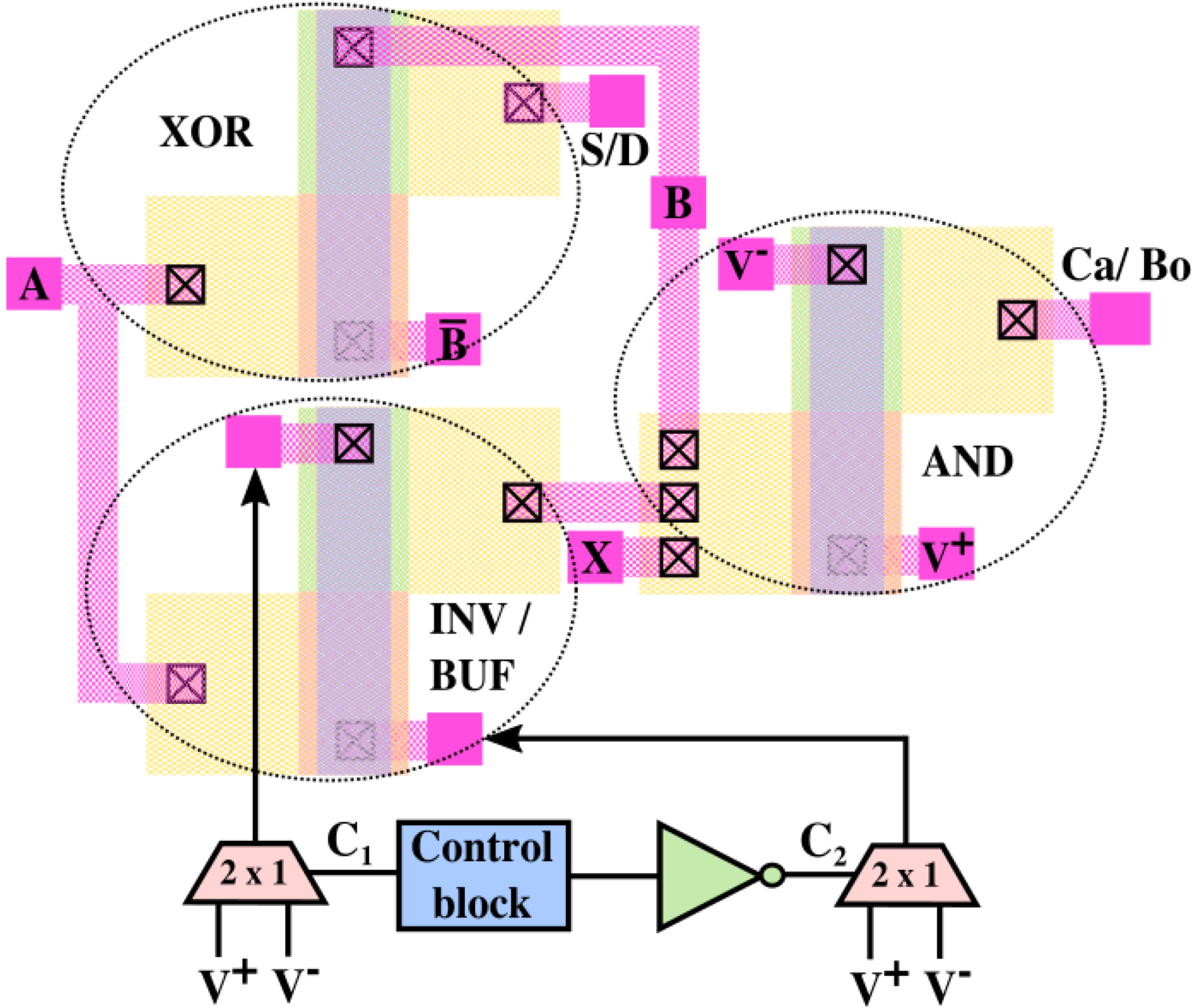}
\caption{MESO adder/subtractor highlighting the capabilities for 
\textit{functional reconfiguration.} 
The XOR and AND gates are implemented as static MESO gates and the INV / BUF is a polymorphic 
MESO gate whose function is derived from control bits ($\text{C}_1$ and $\text{C}_2$) fed by a simple control block. 
\textit{A} and \textit{B} are the inputs, \textit{S} is sum, \textit{D} is difference, \textit{Ca} is carry, 
and \textit{Bo} is borrow. 
Note that dummy contacts are omitted here for the sake of simplicity.}
\label{fig:MESO_adder_subtractor}
\end{figure}

Illustrative waveforms showing the device operation and functional reconfiguration between AND/OR and NAND/NOR, on flipping the control signal $X$, are shown in Fig.~\ref{fig:MESO_Timing}. 
The MESO device with additional peripheral 
circuitry is shown in Fig.~\ref{fig:MESO_peripherals}. 
The \textit{control bits} deciding the input and control signals can either be derived from a control block
(Fig.~\ref{fig:MESO_adder_subtractor}), 
or even from a true random number generator (TRNG),
if random reconfiguration is 
applicable, e.g., for error-tolerant applications 
such as image (video) processing, machine learning, etc.
Configuring the MESO device via different supply voltages and 
electric currents allow us to dynamically implement 
all basic Boolean gates within a single structure. 
This essential feature is used for dynamic camouflaging in this work.

The design house can either provide a fully-camouflaged layout composed of only MESO devices, or a camouflaged layout where selected CMOS gates are replaced by MESO gates.
The MESO device is compatible with CMOS processes 
in the BEOL, enabling heterogeneous integration.\footnote{In general, hybrid \textit{spin-CMOS} designs have been explored in a prior work, e.g., ~\cite{yogendra2015domain}.} 
The proportion of the design camouflaged by a designer depends on the scope of application and impact of camouflaging on PPA overheads. 
The replacement of logic gates can also be performed 
in a manner conducive to protecting the critical infrastructure (i.e., design secrets, proprietary IP).

Please note that the MESO-based primitive can also be leveraged for \textit{static camouflaging}.
In such a scenario, the peripheral circuitry (Fig.~\ref{fig:MESO_peripherals}) dictating the functionality of the MESO device shall be fed with fixed 
control bits and control signals.
Static camouflaging using spin devices has been explored in prior works; interested readers are referred to~\cite{alasad2017leveraging,patnaik18_GSHE_DATE,patnaik2019spin} for further details.

\section{Security Analysis: Untrusted Foundry}
\label{sec:security_analysis_foundry}

\begin{figure}[tb]
\centering
\includegraphics[width=0.7\columnwidth]{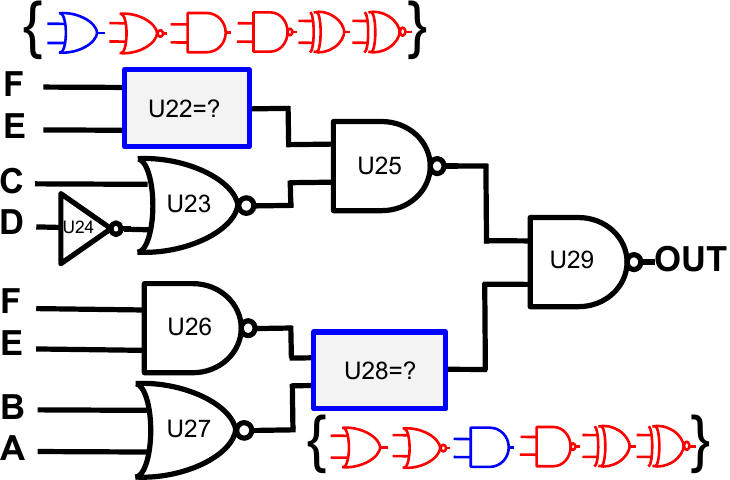}
\caption{An example circuit where two gates (U22 and U28) are camouflaged. The 
camouflaged gates can assume any one of the outlined six 2-input functions. 
The correct functionality of these camouflaged gates is shown in blue.
As the functionality of MESO gates can be reconfigured \textit{post-fabrication}, an attacker's effort of inferring the exact functionality is hindered.
With a random gate-guessing attack, an attacker has 36 possible netlists to consider, with only one amongst them being correct.}
\label{fig:Camo_example}
\end{figure}

An attacker in the foundry can readily infer the IP implemented in CMOS, whereas the MESO gates appear as white-box devices, albeit \textit{without any fixed functionality.}
The MESO implementation of Boolean gates is optically indistinguishable concerning
their physical layout (Fig.~\ref{fig:MESO_adder_subtractor}), which renders optical inspection-guided RE difficult. 
\ul{Since our approach here relies on \textit{post-fabrication reconfigurability}, it is intuitive 
that our scheme is resilient to ``inside foundry'' attacks}. 
As shown in Fig.~\ref{fig:Camo_example}, the \textit{post-fabrication reconfigurability} of MESO gates hinders the attacker's effort to infer the exact functionality. 
A random gate-guessing attack on the circuit shown in Fig.~\ref{fig:Camo_example} has a solution space of 36 possible netlists, 
with only one amongst them being correct.

\subsection{Threat Model}
\label{sec:foundry_threat_model}

The threat model which we adopt for security analysis for an untrusted foundry is outlined as follows:

\begin{itemize}

\item A malevolent employee in the foundry has access to the physical design, including 
material and layout parameters of the MESO gates 
and the chip interconnects.
While an adversary in a foundry can readily obtain 
the dimensions and material composition of the nanomagnet in each MESO gate and, hence,
understand its magnetic properties including saturation magnetization, energy barrier, and critical ME field for switching, 
these design details do not leak any information about the intended 
functionality implemented by the gate.

\item He/she is aware of the underlying gate selection algorithm, number, and type of 
camouflaged gates, but is oblivious to the actual functionality implemented by the camouflaged gate.

\item For security analysis, we assume that the working chip is not yet available in the open market.
Thus, he/she has to apply ``inside foundry'' attacks which are explained briefly next.
    
\end{itemize}

\begin{figure*}[tb]
\centering
\subfloat[]{\includegraphics[width=.24\textwidth]{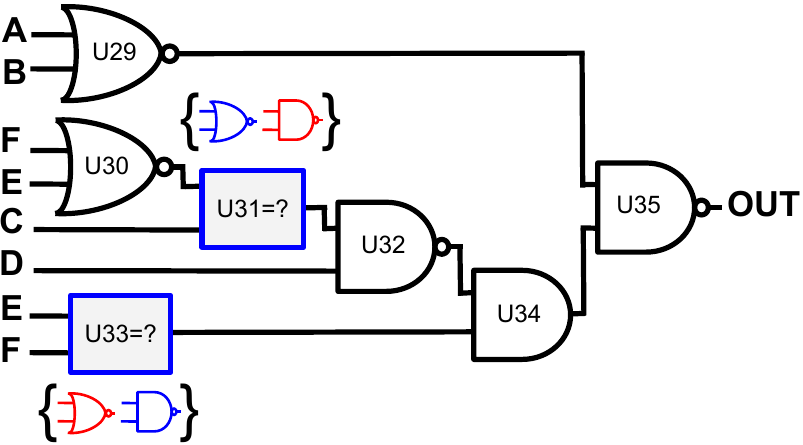}}
\hfill
\subfloat[]{\includegraphics[width=.24\textwidth]{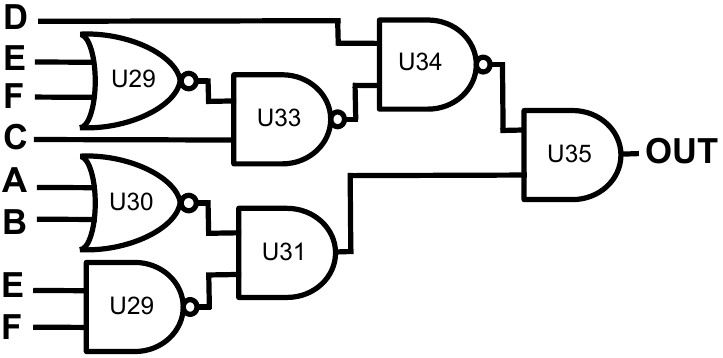}}
\hfill
\subfloat[]{\includegraphics[width=.24\textwidth]{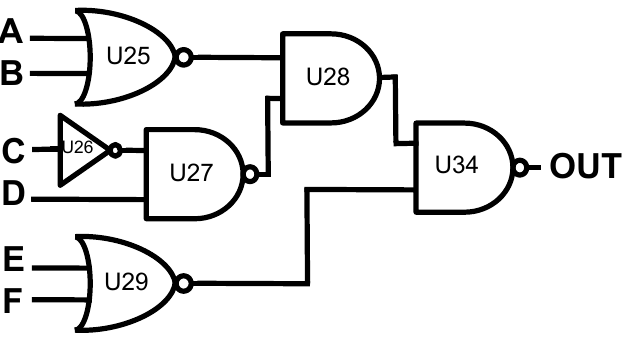}}
\hfill
\subfloat[]{\includegraphics[width=.24\textwidth]{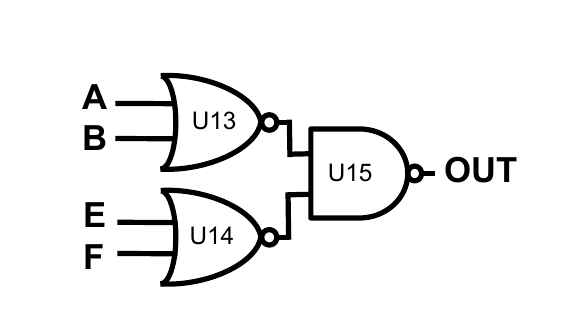}}
\caption{An illustration of an incorrect gate assignment leading to logic redundancy. 
In (a), gates U31 and U33 are camouflaged using a simple NAND/NOR primitive, giving rise to four possible options. 
Correct assignment of camouflaged gates is shown in blue.
Incorrect assignment of gates leads to circuit configurations (b), (c), and (d), respectively.
Note the reduction of gates in (c) and (d) compared to (a), while the gate count is identical in 
(a) and (b), albeit (b) functions differently than (a).}
\label{fig:redundancy_example}
\end{figure*}

\subsection{Attack Model}
\label{sec:foundry_attack_model}

Recently, researchers have proposed attacks~\cite{massad17_CoRR,li2019piercing}, which can be carried out within the confines of an untrusted foundry.
These attacks do not leverage an \textit{activated working chip} as an oracle, which is in contrast 
with algorithmic SAT-based attacks~\cite{subramanyan15,massad15,shamsi17}.
Though these attacks have been primarily tailored toward LL, we believe these would readily apply on LC schemes as well, given that any LL problem can be modeled as an LC scheme and vice-versa.
Besides, for the attacks proposed in~\cite{massad17_CoRR,li2019piercing}, the basic premise is that an incorrect assignment of key-bits
involves significant 
logic redundancies compared to the correct assignment of key-bits.
The attack in~\cite{li2019piercing} determines the likely value of key-bits individually by 
comparing the levels of logic redundancy for each logic value.

\textbf{Example:} We illustrate the effect of an incorrect assignment of key-bits (gates), leading to logic redundancy using a simple example. 
Consider the circuit shown in Fig.~\ref{fig:redundancy_example}(a), logic gates U31 and U33 are camouflaged using a NAND/NOR camouflaging primitive, which leads to four combinations for [U31, U33].
The circuits are shown in Fig.~\ref{fig:redundancy_example}(b--d), 
and they correspond to scenarios 
[U31 = NAND, U33 = NAND], 
[U31 = NAND, U33 = NOR], and 
[U31 = NOR, U33 = NOR], respectively.
After re-synthesis, an incorrect combination of gates deciphered by an attacker leads to circuits with fewer gates (Fig.~\ref{fig:redundancy_example}(c) and Fig.~\ref{fig:redundancy_example}(d)) when compared to the original circuit.
We also note that an attacker might end up with cases like that of Fig.~\ref{fig:redundancy_example}(b), where the total number of gates is same as the original circuit; however, these circuits differ in functionality.

Having no access to these attacks~\cite{massad17_CoRR,li2019piercing},
we refrain from a direct, independent comparison.
However, for the sake of completeness of the security analysis, we perform quantitative experiments, based
on the essence of findings quoted in the respective works of~\cite{massad17_CoRR,li2019piercing}.
For example, the \textit{desynthesis} attack~\cite{massad17_CoRR} can 
correctly infer 23 (up to 29) and 
47 (up to 59) key-bits for 32 and 64 key-gates, respectively, while
the authors of ~\cite{li2019piercing} report success rate in 25\%--75\% percentile distribution.
For a fair comparison, we consider similar ranges of correctly inferred gates.\footnote{Note that this is a powerful assumption for the attacker's capabilities.
This is because the respective attacks~\cite{massad17_CoRR,li2019piercing} tackle LL schemes,
where modeling a locked gate 
requires only one key-bit, whereas for multi-function camouflaging schemes like ours, multiple key-bits are required for modelling one camouflaged gate (Fig.~\ref{fig:Camo_MUX_model}).}

\subsection{Experimental Setup}
\label{sec:foundry_setup}

\begin{figure}[ht]
\centering
\includegraphics[width=0.6\columnwidth]{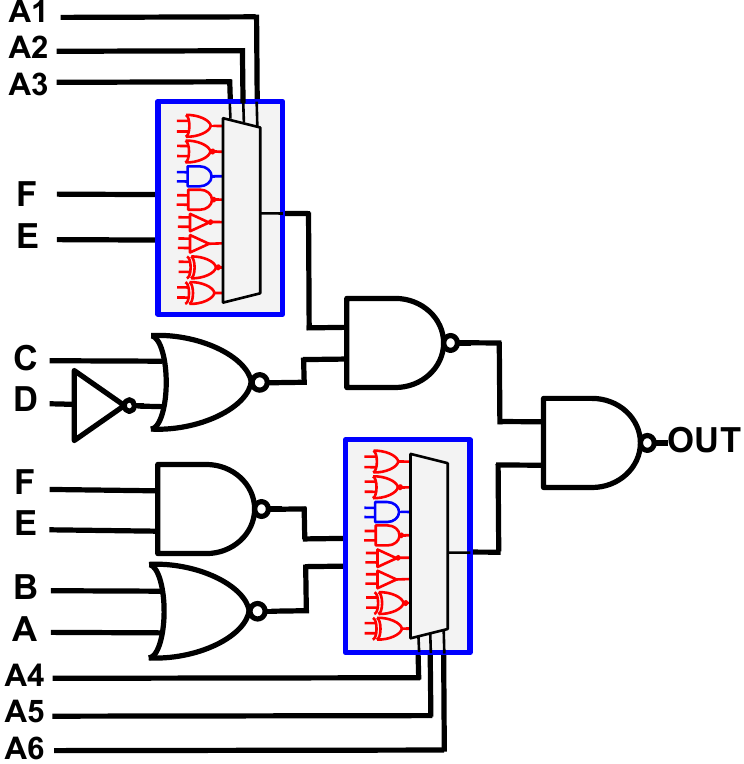}
\caption{Modeling of camouflaged circuits using MUXes.
Each camouflaged gate is replaced with a corresponding MUX which dictates the functionality based on the 
value assigned to the select inputs (A1--A6).
In this example, the 
camouflaged cell can implement any one of the 
eight functions viz. OR, NOR, AND, NAND, INV, BUF, XOR, and XNOR.
This modeling has been used throughout the paper.}
\label{fig:Camo_MUX_model}
\end{figure}

We model the MESO primitive, as 
shown in Fig.~\ref{fig:Camo_MUX_model}.
The logical inputs $a$ and $b$ are fed in parallel into all eight possible Boolean functions, 
and outputs of those gates are connected to an
8-to-1 MUX with three select lines/key-bits.
For a fair evaluation, we camouflage the same set of gates 
for the ISCAS-85 benchmarks c5315 and c7552.
Gates are chosen \textit{randomly} at the beginning and then memorized. Ten such sets are created for each benchmark.
To emulate the attack results from~\cite{massad17_CoRR,li2019piercing}, we employ the following procedure.
We implement a script which randomly picks the correct assignment amongst the camouflaged gates 
such that we obtain three sets, each corresponding to 50\%, 70\%, and 90\% correctly inferred gates.
This procedure is repeated ten times, each for ten different iterations of camouflaged gates, giving us 100 unique trials.
Our camouflaging scheme has been implemented using \textit{Python} scripts operating on \textit{Verilog} files. 
Hamming distance (HD) is computed leveraging \textit{Synopsys VCS} 
with 100,000 input patterns and, functional correctness is ascertained by \textit{Synopsys Formality}.

\subsection{Results}
\label{sec:foundry_exp_results}

\begin{figure*}[tb]
\centering
\subfloat[]{\includegraphics[width=.32\textwidth]{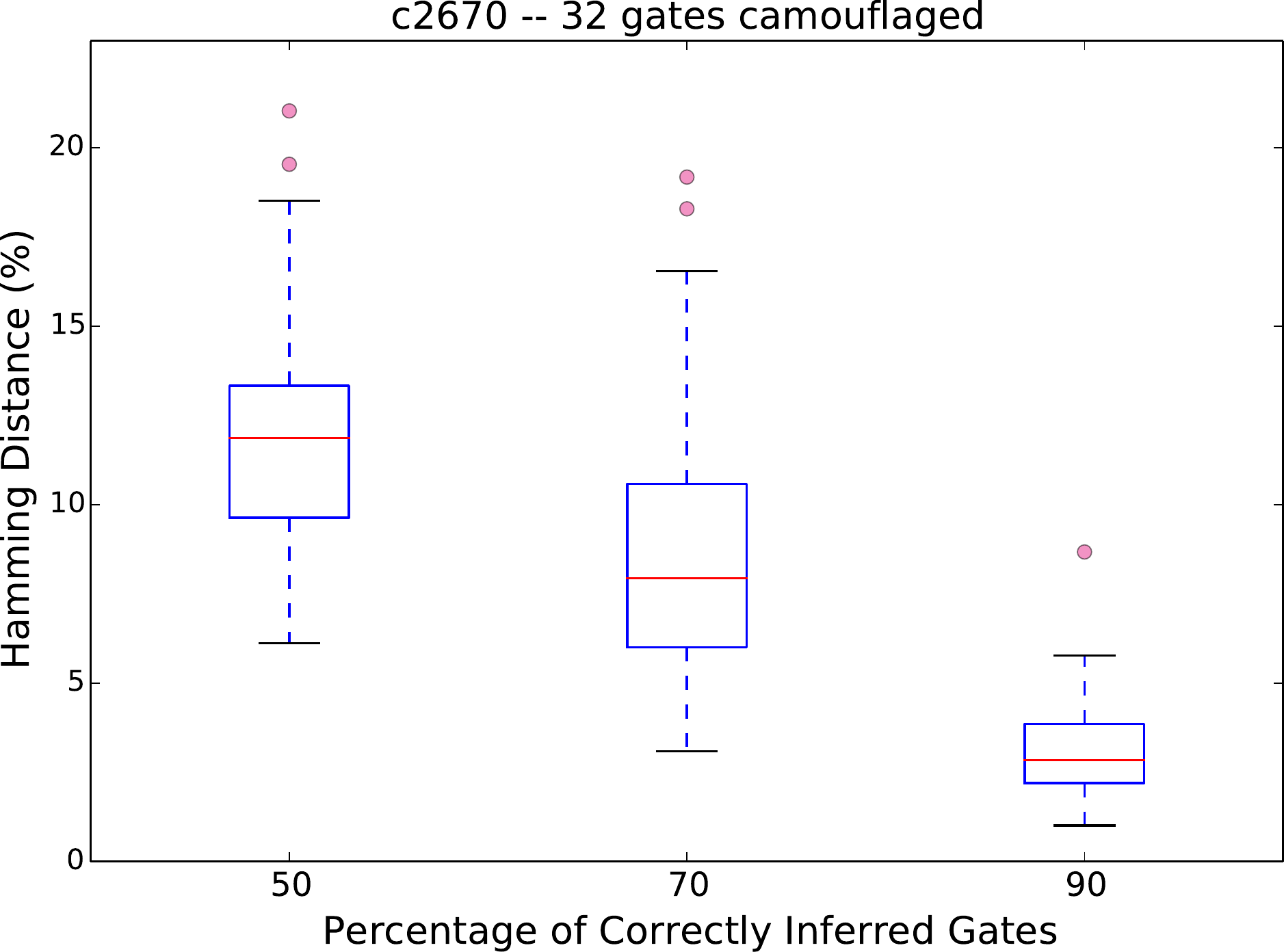}}
\hfill
\subfloat[]{\includegraphics[width=.32\textwidth]{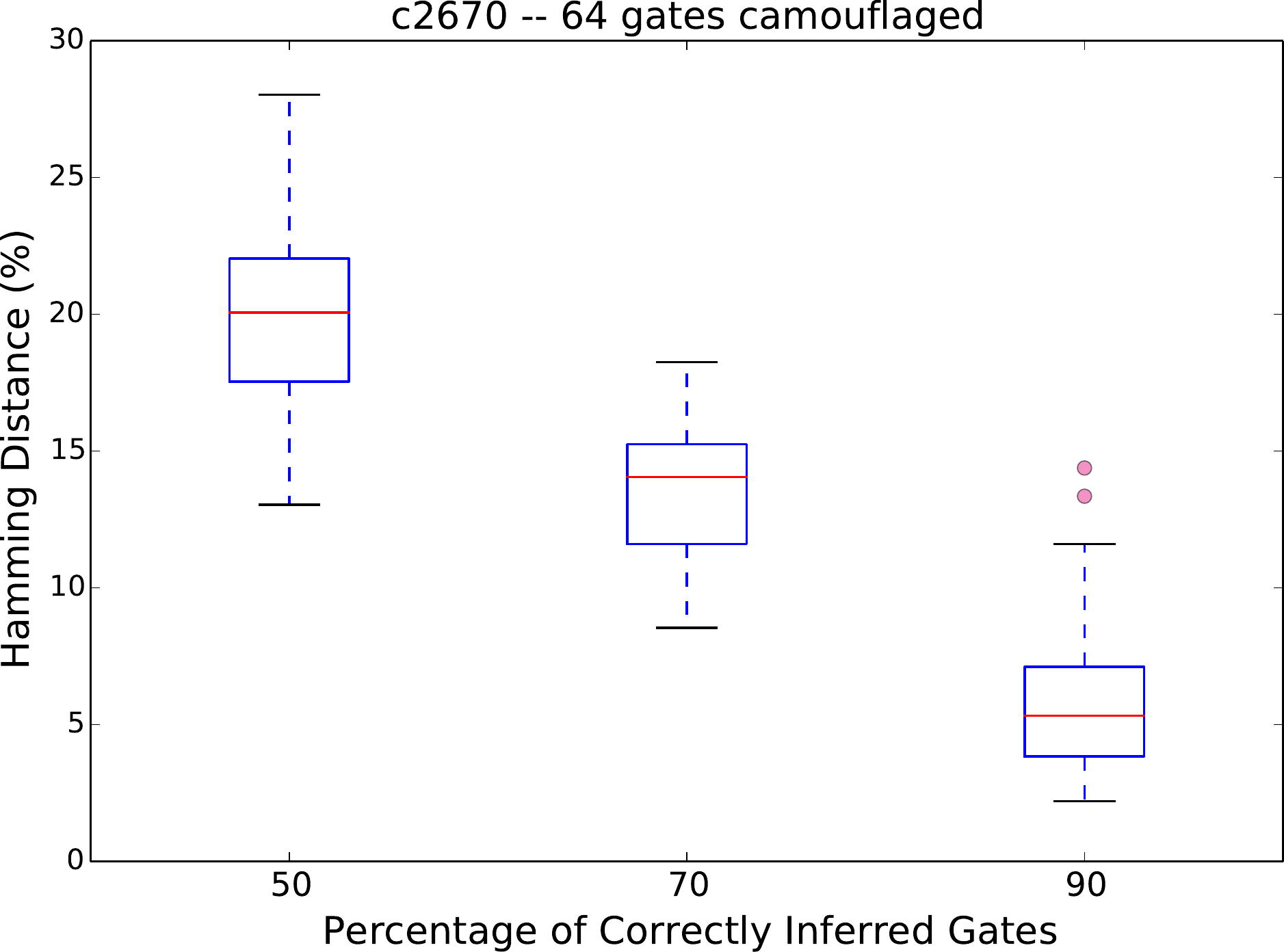}}
\hfill
\subfloat[]{\includegraphics[width=.32\textwidth]{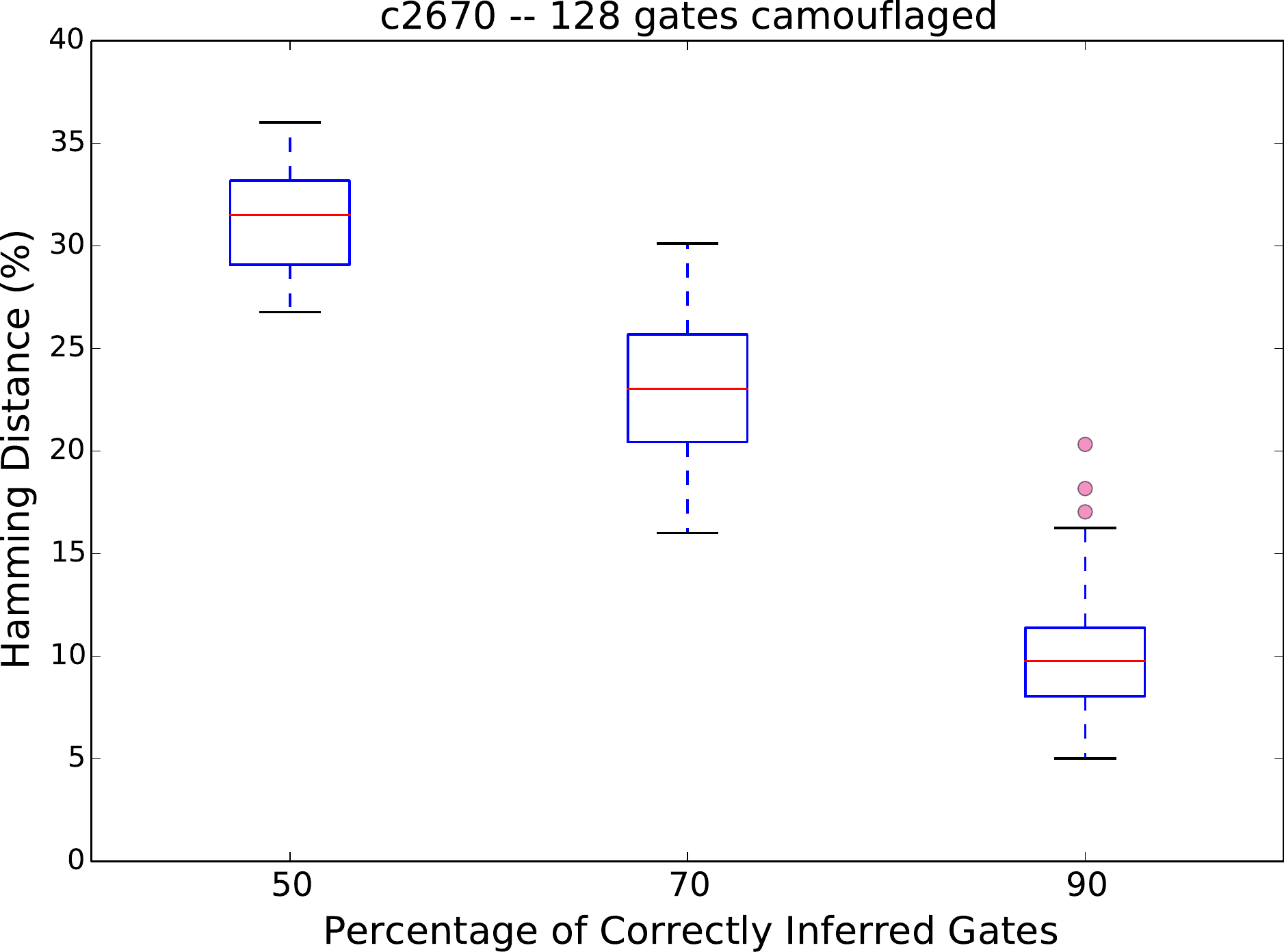}}
\\
\subfloat[]{\includegraphics[width=.32\textwidth]{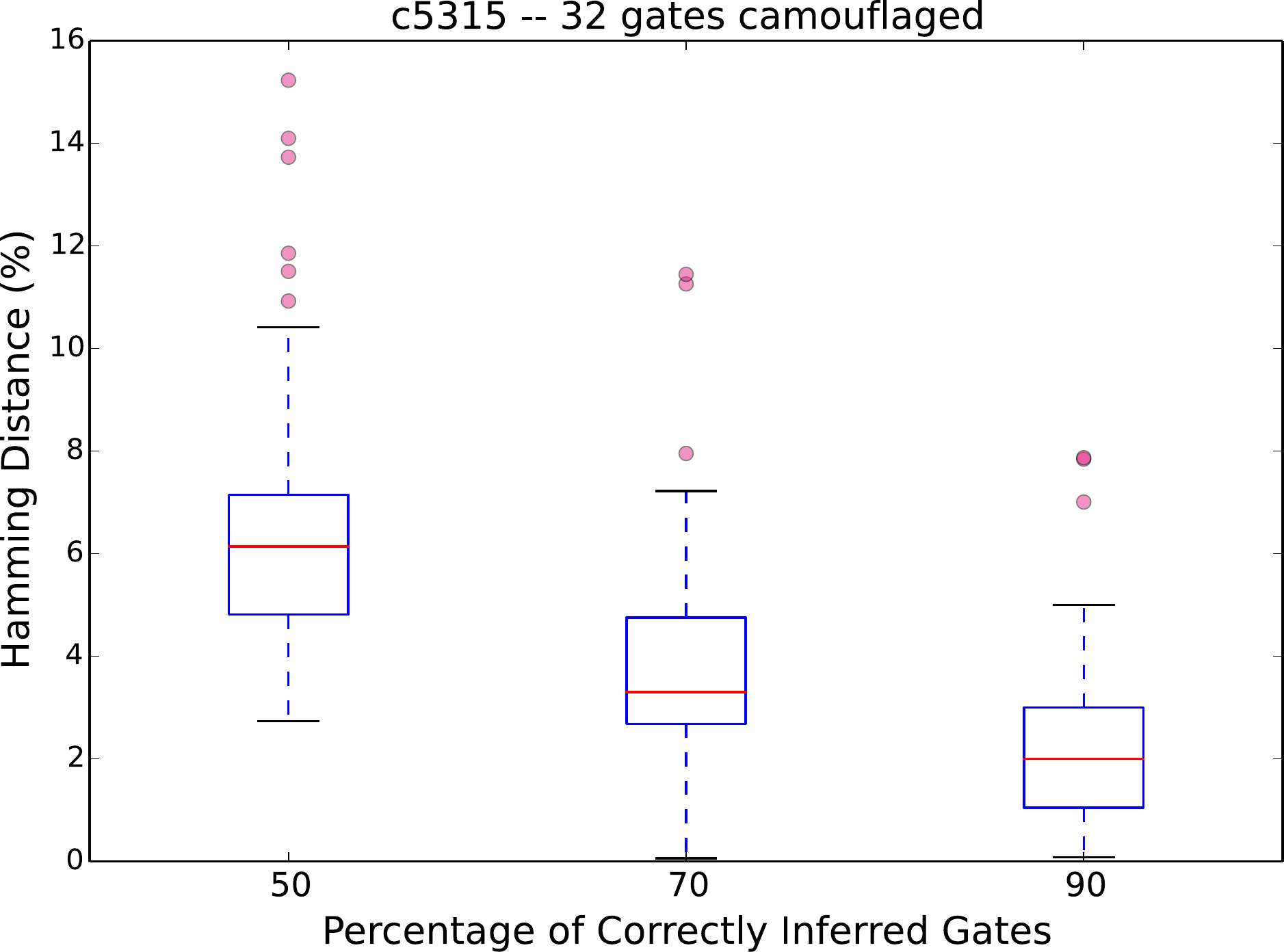}}
\hfill
\subfloat[]{\includegraphics[width=.32\textwidth]{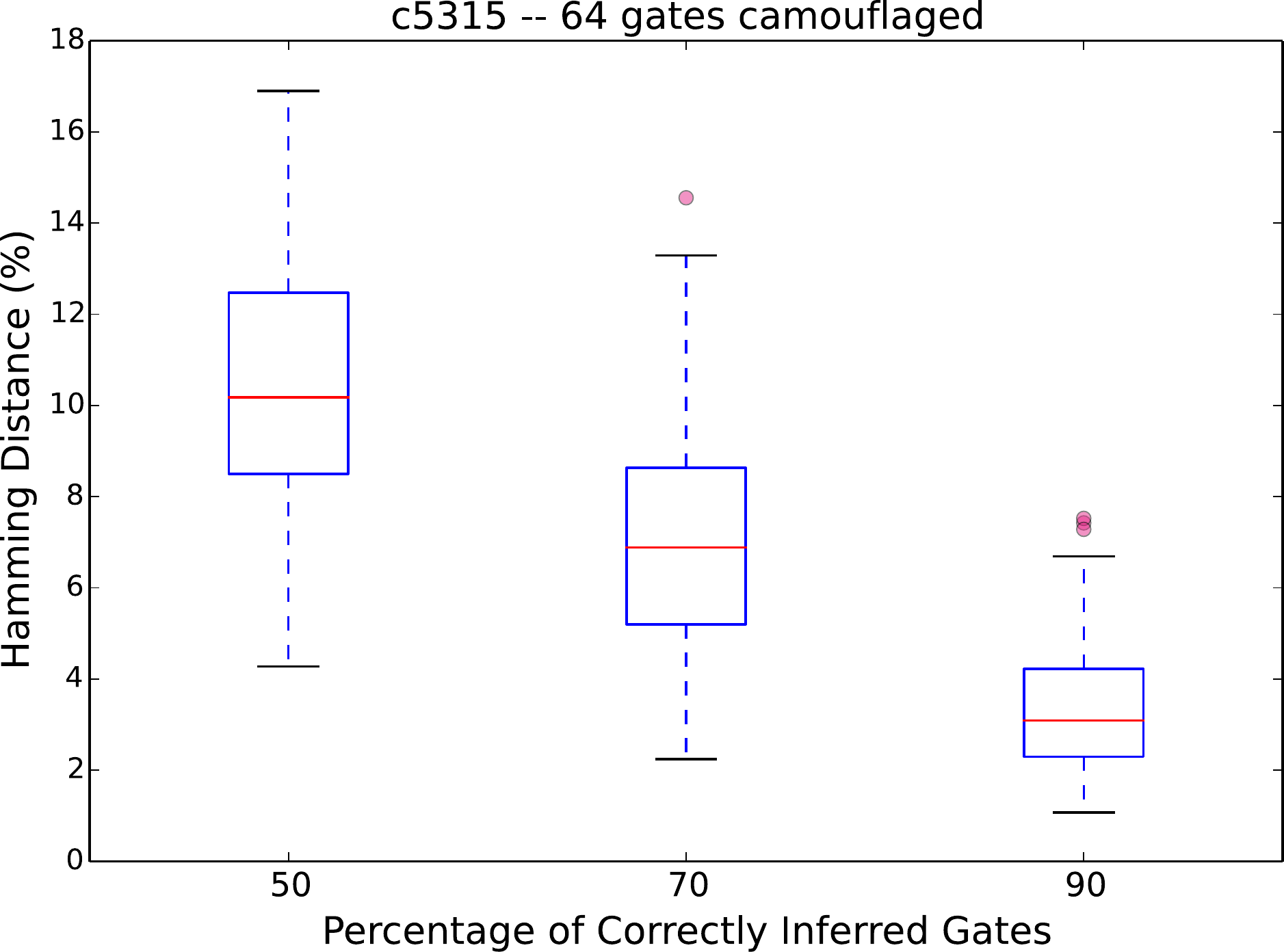}}
\hfill
\subfloat[]{\includegraphics[width=.32\textwidth]{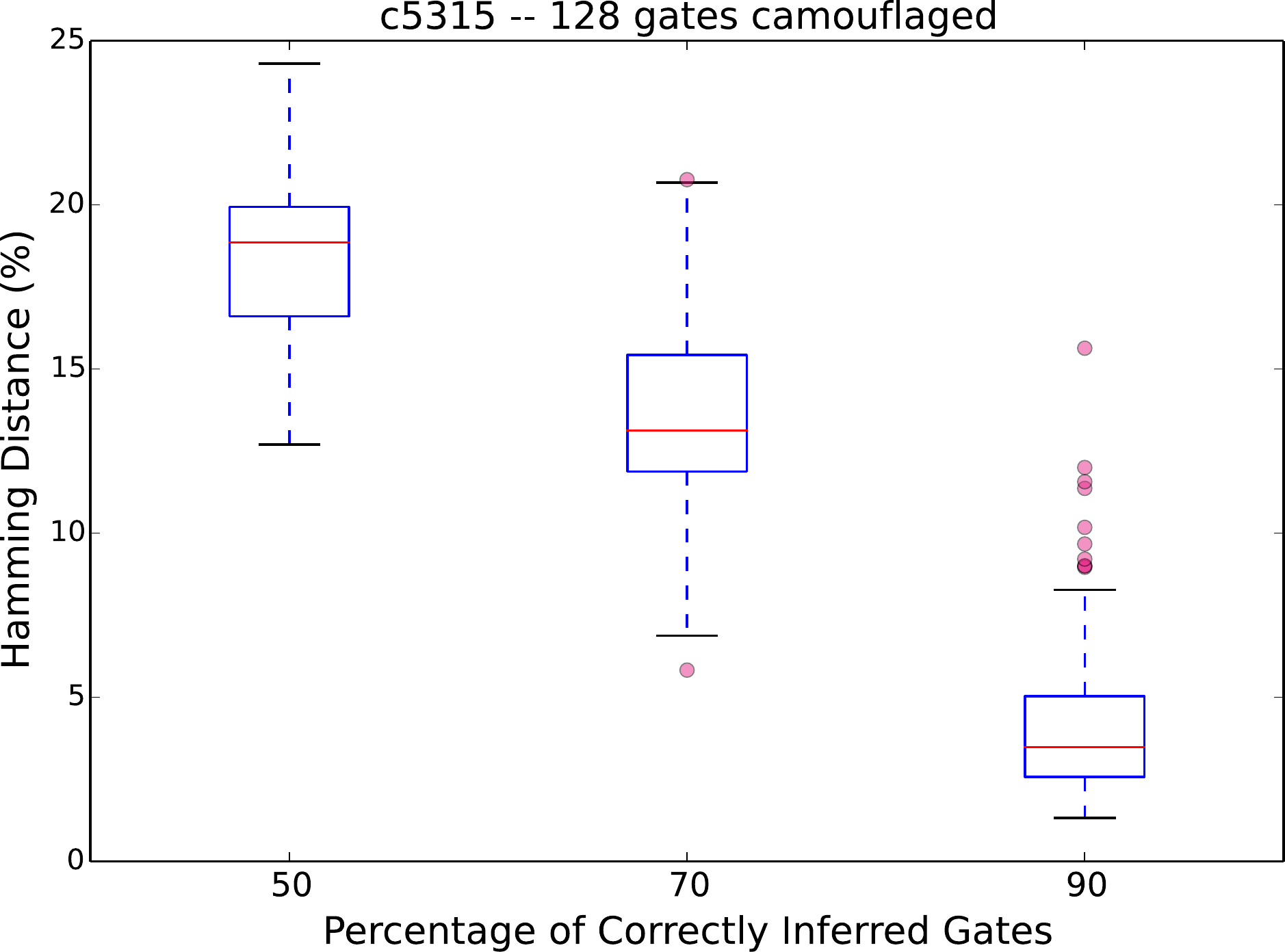}}
\\
\subfloat[]{\includegraphics[width=.32\textwidth]{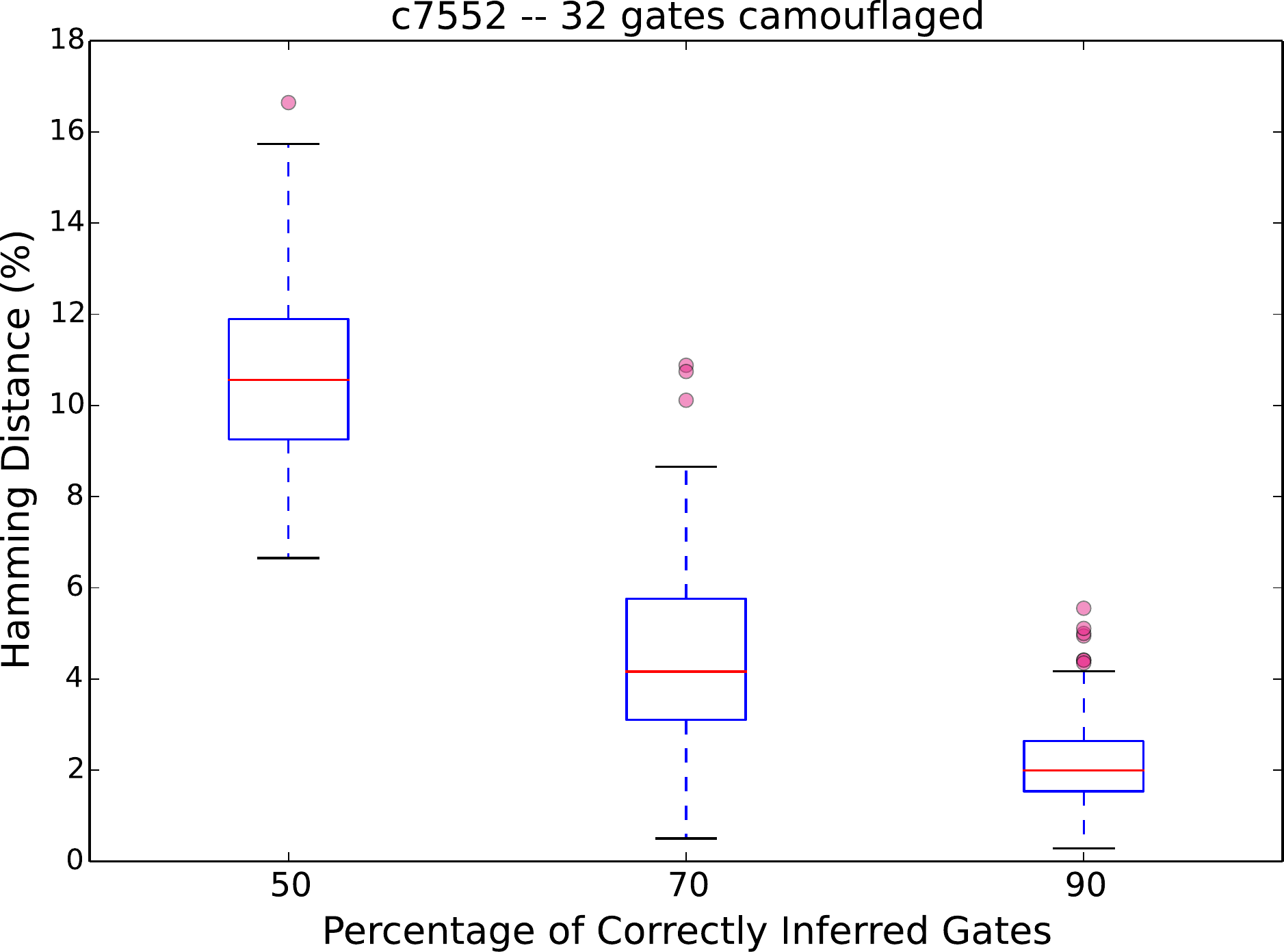}}
\hfill
\subfloat[]{\includegraphics[width=.32\textwidth]{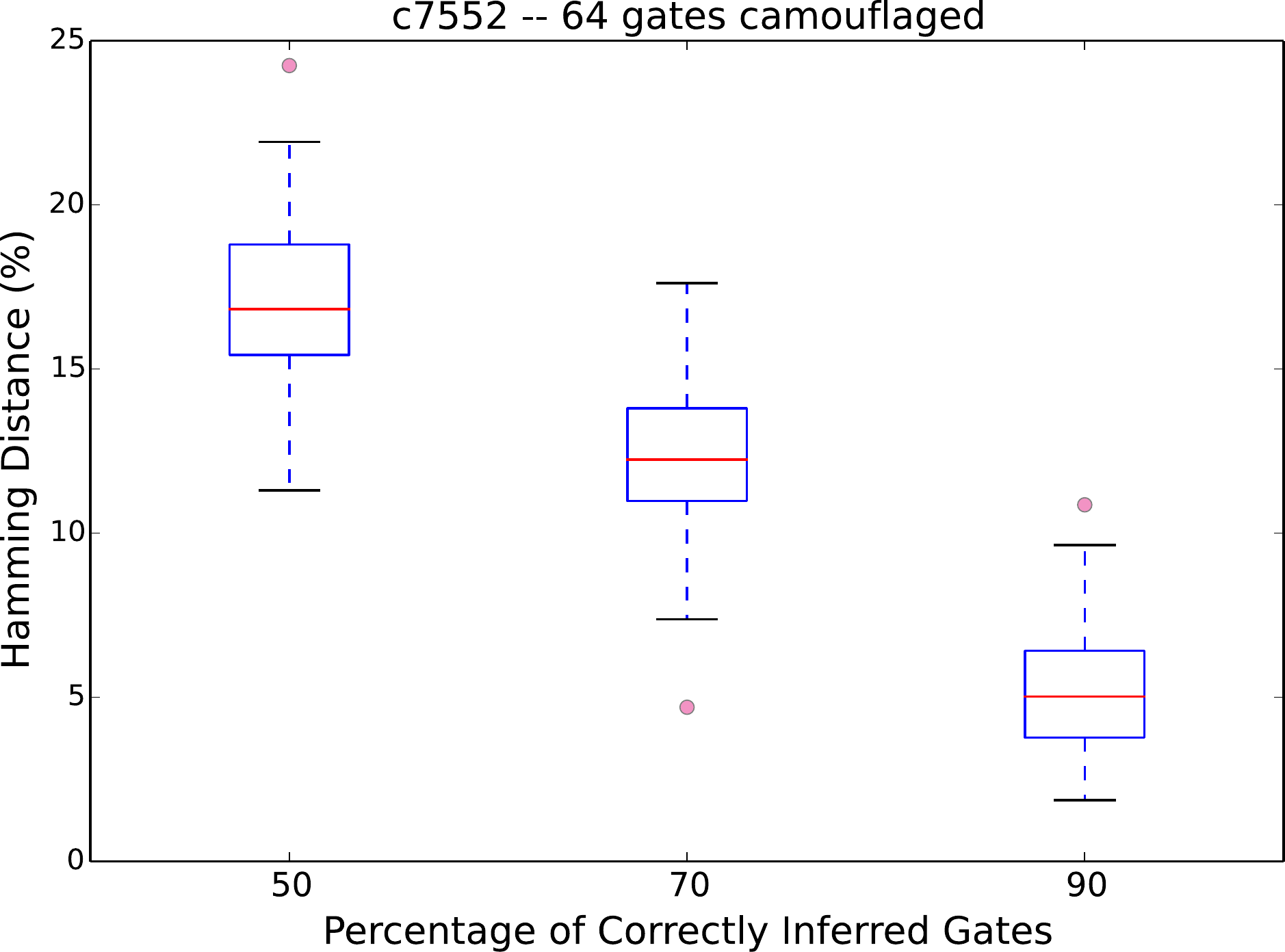}}
\hfill
\subfloat[]{\includegraphics[width=.32\textwidth]{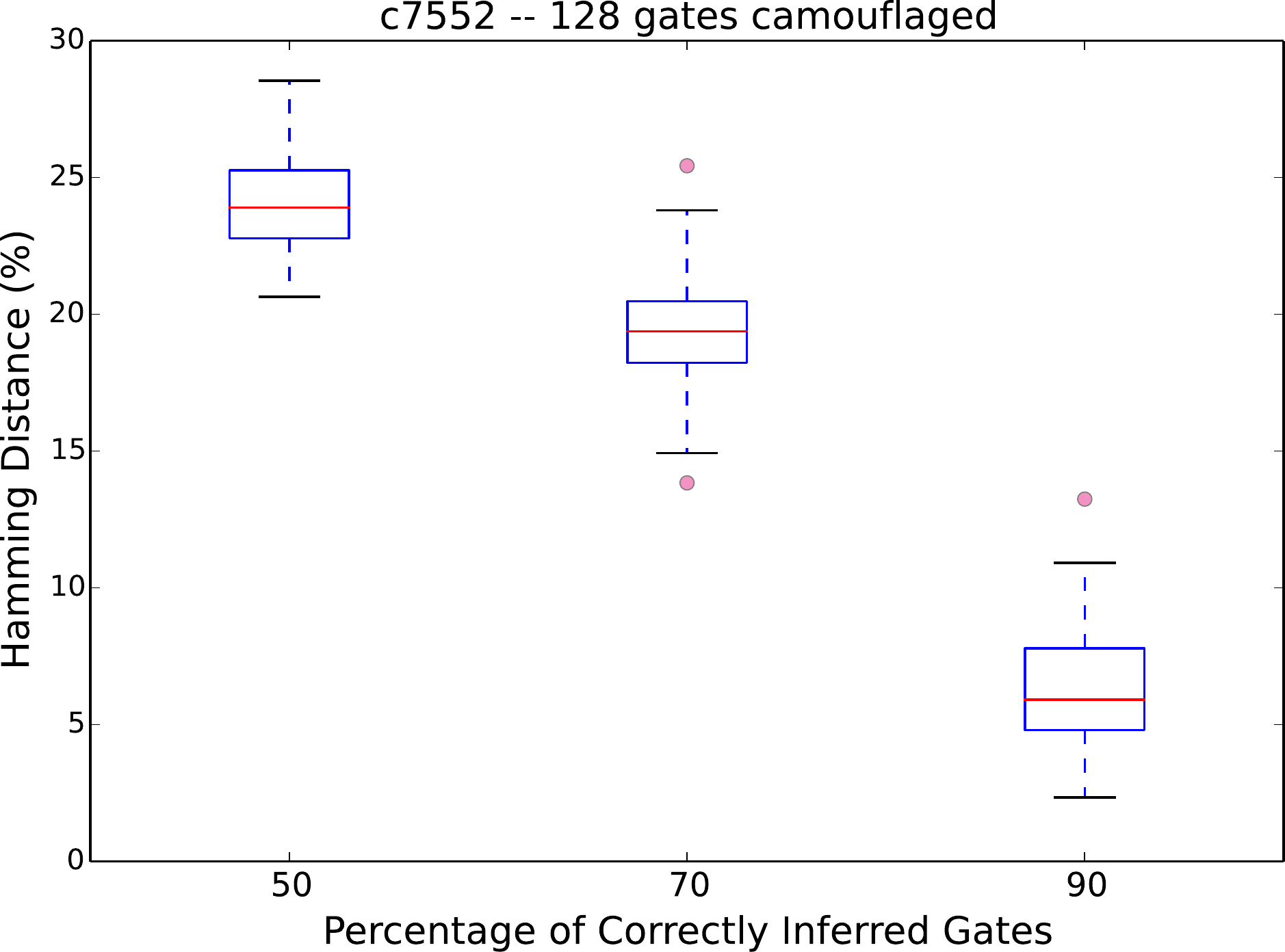}}
\\
\caption{Hamming distance (HD) plotted against the percentage of correctly inferred gates 
(50\%, 70\%, and 90\% of total camouflaged gates) of different sets of camouflaged gates, on selected
ISCAS-85 benchmarks c2670, c5315 and c7552.
Mean HD is proportional to the number of correctly inferred gates amongst the total number of camouflaged gates.
Each box comprises data for 100 trials of random selection of gates to camouflage.}
\label{fig:untrusted_foundry_exps}
\end{figure*}

Once we ascertain the percentage of correctly inferred gates for different levels of attack accuracy (50\%, 70\%, and 90\%), 
we calculate the HD between the reconstructed and the golden netlist. 
The results are shown as box-plots in Fig.~\ref{fig:untrusted_foundry_exps} for two ISCAS-85 benchmarks c5315 and c7552. 
It is intuitive to note that, as the percentage of 
the correctly inferred gates is increased, 
there is a steady reduction in the HD, which also hints that the reconstructed netlist becomes functionally similar to the original circuit.
For the ISCAS-85 benchmark c7552, assuming an attack accuracy of 90\%, the mean HD increases from about 2\% when 32 gates are camouflaged (29 are
inferred correctly) to 5\% when 128
gates are camouflaged (115 are inferred correctly).
Note that such HD numbers could already 
suffice for an attacker recovering an 
approximate version of the original functionality.
However, for attacks which can only recover 50--70\% of the total camouflaged gates, 
the HD for the reconstructed circuit is between 6\% to 25\%,
depending on the size and type of the benchmark,
the number of gates being camouflaged, and the number of gates correctly inferred.
These findings also imply that camouflaging large parts of a design might suffice to thwart ``inside foundry'' attacks, which 
we confirm by a simple experiment as discussed next.
For a larger ITC-99 benchmark like b22\_C, we camouflage 50\% of the total logic gates (7,228 gates) present in the overall design.
Assuming that an attacker can identify 90\% of these gates correctly, this still leaves 722 gates wrongly inferred, which yields an HD of 43\% (across ten random trials).
\ul{Overall, the property of \textit{post-fabrication reconfigurability} for the MESO gates 
allow us to change the functionality, enabling superior security through dynamic camouflaging.}

\section{Security Analysis: Untrusted Test Facility}
\label{sec:untrusted_test}

Attackers present in the test facility having access to test patterns and corresponding output responses (generated and supplied by the trusted design house), can jeopardize the security guarantees offered by LL and LC.
Modern Automatic Test Pattern Generation (ATPG) algorithms have been designed to maximize the fault coverage (FC) with minimal test pattern count, which directly translates to a lower test cost. 
Such an approach, however,
divulges critical information pertaining to the internal circuit specifics~\cite{yasin17_TIFS}.

\begin{figure}[ht]
\centering
\includegraphics[scale=0.85]{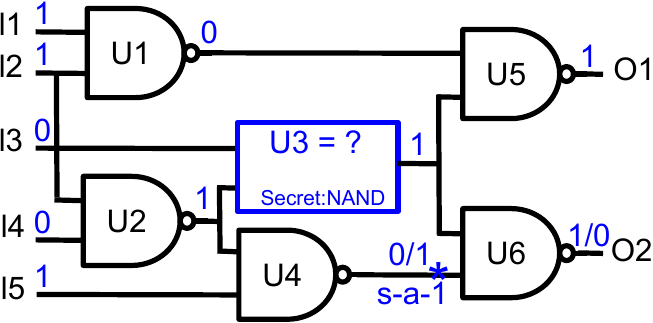}
\caption{An input pattern which helps in the detection of stuck-at-1 fault at the output of U4. The circuit output `1/0' at output O2 indicates the response for fault-free/faulty circuit are 1 and 0,
respectively. 
The input pattern 11001, along with the expected output response 11 is
provided to the test facility for testing manufactured ICs. 
The test data hints that U3 cannot be NOR.
Note that, here we assume 
that the camouflaged gate can function only as NAND/NOR.}
\label{fig:example_Test_Camo}
\end{figure}

In the context of VLSI testing principles, 
detection of a stuck-at-fault
involves two principal 
components, namely (i)~fault activation and 
(ii)~fault propagation.
In \textit{fault activation}, a faulty node is assigned a value opposite to the fault induced on that particular node. 
Consider the example shown in Fig.~\ref{fig:example_Test_Camo}; here, the output
of logic gate U4 is \textit{s-a-1} (stuck-at-1).
In order to detect this fault, fault activation is achieved by setting this node 
to logic '0'.
Next, \textit{fault propagation} entails
propagating the effect of the fault along a sensitization path to one of the primary outputs.
To achieve fault propagation (here to O2), the output of U3 must be '1'.
An input pattern which can detect a fault at a 
given node by achieving the effects mentioned above is defined as a \textit{test pattern}.
In Fig.~\ref{fig:example_Test_Camo}, the input pattern \textit{11001} and the corresponding output response \textit{11} is supplied to the test facility, among others.

\subsection{Threat Model}
\label{sec:testing_threat_model}

Apart from outsourcing of chip fabrication, many design companies also outsource the testing phase to off-shore companies such as Amkor, ASE, SPIL, etc.~\cite{yasin17_TIFS}. 
The implications of
an untrusted test facility in the supply chain have been explored 
in the context of 
LL~\cite{yasin16_test} and static LC~\cite{yasin17_TIFS}.
However, there has been no thorough analysis yet on the efficacy of test data-based attacks~\cite{yasin17_TIFS} for different 
static camouflaging schemes as well as 
dynamic camouflaging.
In our threat model, the attacker
resides in the test facility and has access to
the following assets:

\begin{itemize}

\item Gate-level camouflaged netlist, e.g., obtained by RE.

\item Knowledge of the test infrastructure, which includes identification of 
scan chains, compressor, decompressor, et cetera, on the target chip.

\item Test patterns and their corresponding output responses, which have been provided by the design house.
He/she also has access to ATPG tools used to generate the test patterns.

\end{itemize}

\subsection{Attack Model}
\label{sec:testing_attack_model}

Yasin \textit{et al.}~\cite{yasin17_TIFS} proposed \textit{HackTest}, which revealed the true identity of camouflaged gates within minutes 
by exploiting test data.
The attack leverages the fact that the generation of test patterns is typically tuned to obtain the highest possible FC.
Hence, given the test stimuli and responses,
an attacker can search over the key space (using optimization techniques) to infer the correct assignment of camouflaged gates which maximizes the FC.
Arguably, such an attack is more powerful than SAT-based attacks~\cite{subramanyan15,massad15} which require access to a working chip.

The process of ATPG is highly dependent on the internal specifics of the underlying circuit, which include the type and count of gates, the inter-connectivity amongst these gates, etc.
Years of research have yielded powerful algorithms which lower the test pattern count 
while achieving a high FC. 
However, these algorithms do not factor in security (yet), and thereby, test patterns
become a rich source of information for an opportunistic attacker.
Next, we briefly explain the notion of \textit{HackTest} with a simple example; interested readers are kindly referred to~\cite{yasin17_TIFS} for further details.

\textbf{Example:} Upon performing ATPG for the circuit shown in Fig.~\ref{fig:Camo_example}, for the correct assignment of two camouflaged
gates (U22 = OR and U28 = AND), eight test patterns are generated by \textit{Synopsys Tetramax}, providing a fault and test
coverage of 100\%.
Camouflaging two gates with two functions each gives rise to four possible circuit configurations;
Table~\ref{tab:fault_covergae_init} denotes the 
FC for these configurations.
Armed with input patterns and 
corresponding output responses, both tailored for the correct assignment, an attacker calculates 
the FC for all possible circuit configurations.  
As shown in Table~\ref{tab:fault_covergae_init},
maximal FC is observed only for the correct assignment of camouflaged gates.
This is because, for \textit{static camouflaging}, test patterns have to be generated for the correct assignment of camouflaged gates.  
An attacker can easily use FC to guide his/her attack to identify the correct functionality of camouflaged gates.

\begin{table}[ht]
\centering
\footnotesize
\caption{Fault coverage achieved for different assignments to the camouflaged netlist in Fig.~\ref{fig:Camo_example}. 
Here we assume U22 and U28 are implementing either AND/OR.
The correct assignment
is OR and AND, respectively;
note that other assignments
result in significantly lower fault coverage.}
\label{tab:fault_covergae_init}
\begin{tabular}{ccc}
\hline
\textbf{U22} & 
\textbf{U28} & 
\textbf{Fault Coverage (\%)} \\ 
\hline 
AND & AND & 63.33 \\ \hline
AND & OR &  38.33 \\ \hline
OR & AND & 100 \\ \hline
OR & OR &  78.33 \\ \hline
\end{tabular}
\end{table}

\subsection{Experimental Setup}
\label{sec:testing_setup}

We launch \textit{HackTest} on selected benchmarks of ISCAS-85 and ITC-99 suite.
Statistics of benchmarks like the
number of logic gates (\# Gates),
number of faults (\# Faults), 
number of test patterns generated by \textit{Synopsys Tetramax} (\# Test patterns), and 
corresponding FC are shown in Table~\ref{tab:stats_BM_testing}.
We implement the MESO-based camouflaging primitive along with some selected prior art~\cite{rajendran13_camouflage,bi16_JETC}.
As \textit{HackTest} requires a \textit{BENCH} file format, we employ custom scripts to convert \textit{Verilog} files to required formats.
For the small-scale ISCAS-85 benchmarks,
we prepare ten random sets each for camouflaging 32, 64, and 128 gates, respectively.
For the large-scale ITC-99 benchmarks, we camouflage 350 gates. 
For the sake of uniformity in comparison, the selection of camouflaged gates is random but fixed, i.e., they are common and maintained across all benchmarks for any given camouflaging scheme.
The attack is implemented using custom \textit{Python} scripts executing within 
\textit{Synopsys Tetramax}.
All attack experiments are carried out on an
Intel Xeon E5-4660 @ 2.2 GHz with \textit{CentOS 6.9} and the time-out (t-o) is set to 24 hours.

\begin{table}[ht]
\centering
\footnotesize
\caption{Statistics of ISCAS-85 and 
ITC-99 benchmarks used in this work. 
All benchmarks achieve 100\% test coverage and achieve exactly/close to 100\% fault coverage.}
\label{tab:stats_BM_testing}
\setlength{\tabcolsep}{1mm}
\begin{tabular}{ccccc}
\hline
\textbf{Benchmark} & 
\textbf{\# Gates} & 
\textbf{\# Faults} & 
\textbf{\# Test Patterns} &
\textbf{Fault Coverage (\%)} \\ 
\hline 
c880 & 273 & 1,764 & 63 & 100 \\ \hline
c1908 & 230 & 1,462 & 80 & 100 \\ \hline
c2670 & 433 & 2,936 & 134 & 100 \\ \hline
c3540 & 814 & 5,472 & 177 & 100 \\ \hline
c5315 & 1,232 & 7,708 & 124 & 100 \\ \hline
c7552 & 1,197 & 7,474 & 167 & 100 \\ \hline
b14\_C & 4,125 & 24,668 & 470 & 99.99 \\ \hline
b15\_C & 6,978 & 42,310 & 812 & 99.96 \\ \hline
b20\_C & 9,226 & 54,894 & 897 & 99.89 \\ \hline
b22\_C & 14,457 & 85,852 & 1,356 & 99.95 \\ \hline
\end{tabular}
\end{table}

\subsection{Results}
\label{sec:testing_exps_results}

Next, we detail our observations on employing \textit{HackTest} for various test cases.
We begin by examining the impact of \textit{HackTest}
on various static camouflaging schemes. 
Finally, we enumerate our findings concerning the resiliency of \textit{dynamic camouflaging}, which is the main focus of this work.

\subsubsection{HackTest on Static Camouflaging}

\begin{figure*}[tb]
\centering
\subfloat[]{\includegraphics[width=.32\textwidth]{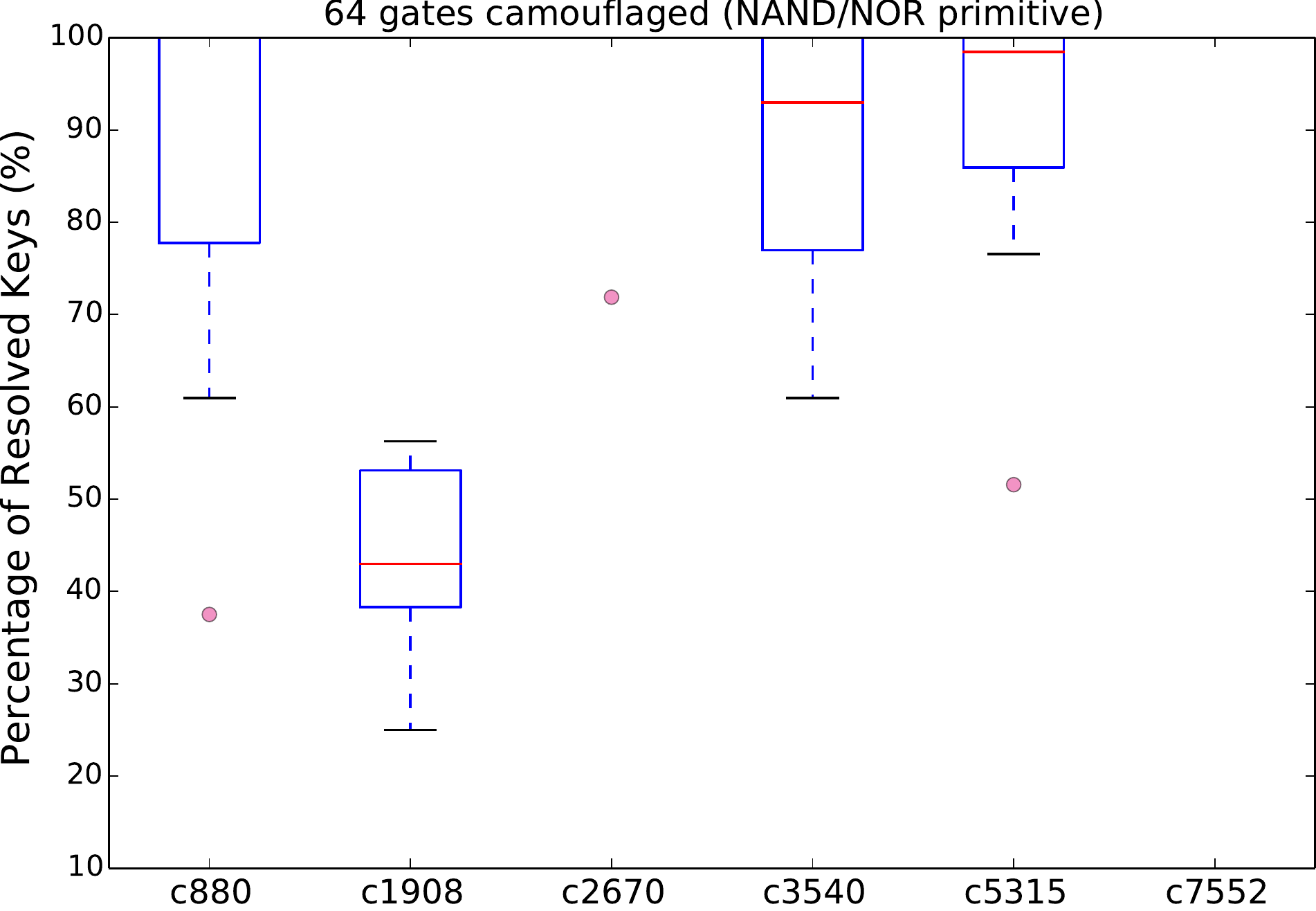}}
\hfill
\subfloat[]{\includegraphics[width=.32\textwidth]{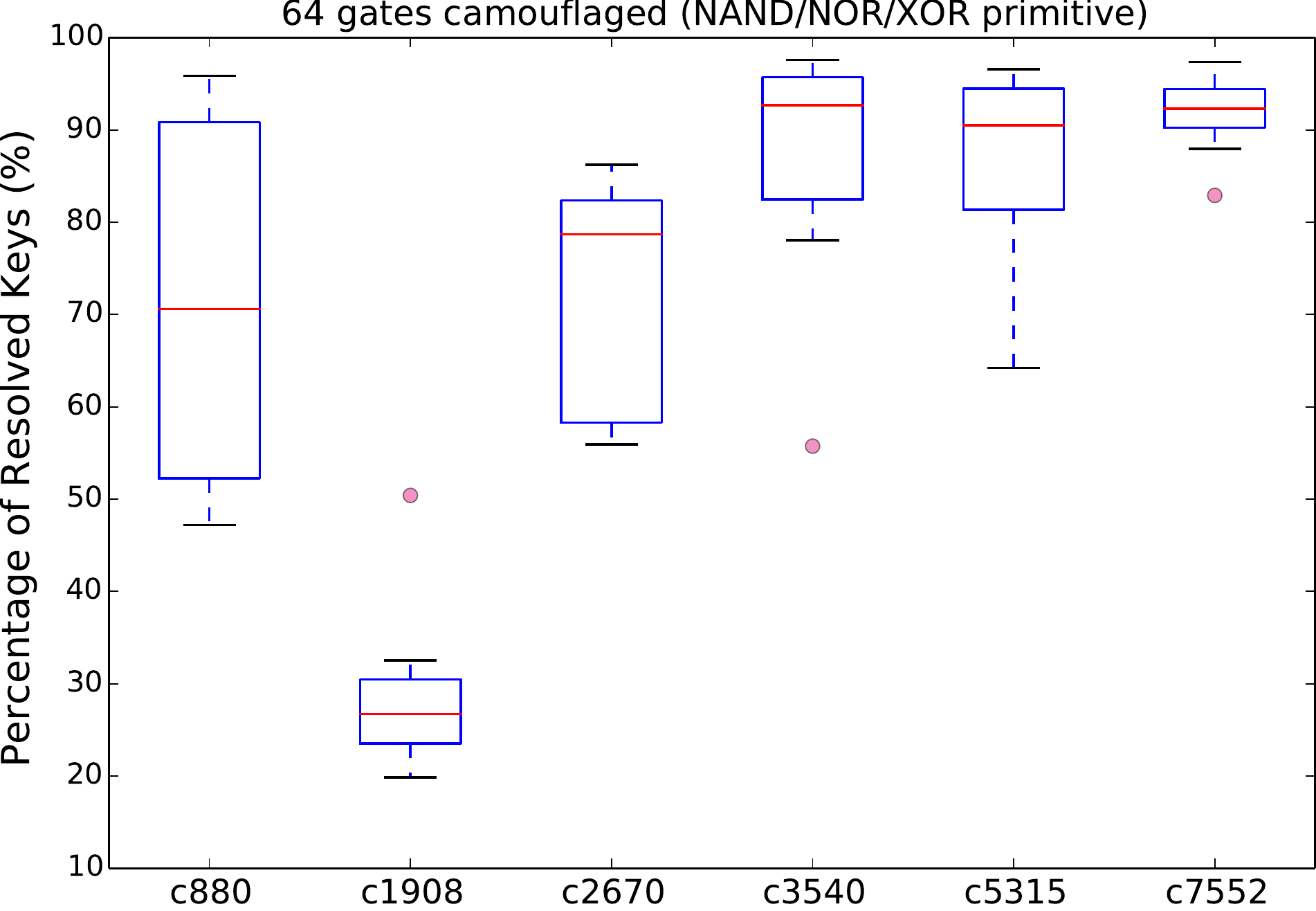}}
\hfill
\subfloat[]{\includegraphics[width=.32\textwidth]{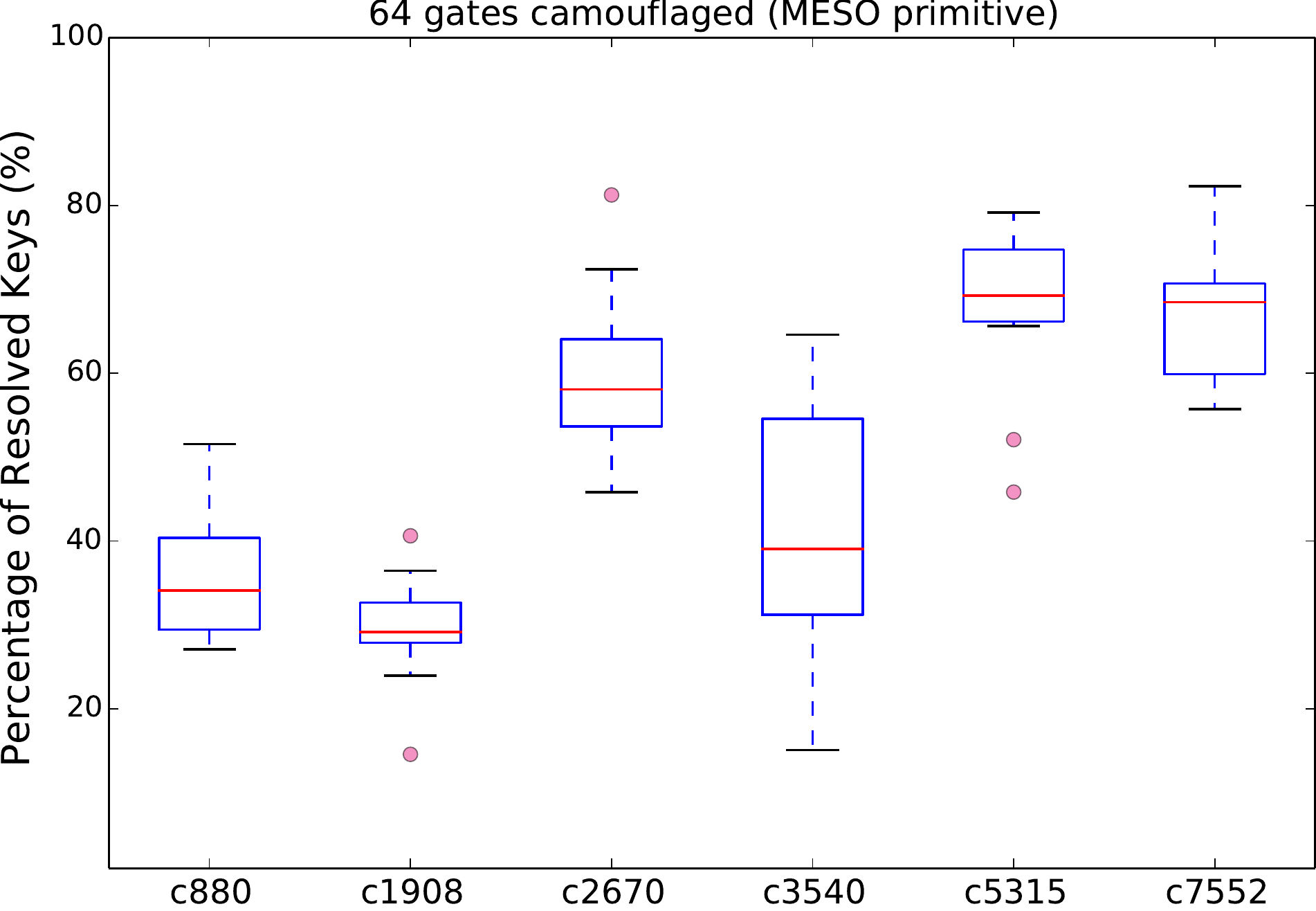}}
\\
\subfloat[]{\includegraphics[width=.32\textwidth]{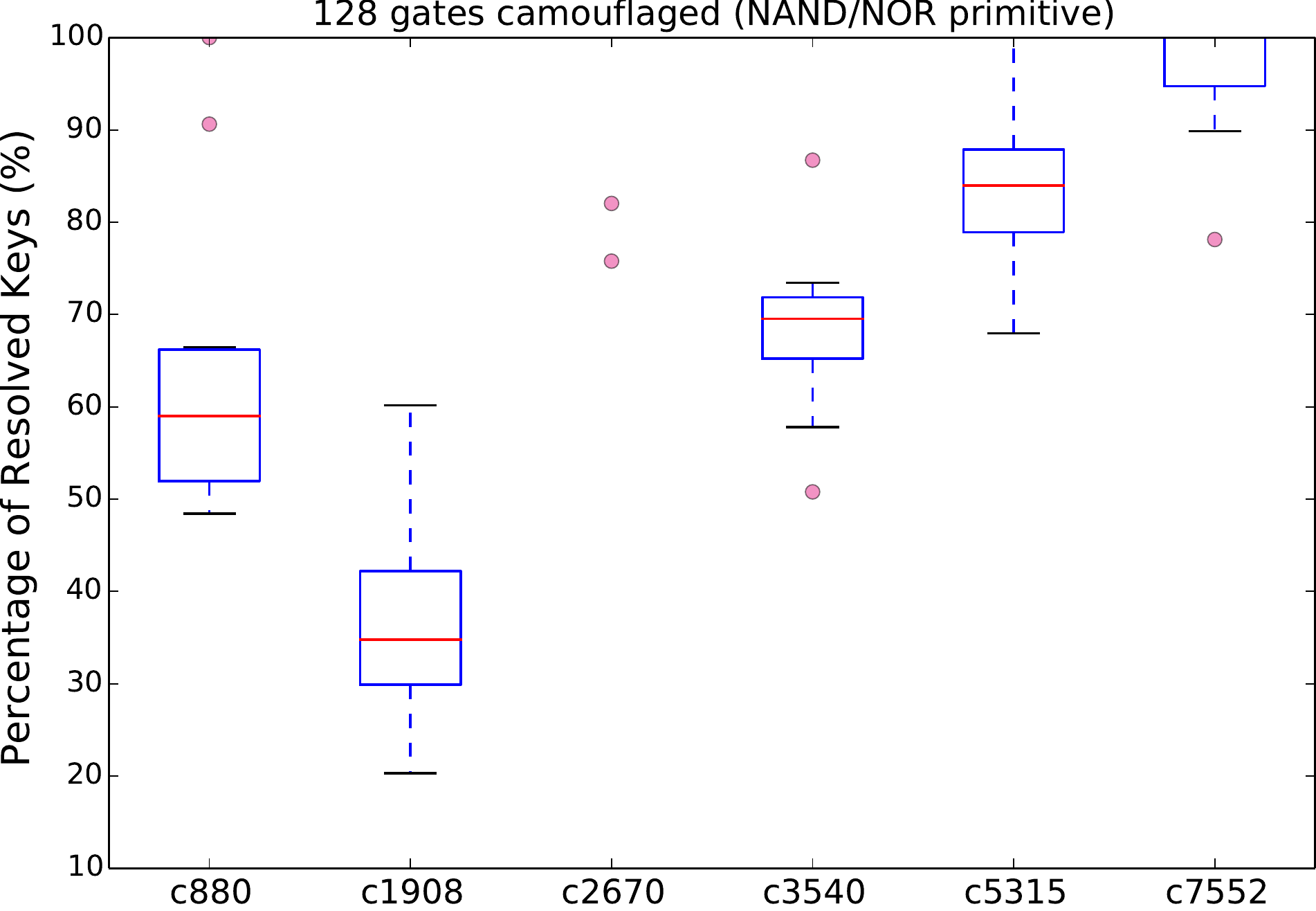}}
\hfill
\subfloat[]{\includegraphics[width=.32\textwidth]{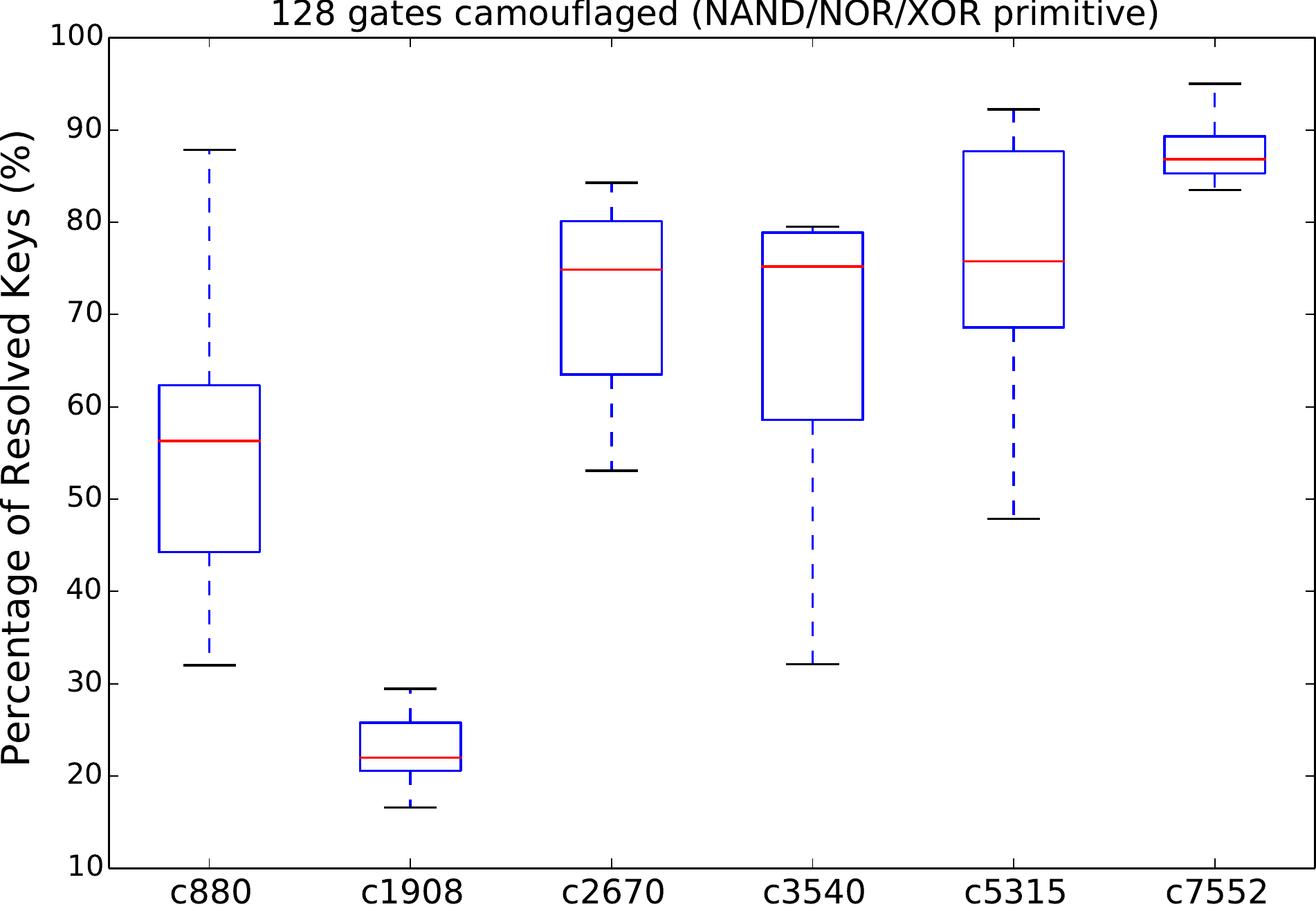}}
\hfill
\subfloat[]{\includegraphics[width=.32\textwidth]{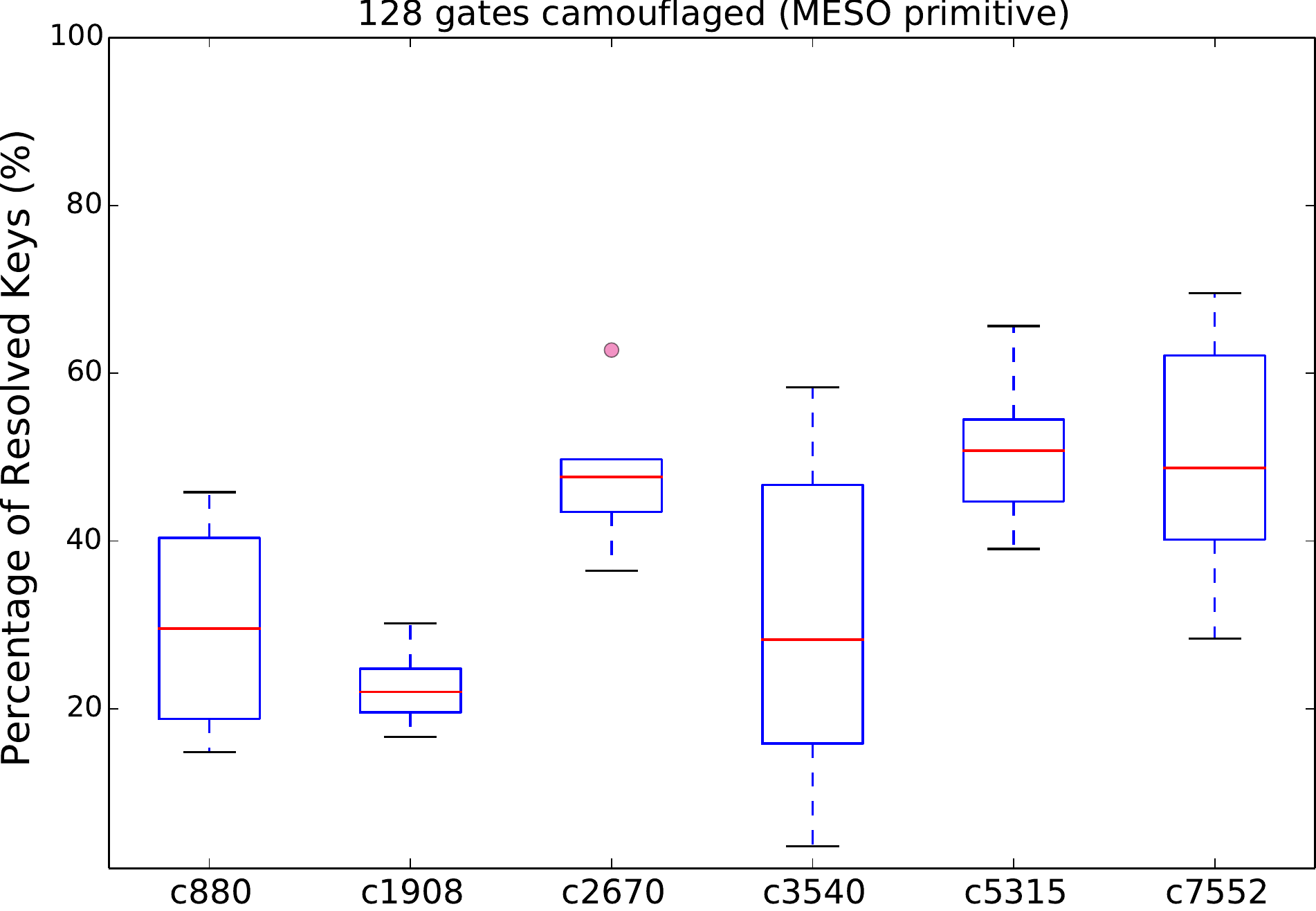}}
\caption{Percentage of key-bits resolved by \textit{HackTest}
for static LC of different sets of camouflaged gates, on selected ISCAS-85 benchmarks. 
The top three box-plots denote 64 camouflaged gates, while bottom three box-plots
denote 128 camouflaged gates for three camouflaging schemes: NAND/NOR, NAND/NOR/XOR, and MESO primitive (implementing 8
functions, Fig.~\ref{fig:primitive}).
Each box comprises data for 10 trials of random selection of gates to camouflage.}
\label{fig:HackTest_basic}
\end{figure*}

Yasin \textit{et al.}~\cite{yasin17_TIFS} demonstrated the efficacy of \textit{HackTest} on benchmarks camouflaged with 32/64 gates employed with NAND/NOR camouflaged cells.
For the sake of completeness, we implement this scheme along with a few others. 
The success rate for \textit{HackTest}~\cite{yasin17_TIFS} is reported as the percentage of key-bits inferred by the attack.

From the box-plots with 64 gates camouflaged (Fig.~\ref{fig:HackTest_basic} (a), (b), and (c)), the attack's complexity is demonstrated with the 
increase in the number of functions
implemented by a single camouflaged gate.
For the NAND-NOR camouflaging scheme, 
\textit{HackTest} performs extremely well; all ten random iterations of ISCAS-85 benchmark c7552 can be decamouflaged correctly.
We observe a high accuracy rate for other benchmarks as well, except for \textit{c1908}. 
Though the overall accuracy remains high for the NAND-NOR-XOR camouflaging scheme~\cite{rajendran13_camouflage}, it fails to compete, especially when compared to the NAND-NOR camouflaging scheme. 
For the MESO-based static camouflaging
scheme, which supports eight functions, 
a stark reduction in attack's efficiency is observed.
From the box-plots with 128 gates camouflaged 
(Fig.~\ref{fig:HackTest_basic} (d), (e), and (f)), the attack's complexity is demonstrated with an increase in the number of camouflaged gates (w.r.t. Fig.~\ref{fig:HackTest_basic} (a), (b), and (c)).
The overall success rate is lower for all the
camouflaging schemes, hinting on the fact that
the attack's success rate is proportional to the total number of camouflaged gates and the number of correctly/incorrectly inferred gates.

\begin{table}[tb]
\centering
\footnotesize
\caption{Impact of \textit{HackTest} on MESO-based static camouflaging on Hamming distance (HD) and Output error rate (OER) for selected ISCAS-85 benchmarks with 128 camouflaged gates.
HD and OER are averaged across 10 random trials of camouflaged gates.}
\label{tab:static_camo_TEST_HD_OER}
\setlength{\tabcolsep}{1mm}
\begin{tabular}{ccccc}
\hline
\textbf{Benchmark} & 
\textbf{\# Camo. Gates} & 
\textbf{\# Correctly Inferred} & 
\textbf{HD (\%)} &
\textbf{OER (\%)} \\ 
\hline 
c880 & 128 & 23 & 47 & 100 \\ \hline
c1908 & 128 & 29 & 46 & 100 \\ \hline
c2670 & 128 & 51 & 28 & 100 \\ \hline
c3540 & 128 & 39 & 47 & 100 \\ \hline
c5315 & 128 & 65 & 23 & 100 \\ \hline
c7552 & 128 & 64 & 24 & 100 \\ \hline
\textbf{Average} & \textbf{128} & \textbf{45} & \textbf{35.8} & \textbf{100} \\ \hline
\end{tabular}
\end{table}

We also analyze the effect of incorrect key-bits on security metrics
HD and Output Error rate (OER) for MESO-based static camouflaging primitive; results are shown in Table~\ref{tab:static_camo_TEST_HD_OER}.
The HD for benchmarks c880, c1908, 
and c3540 approach the ideal value of 50\%, 
while the values are around 25\% for benchmarks c2670, c5315, and c7552.
The OER, however, is 100\% for all the designs.
The decrease in HD for benchmarks c5315 and c7552 can be ascertained to the fact that
the number of wrongly inferred gates form a very small portion of the overall design.
For example, we camouflage 128 gates out of 1,197 gates for c7552 which forms about 10.69\% of the overall design.
\textit{HackTest} resolves 64 gates correctly, bringing the percentage of wrongly inferred gates 
to 5.35\%.
Similarly, camouflaging 128 gates out of 273 for c880 forms 46.89\% of the design.
\textit{HackTest} resolves only 23 gates correctly, which increases the proportion of 
wrongly inferred gates to 38.46\%, which is higher than c7552.

To summarize, we observe that the efficiency of \textit{HackTest} is directly proportional to 
(i)~size and type of the benchmark, 
(ii)~number and type of camouflaged gates, and 
(iii)~number of functions implemented by a 
camouflaged gate.

\subsubsection{HackTest on Dynamic Camouflaging}

\begin{table}[tb]
\centering
\scriptsize
\caption{
Impact of increasing the number of possible functions implemented by the MESO-based primitive on \textit{HackTest}'s accuracy for selected ITC-99 benchmarks.
Test patterns are generated by \textit{Tetramax ATPG} for fault coverage and test coverage 
of 99\% and 100\%, respectively.
Results are averaged across 10 random trials of camouflaged gates.}
\label{tab:dynamic_camo_testing_diff_functions}
\setlength{\tabcolsep}{0.5mm}
\begin{tabular}{cccccc}
\hline
\textbf{Benchmark} & 
\textbf{\# Camo. Gates} & 
\textbf{3 functions} & 
\textbf{4 functions} & 
\textbf{8 functions} &
\textbf{16 functions} 
\\ \hline
b14\_C & 350 & 20.37 & 15.13 & 14.69 & 11.49 
\\ \hline
b15\_C & 350 & 11.47 & 10.4 & 8.58 & 7.23 
\\ \hline
b20\_C & 350 & 17.03 & 14.03 & 11.11 & 8.51 
\\ \hline
b22\_C & 350 & 27.03 & 21.52 & 15.33 & 12.48 
\\ \hline
\textbf{Average} & 350 & 18.98 & 15.27 & 12.43 & 9.93 \\ \hline
\end{tabular}
\end{table}

\begin{table}[tb]
\centering
\scriptsize
\setlength{\tabcolsep}{0.4mm}
\renewcommand{\arraystretch}{1.07}
\caption{Impact of \textit{HackTest} for MESO-based static (S. Camo) and dynamic camouflaging (D. Camo) schemes on HD and OER for selected ITC-99 benchmarks.
The number of wrongly inferred gates for dynamic camouflaging is higher on average when compared to static camouflaging, which translates to an 
improved HD.
HD and OER are calculated by averaging across 
10 random trials.}
\begin{tabular}{ccccccccc}
\hline
\multirow{2}{*}
{\textbf{Benchmark}} & 
\multicolumn{2}{c}{\textbf{\# Wrongly Inferred Gates}} & 
\multicolumn{2}{c}{\textbf{HD (\%)}} & \multicolumn{2}{c}{\textbf{OER (\%)}} \\
\cline{2-7} &
\textbf{S. Camo.} &
\textbf{D. Camo.} &
\textbf{S. Camo.} &
\textbf{D. Camo.} &
\textbf{S. Camo.} &
\textbf{D. Camo.} 
\\ \hline
b14\_C & 249 & 298 & 36.04 & 42.09 & 100 & 100 
\\ \hline
b15\_C & 296 & 320 & 32.07 & 34.23 & 100 & 100 
\\ \hline
b20\_C & 279 & 310 & 32.15 & 35.03 & 100 & 100 
\\ \hline
b22\_C & 212 & 296 & 20.09 & 29.57 & 100 & 100 
\\ \hline
\textbf{Average} & 259 & 306 & 30.09 & 35.23 & 100 & 100 
\\ \hline
\end{tabular}
\label{tab:comparison_Static_Dynamic_TEST_ITC}
\end{table}

As elucidated before, none of the static camouflaging approaches~\cite{rajendran13_camouflage,li16_camouflaging} 
allow for post-fabrication reconfiguration 
and, hence, test patterns are generated 
for the correct assignment of camouflaged gates. 
\ul{MESO-based dynamic camouflaging
circumvents this threat by allowing for \textit{post-test configuration}.
That is, the fabricated IC can be initially configured with an incorrect I/O mapping and functionality.}
The ``falsely configured'' IC and related test data are then 
sent to the test facility.\footnote{
Testing for structural defects does not require the chip to be functional; chips can be configured to any function and tested
with no loss in test quality~\cite{yasin16_test,yasin17_TIFS}.}
Accordingly, an attacker will end up with an \textit{incorrect} IP when
mounting \textit{HackTest} on the IC.\footnote{This resonates with the idea of \textit{post-test activation}~\cite{yasin16_test}, which is the adopted strategy for safeguarding against
untrusted test facilities in logic locking.}
After testing is finished,
the MESO gates are reconfigured (by the design 
house or some trusted entity) to reflect the
true, intended functionality.

Table~\ref{tab:dynamic_camo_testing_diff_functions} details the effect of increasing the number of possible functions implemented by the MESO-based primitive on \textit{HackTest}'s accuracy.
We observe that the attack accuracy reduces (for the same set of camouflaged gates) when the number of functions implemented by the MESO-based primitive is increased. 
This can be reasoned from the fact that, with an increase in the number of possible functions,
the attack has a larger solution space to tackle. 

\begin{figure}[ht]
\centering
\includegraphics[width=0.7\columnwidth]{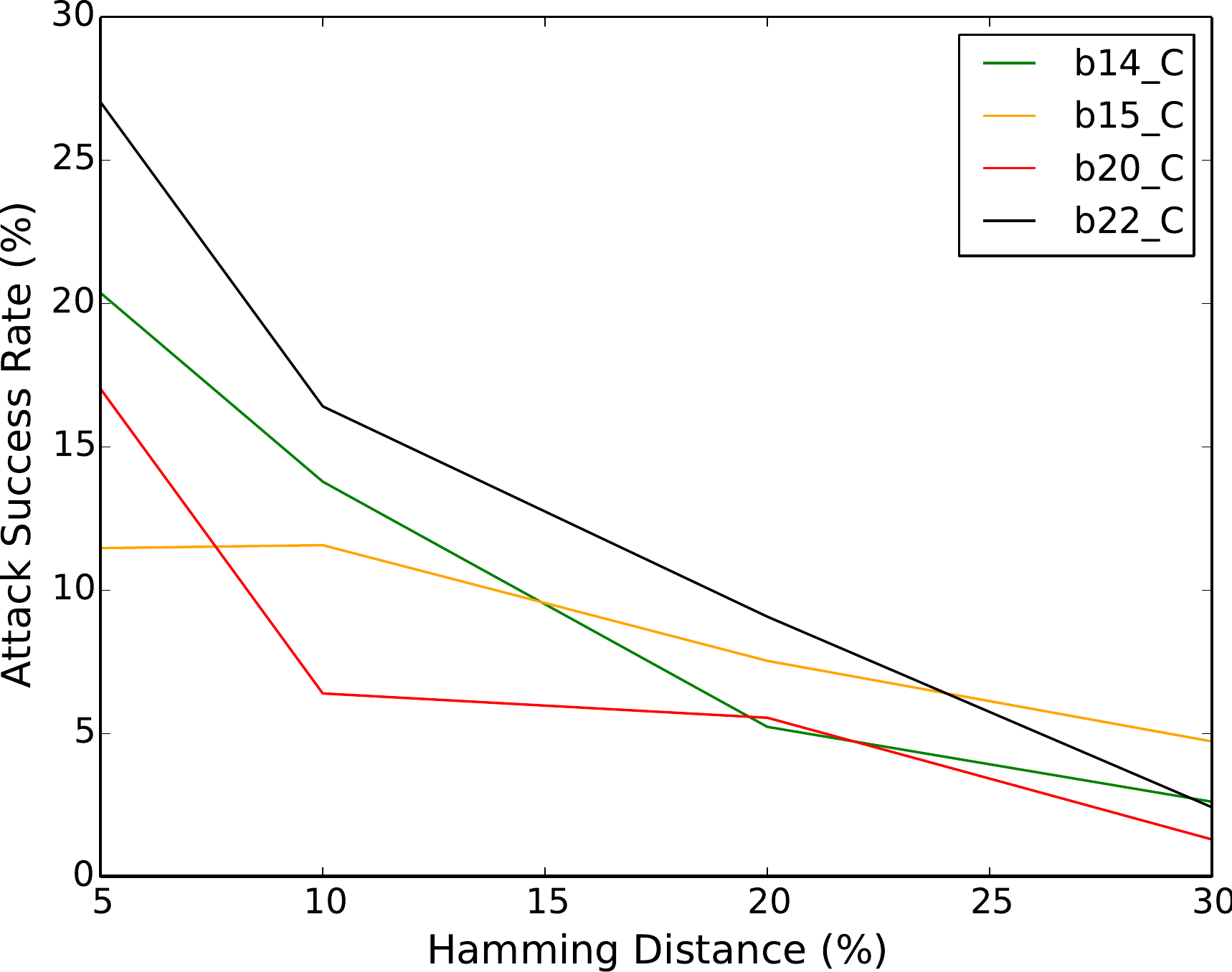}
\caption{Success rate of \textit{HackTest} as a function of HD for selected ITC-99 benchmarks. It is evident that the degree of
functional reconfiguration, also expressed by HD, can be leveraged by a designer to influence the overall success rate for \textit{HackTest}.}
\label{fig:line_graph_FC_HD}
\end{figure}

Finally, we also examine the security promises for both static and dynamic camouflaging for the MESO-based primitive; results are shown in Table~\ref{tab:comparison_Static_Dynamic_TEST_ITC}.
It can be seen that the number of wrongly inferred gates is higher for dynamic camouflaging. 
This increase 
also translates to a higher HD (about 5.14\%). 
The OER, however, remains at 100\% for both the schemes.
Figure~\ref{fig:line_graph_FC_HD} shows the dependence of \textit{HackTest}'s success rate as a function of HD
for selected ITC-99 benchmarks.
This plot reiterates that the degree of functional reconfiguration (measured as HD) has a strong impact on the accuracy of \textit{HackTest}.

\section{Security Analysis: Untrusted End-User}
\label{sec:untrusted_user}

\subsection{Threat Model}
\label{sec:end_user_threat_model}

The threat model which we employ for security analysis for an untrusted end-user follows very closely to the ones described in the literature~\cite{subramanyan15,massad15,li16_camouflaging}.

\begin{itemize}

\item The attacker has access to advanced, specialized equipment for reverse engineering an IC,
which includes setup to depackage an IC, delayer it, imaging of individual layers, and
image-processing tools.
\item Further, he/she can readily distinguish between a camouflaged cell and a regular, standard cell.
If hybrid spin-CMOS circuits are used, it is straightforward to identify the CMOS gates, whereas the complexity is increased manifold, if all the gates are implemented using MESO devices.

\item The attacker is aware of the total number of camouflaged gates, and the number and type of functions implemented by each camouflaged cell.

\item He/she procures multiple chip copies from the open market, uses one of them as an oracle (to observe the input-output mapping), 
and extracts the gate-level netlist of the chip by reverse engineering the others.
This paves the way for algorithmic SAT-based attacks~\cite{subramanyan15,massad15,shamsi17}.

\item Consistent with the most prior art, we assume that an attacker cannot invasively probe the output of a camouflaged cell.\footnote{We
acknowledge that there is an attack proposed by Keshavarz \textit{et al.}~\cite{KeshavarzHOST2018}, where an SAT-based formulation is augmented with probing
and fault-injection capabilities to reverse engineer a relatively small \textit{S-Box}. 
Still, it remains to be seen how this attack would fare when large-scale camouflaging is effected.
Having no access to this attack at the time of writing, we refrain from any empirical analysis.}
It is straightforward to note that once an adversary is allowed probing capabilities, i.e., to probe the output of a camouflaged cell
(or read out contents from a TPM for locking),
then the security guarantees offered by these schemes are substantially weakened, if not even nullified.

\end{itemize}

An attacker may also try to observe various side-channels like power, timing, photonic, acoustic, etc.
However, note that we do not consider the  effect of side-channels emission from the MESO 
switch in this work; this remains part of our future work, once efficient circuit- and/or layout-level models are available for MESO gates.

\subsection{Attack Model and Setup}
\label{sec:end_user_attack_setup}

In 2015, Subramanyan \textit{et al.}~\cite{subramanyan15} and Massad \textit{et al.}~\cite{massad15} independently
demonstrated SAT-based attacks to circumvent security guarantees offered by LL and LC, respectively.
Interested readers are referred to the respective papers for further details.
We leverage the publicly available attack~\cite{subramanyan15} to perform the security analysis for an untrusted end-user.

\begin{figure}[b]
\centering
\includegraphics[scale=0.095]{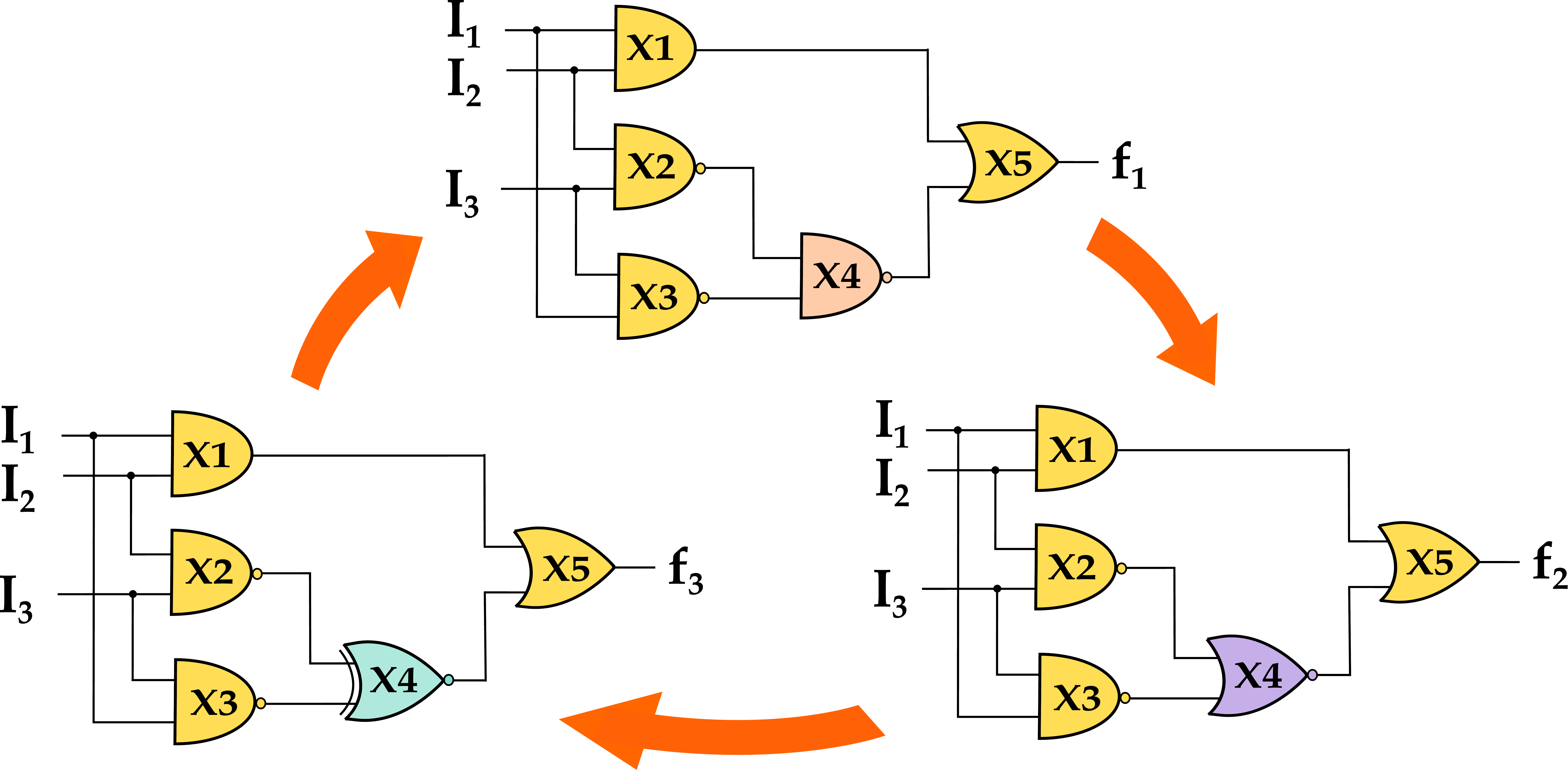}
\caption{Dynamic morphing of gate X4 in a representative circuit. 
Circuit implementing $f_1$ is the original template and $f_2$, $f_3$ are the morphed versions.}
\label{fig:circuits1}
\end{figure}

\begin{figure}[b]
\centering
\includegraphics[scale=0.36]{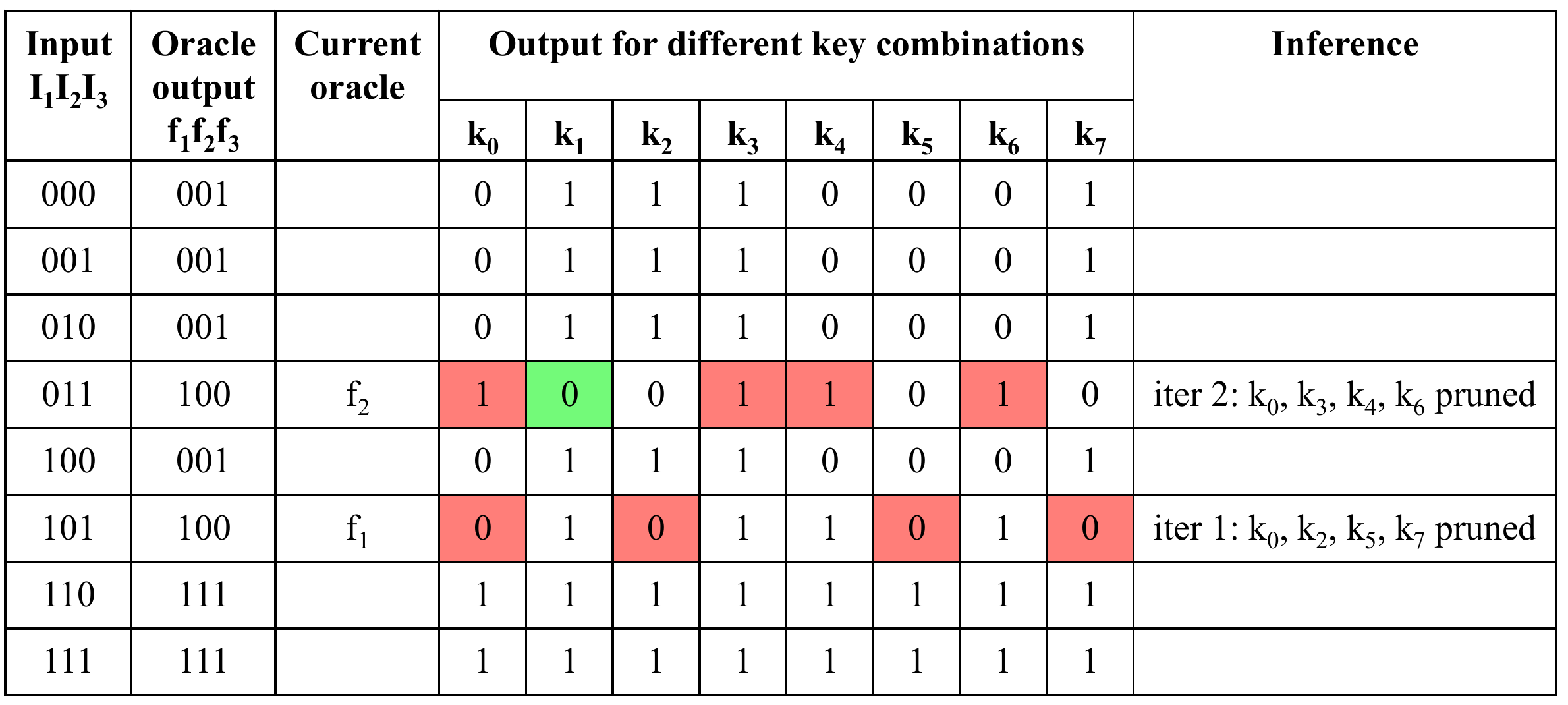}
\caption{SAT-based attack~\cite{subramanyan15} on the polymorphic circuit of Fig.~\ref{fig:circuits1}. 
For $k_0$ and $k_1$, the INV and BUF operations performed on the output of $X2$.}
\label{fig:dynamic_camo_SAT}
\end{figure}

\subsection{Results and Discussion}
\label{sec:dynamic_camo_end_user_results}

Next, we explain how dynamic camouflaging can
thwart attacks arising from the perspective of malicious end-users.\footnote{
The MESO-based primitive can
also be leveraged in the context of static camouflaging to protect against malicious end-users.
It has been shown empirically in prior studies~\cite{massad15,patnaik2020obfuscating,patnaik18_GSHE_DATE,patnaik2019spin} that large-scale camouflaging with cells implementing more functions
pose significant computational complexity for such attackers.}
We illustrate the related concept of \textit{run-time
polymorphism for dynamic morphing}
through a conceptual example; consider the circuit
in Fig.~\ref{fig:circuits1}.
Here, $X4$ is the only camouflaged, polymorphic
gate modeled with three key-bits.
Assuming a key distribution such that 
INV, BUF, AND, OR, NAND, NOR, XOR, and XNOR gates 
correspond to key-bits $\{000, 001,...., 111\}$, respectively, the dynamic key of the circuit cycles from $100$ to $101$, and then to $111$, as per the outlined 
functional reconfiguration in Fig.~\ref{fig:circuits1}.

The application of the SAT-based attack~\cite{subramanyan15,massad15} for the simple scenario in
Fig.~\ref{fig:circuits1} is
explained next (Fig.~\ref{fig:dynamic_camo_SAT}).
Consider that the oracle (i.e., an actual working chip obtained from the market) implements $f_1$ during 
the first iteration of the SAT solver, 
where the input applied is $101$.
Note that the oracle is to be configured for test mode, to provide access to the circuit internals through scan chains, as required when modeling the
whole circuit for the SAT-based attack.
In principle, the oracle may behave differently in the test mode and in the operational (functional) mode.
Naturally, the SAT solver is oblivious to the function being active internally in the oracle during \textit{any} iteration.
Also note that, once inputs are applied to the oracle, the SAT solver has to wait until the oracle provides the corresponding outputs.
Now, that first SAT iteration prunes key combinations 
$k_0, k_2, k_5$, and $k_7$. 
While this is happening, assume that the gate $X4$ 
has morphed into NOR, and the oracle 
is now implementing
function $f_2$. In the second SAT iteration, the input pattern $100$, therefore, eliminates keys $k_3, k_4$, and $k_6$. 
Thereafter, the SAT solver concludes that the correct key bit and identity of gate $X4$ are $001$ and BUF, respectively.

In essence, dynamic camouflaging can deceive
and mislead the SAT solver to converge to an \textit{incorrect key}, leading to an
\textit{incorrect} gate assignment.
For an exploratory study,
we extend the framework from~\cite{subramanyan15} to realize SAT-based attacks on polymorphic versions of ITC-99 benchmarks. 
Even for 100,000 randomized trials,
the related attacks fail due to inconsistent I/O mappings, as these induce \textit{unsatisfiable} (\textit{UNSAT}) scenarios for the attack framework.

Taking this simple example from Fig.~\ref{fig:circuits1} further,
for error-tolerant applications like image/video processing, the circuit may 
indeed be reconfigured randomly, e.g.,
by deriving the control bits of the MESO gates from a TRNG---we present results for \textit{AppSAT}~\cite{shamsi17} on such error-tolerant applications in Section~\ref{sec:MESO_CeNN_results_discussion}.

Besides SAT-based attacks, when concerned about \textit{physical attacks} conduced by an end-user, one has to ensure that the interconnect fabric
which routes the control bits and control signals to the MESO gates is resilient against probing; e.g., shielding may be used toward that end~\cite{ngo17}.
\textit{Removal attacks} targeting the TRNG 
shall result in floating controls for the MESO gates, leading to noisy outputs, and loss of functionality (as well as hindrance of SAT-based attacks discussed above).
Other advanced attacks, e.g., directed at distorting the entropy of the TRNG to change its bias~\cite{bayon2012contactless}, are considered out-of-scope for this work.

\section{Case Study: MESO CeNN-Based Approximate Image-Processing IP}
\label{sec:case_study}

In this section, we demonstrate how dynamic camouflaging can help in protecting 
approximate, error-tolerant circuits. 
We design a cellular neural network (\textit{CeNN}) using MESO gates.
\textit{CeNNs} are a massively parallel neural network-based computing paradigm, which consist of 
an n-dimensional array of locally interconnected cells that communicate within a neighbourhood. 
They are typically used in a variety of applications including image filtering and reconstruction, 
edge detection, solving partial differential equations and optimization problems. The cells of a CeNN are multiple-input single-output processors, characterized by an internal state variable. These processing cells act as neurons that integrate the input currents, and the interconnects between the cells act as synapses that perform weighting of the inputs. 
The dynamical state equation for a CeNN neuronal cell, put forth by Chua and Yang \textit{et al.}~\cite{chua1988cellular}, is as follows:

\begin{equation}
\begin{aligned}
    C\frac{dx_{ij}}{dt} &=-\frac{1}{R}x_{ij}+\sum_{kl}A(i,j;k,l)f(x_{kl})\\
    &+\sum_{kl}B(i,j;k,l)U_{kl}+I_{ij}
    \end{aligned}
\end{equation}
where $x_{ij}$ is the internal state of a neuron, $\{i,j\}$ represent the neighbourhood of the neuron, $A$ and $B$ are the synaptic weights connecting two neighbouring cells, $I$ is a constant bias current, $R$ and $C$ are the resistance and 
capacitance of the cell, $U_{kl}$ and $f(x_{kl})$ are the input and state-dependent output of the cells, respectively. 

As demonstrated in~\cite{pan2016proposal}, 
spintronic devices are able to directly implement a CeNN for image-processing applications, without the need for analog VLSI elements. We chose CeNN as a representative example to highlight the application of dynamic camouflaging in real scenarios owing to the fact that it can function as a relatively simple and low-cost image-processing circuit, with a single layer of input cells. However, the concept of dynamic camouflaging can be extended to any approximate IP, without loss of generality. 
 
\subsection{Construction}

\begin{figure}[tb]
\centering
\includegraphics[scale=0.17]{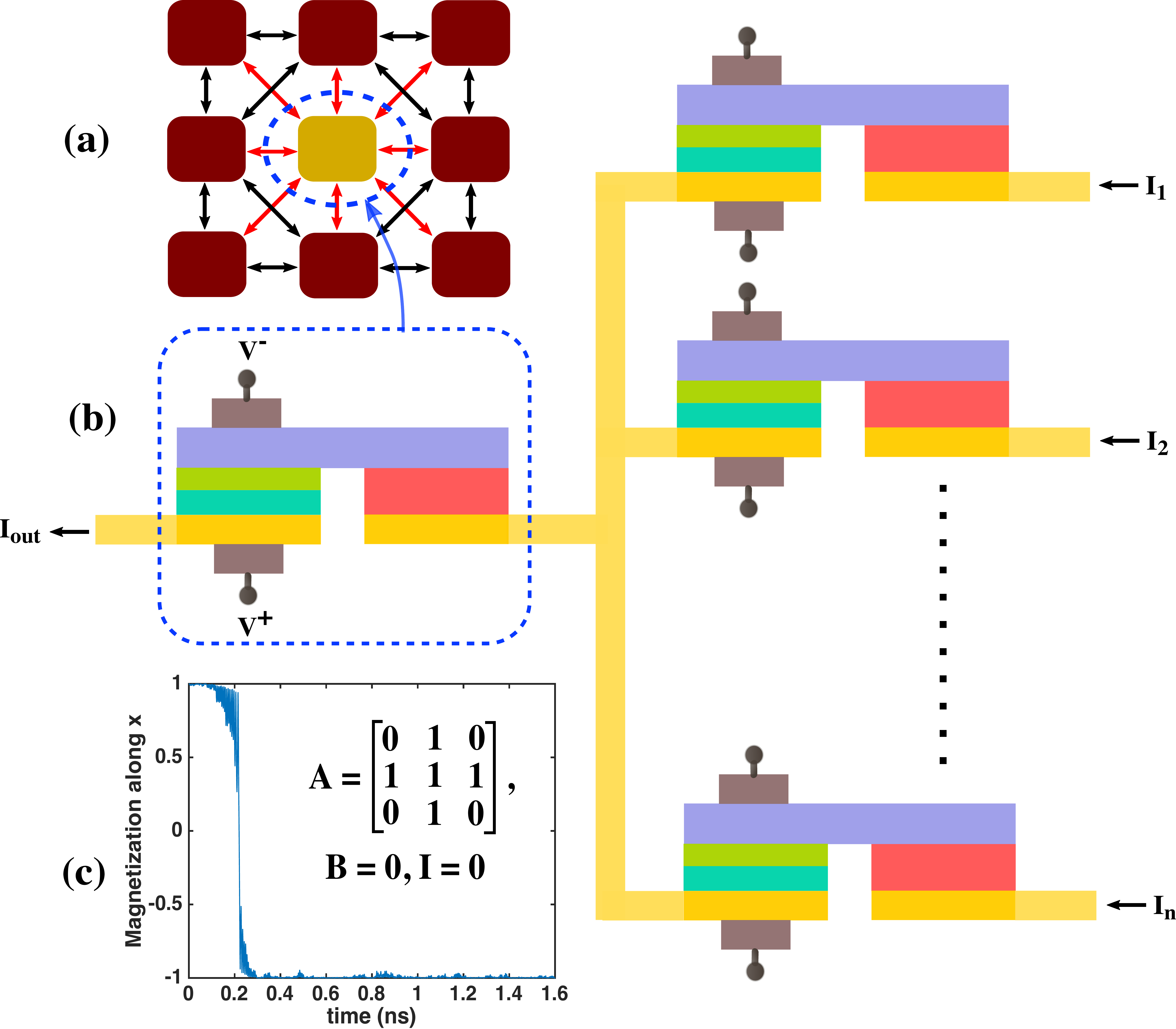}
\caption{(a) Inter-connectivity of cells in the MESO-based Cellular Neural Network (CeNN). 
(b) Construction of the MESO CeNN for image reconstruction. 
Each cell in the network is implemented by 
a MESO device.
(c) Magnetization vs. time shows the 
switching of the central MESO CeNN cell, when inputs from its nearest neighbor cells are applied. 
The switching delay is $\sim200$ ps. 
The MESO CeNN is simulated using a Landau-Lifshitz-Gilbert dynamics framework on \textit{CUDA-C}~\cite{kani2017modeling}, with 1,000 nanomagnet simulations per cell. 
Inset shows templates $\{A, B, I\}$ used to configure the CeNN.}
\label{fig:CeNN}
\end{figure}

We adopt the same methodology for the 
construction of the magnetic synapses and neuron of the CeNN as in~\cite{pan2016proposal}. 
However, we design the neuron cells using the MESO device instead of the all-spin logic device used 
in~\cite{pan2016proposal}.
The parameters for the MESO device used for the CeNN cells are obtained from~\cite{manipatruni2019scalable}. 
Fig.~\ref{fig:CeNN} (a) highlights the connectivity of cells in the MESO CeNN and Fig.~\ref{fig:CeNN} (b) shows the construction of the MESO CeNN.
The transient switching of a MESO CeNN cell along with the CeNN templates $\{$A, B, I$\}$ used for simple image reconstruction (from~\cite{pan2016proposal}) are portrayed in Fig.~\ref{fig:CeNN} (c). 
The central MESO device is connected to eight other MESO devices in a $3\times3$ grid. 
The weighting operation can be realized by (i)~using a layer of CMOS transistors with different driving strengths, in between the input and output layers, as demonstrated in~\cite{pan2016proposal}, or by (ii)~inserting multi-terminal magnetic domain wall (DW) weighting devices~\cite{he2017energy} in the interconnects between the input and output layers. 
Both these weighting mechanisms are able to implement several levels of weights for precise image-processing applications. 
We note that the former approach also requires additional transduction circuitry for converting the current signals from the input layers, into voltage signals that can be fed to the CMOS driving transistors. 
In the case of the DW weighting devices, dedicated programming terminals are used to set the position of the DW and control the conductance (weight) of the device, which then scales the current passing through its input terminals. 
Readers are referred to the respective papers for further details. 
The weighting units are omitted from Fig.~\ref{fig:CeNN} (b) for simplicity.

\begin{figure*}[tb]
\centering
\includegraphics[width=1\textwidth]
{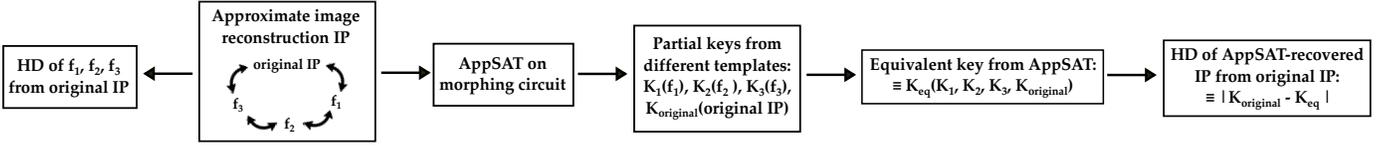}
\caption{Flowchart illustrating the application of \textit{AppSAT} on a dynamically camouflaged approximate image-processing IP. 
\textit{AppSAT} recovers partial keys from the 
different circuit templates, and the equivalent stitched key deviates significantly from the original IP.}
\label{fig:AppSAT_flowchart}
\end{figure*}

\subsection{Experimental Setup}

We investigate the implications of attacking an 
approximate image-processing IP with \textit{AppSAT}~\cite{shamsi17}. 
Approximate circuits are vulnerable to such attacks since the attacker can recover a functionally-similar IP. 
Since the original circuit is approximate to begin with, an attacker might be satisfied by 
obtaining an IP which has, say, 95$\%$ fidelity compared to the original design. 
For example, consider the case of an approximate image reconstruction hardware module. 
The application of \textit{AppSAT} on this module may give an attacker a functionally-similar circuit and, if needed, he/she could then augment this reverse engineered module with software ML models to obtain a precision equivalent to the original image reconstruction hardware. 
In this section, we show that the \textit{AppSAT}-recovered IP of an approximate circuit can deviate significantly from the original IP, enough to render such an attack futile.

\begin{figure}[ht]
\centering
\includegraphics[scale=0.6]{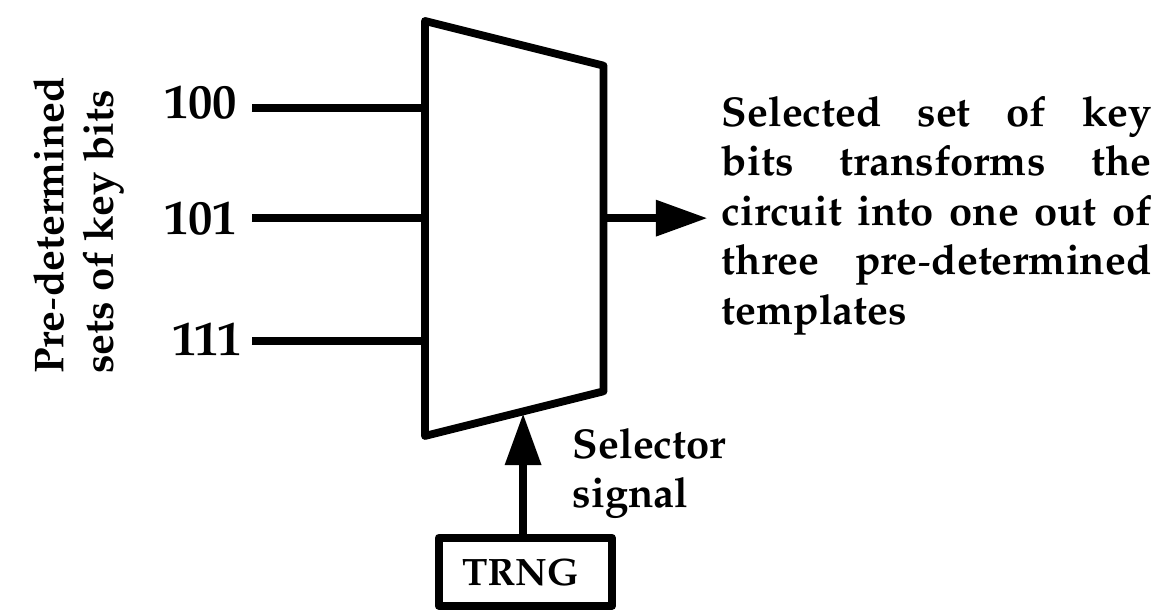}
\caption{Generation of control/selector signal for randomized functional transformation of the circuit between a set of pre-determined templates.}
\label{fig:random_reconf}
\end{figure}

To safeguard the MESO-based CeNN against 
\textit{AppSAT}, we use \textit{run-time polymorphism} 
for dynamic morphing.
This means that in the MESO CeNN circuit of Fig.~\ref{fig:CeNN} (b), certain MESO gates will be polymorphic, enabling a circuit-level polymorphic
reconfiguration between the original circuit and, say, three templates f$_1$, f$_2$, and f$_3$. 
Hence, the image-processing IP will work at a \textit{sub-optimal accuracy} which is inversely proportional
to the HD between the different morphing circuit templates and the original function.\footnote{Here we use HD as a representative metric for image quality. 
However, HD can be translated to other image processing relevant metrics and the conclusions of this study do not depend on the choice of this metric. } 
By tuning this HD through system-level design, i.e., selecting gates that need to be polymorphic, one can control how similar the IP recovered by
\textit{AppSAT} will be when compared to the original IP. 
Reconfiguration between these circuit templates is controlled by using a TRNG to drive a selector circuit,
which selects one set of key-bits, as shown in a simple example in Fig.~\ref{fig:random_reconf}. 
Note that, in this scenario, the TRNG does not control the distinct key-bits of the MESO gate individually;
rather, the control signal is derived from the TRNG such that it randomly cycles between pre-determined sets,
which will transform the gate/circuit into one out of several pre-determined templates.
\ul{When \textit{AppSAT} is mounted on 
such a dynamically morphing circuit,
the constant functional reconfiguration results in the attack recovering parts of the key from different circuit templates at different instances of time.
Therefore, the overall stitched key recovered by \textit{AppSAT} from all the circuit versions may have an HD significantly different from the original IP.} Please note that, the HD between the polymorphic templates and the original IP, which dictates the accuracy at the system-level, is \textit{different} 
from the HD between the \textit{AppSAT}-recovered IP 
and the original IP. 
The application of \textit{AppSAT} on the approximate IP, along with the key recovery and HD calculation, is represented as a flowchart in Fig.~\ref{fig:AppSAT_flowchart}. 

Since CAD tools and synthesizable \textit{Verilog} libraries for emerging spin devices like MESO are 
under development,
we present proof-of-concept simulations on large-scale ITC-99 benchmarks. 
In experiments on b14\_C benchmark, $\sim11\%$, $\sim9\%$ and $\sim14\%$ HDs between the original IP and the polymorphic templates f$_1$, f$_2$, and f$_3$, respectively,  translate to $\sim28\%$ HD between the \textit{AppSAT}-recovered IP and the original IP. 
In Table~\ref{tab:HD_comparison}, templates 1-3 are approximate versions of each benchmark, with their respective HD from the original design. 
We execute \textit{AppSAT} (setup details same as in~\cite{shamsi17}) considering that the benchmark morphs between its original form and 
three approximate templates.
\textit{AppSAT} provides an \textit{approximate-key} after time-out, unlike the SAT-based attack~\cite{subramanyan15}. 
With this approximate key, HD is computed between the \textit{AppSAT}-recovered
IP and the original IP, whereas results are quoted in the last column of Table~\ref{tab:HD_comparison}. 

\begin{table}[ht]
\centering
\footnotesize
\setlength{\tabcolsep}{1mm}
\renewcommand{\arraystretch}{1.3}
\caption{Comparison of HD (in \%)
between various polymorphic templates and original function, and HD inferred between \textit{AppSAT} recovered IP and original IP
for selected ITC-99 benchmarks. 
HD is calculated using \textit{Synopsys VCS} for 100,000 patterns} 
\label{tab:HD_comparison}
\begin{tabular}{*{5}{c}}
\hline
\textbf{Benchmark}
& \multicolumn{3}{c}{\textbf{HD (in \%) from the original design}}
& \multicolumn{1}{c}{\textbf{HD inferred }}
\\
\cline{2-4}
& \textbf{template-1} 
& \textbf{template-2} 
& \textbf{template-3} 
& \textbf{after \textit{AppSAT}}
\\
\hline

b14\_C
& 11.22 & 9.26 & 13.78 
& 28.81 
\\ \hline

b15\_C 
& 12.35 & 9.62 & 12.88 
& 32.15 
\\ \hline

b17\_C 
& 11.14 & 10.62 & 15.24 
& 36.22 
\\ \hline

b20\_C
& 12.51 & 14.37 & 17.86 
& 34.34 
\\ \hline

\end{tabular}
\end{table}

\subsection{Results and Discussion}
\label{sec:MESO_CeNN_results_discussion}

\begin{figure}[ht]
\centering
\subfloat[Original]{\includegraphics[scale=0.3]{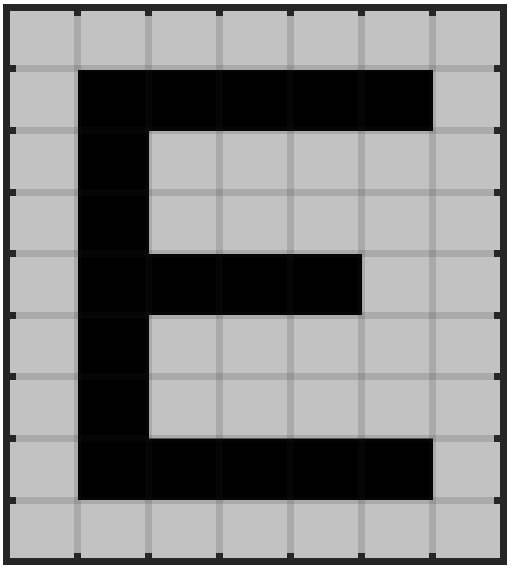}\label{Image1}} 
~
\subfloat[$10\%$ HD]{\includegraphics[scale=0.3]{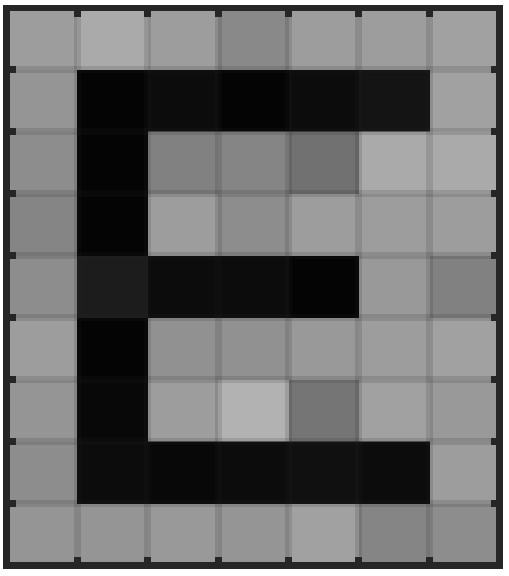}\label{Image2}} 
~
\subfloat[$15\%$ HD]{\includegraphics[scale=0.3]{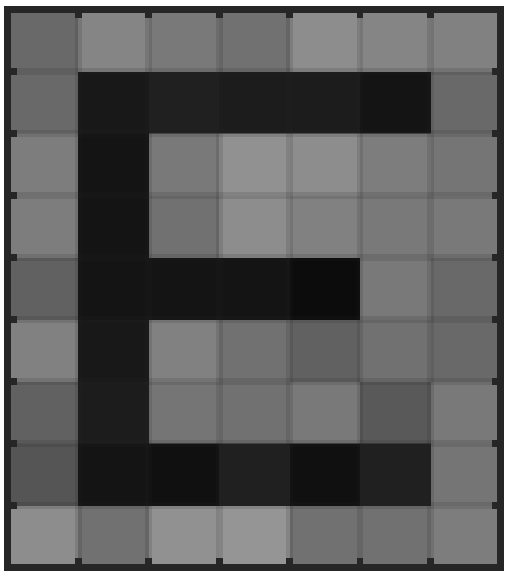}\label{Image3}} 
~
\subfloat[$20\%$ HD]{\includegraphics[scale=0.3]{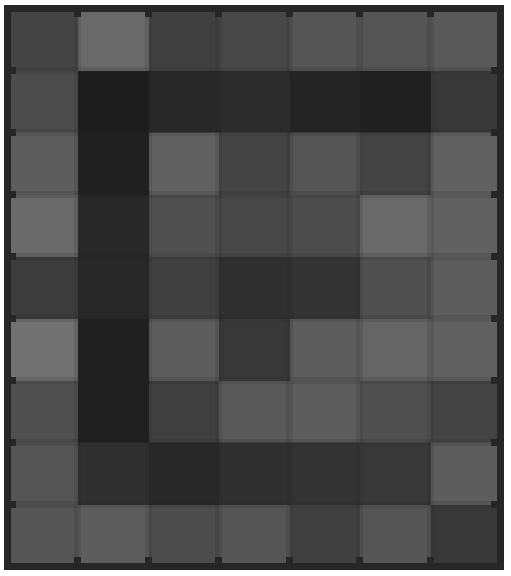}\label{Image4}} 
~
\subfloat[$25\%$ HD]{\includegraphics[scale=0.3]{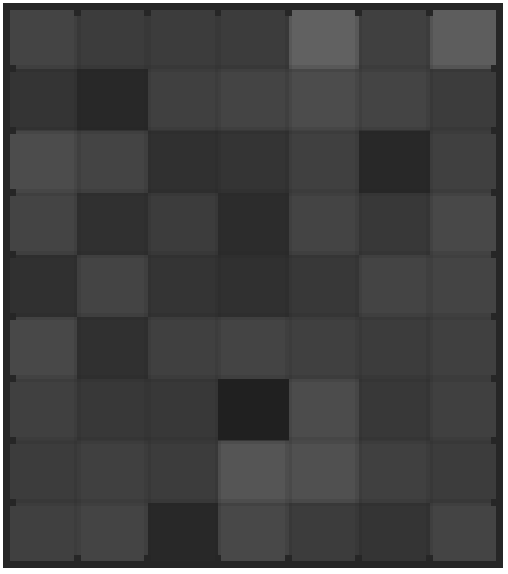}\label{Image5}}
\caption{(a) Original image to the MESO-based CeNN image reconstruction. 
(b-e) Images reconstructed with approximate IP of the CeNN recovered from \textit{AppSAT}, for HD of $10\%$, $15\%$, $20\%$, and $25\%$, respectively between \textit{AppSAT}-recovered IP and the original IP. 
It is essential to note that this HD is different from the accuracy of the approximate circuit reported in Table~\ref{tab:HD_comparison}.}
\label{fig:Image_Processing}
\end{figure}

The image reconstructed by the CeNN IP as recovered by
\textit{AppSAT} (at various representative values of HD between
the \textit{AppSAT}-recovered IP and original IP) is shown in Fig.~\ref{fig:Image_Processing}.
Although the average HD numbers for the proof-of-concept simulations and attacks above are between 28--36\% (see last column of
Table~\ref{tab:HD_comparison}), here we assume an even more powerful attack.
That is, we gauge the resilience offered by dynamic morphing when trying to reconstruct images for representative HD values of 10--25\%.
As can be seen in Fig.~\ref{fig:Image_Processing} (e), 
at a sufficiently large HD of $25\%$, the \textit{AppSAT}-recovered CeNN IP fails to faithfully reconstruct the original image.
For the \textit{AppSAT}-recovered IPs incurring even larger HDs in Table~\ref{tab:HD_comparison},
the reconstructed image will naturally be even more noisy. 

\begin{figure}[ht]
\centering
\includegraphics[scale=0.25]{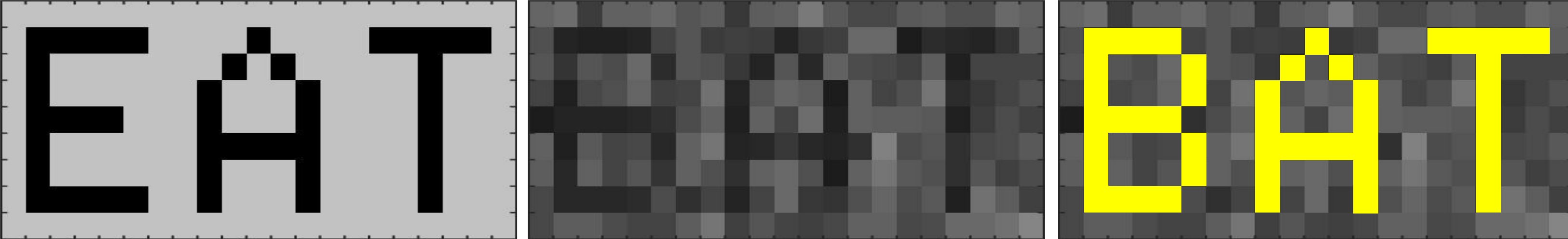}
\caption{Incorrectly inferred text, reconstructed using an approximate IP of the CeNN recovered by \textit{AppSAT}, for HD of $20\%$ between \textit{AppSAT}-recovered IP and the original IP. }
\label{fig:CeNN_word}
\end{figure}

\begin{figure}[ht]
\centering
\includegraphics[scale=0.4]{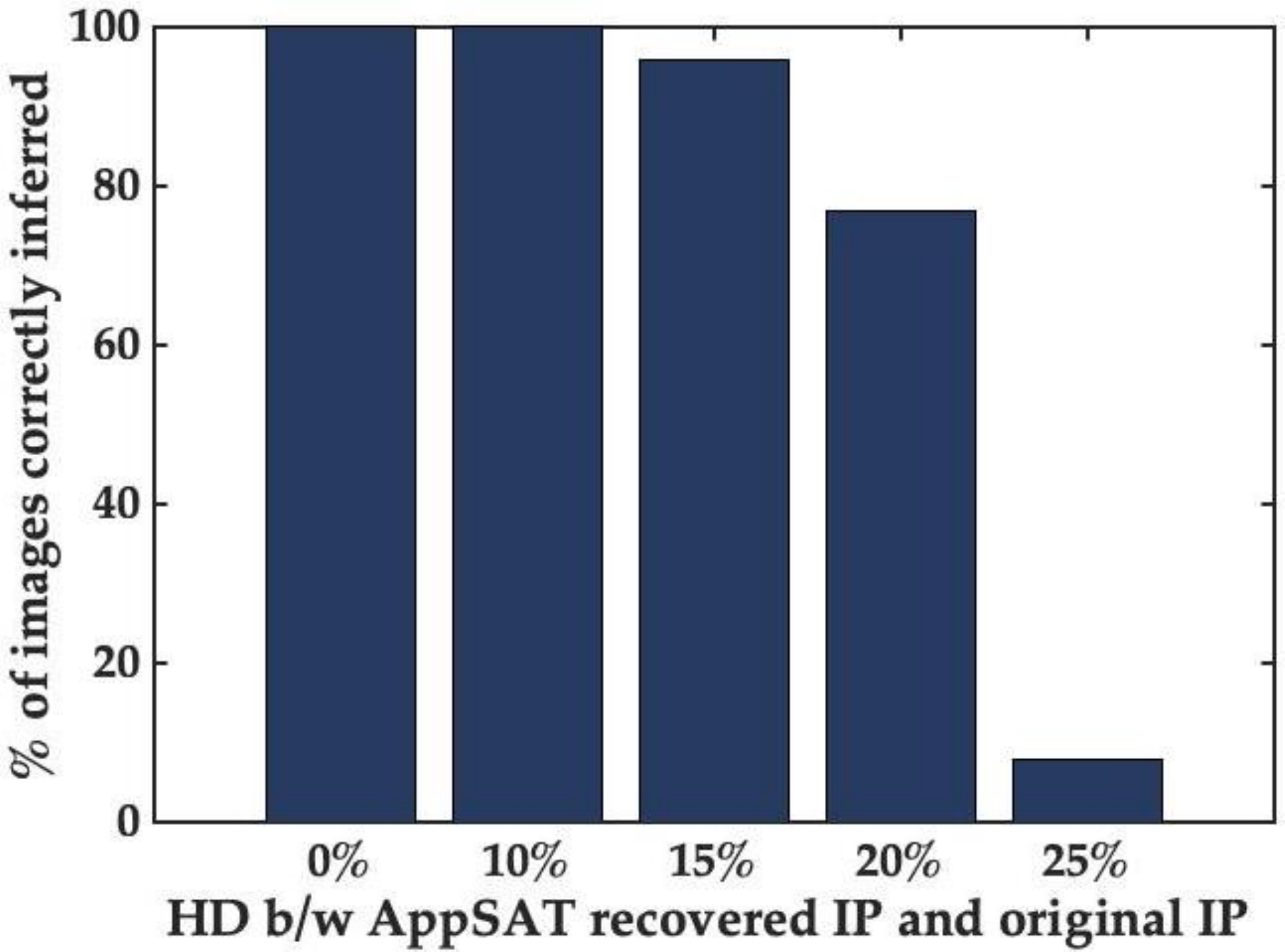}
\caption{Proportion of alphabets correctly 
identified by neural net-based \textit{Tesseract 4.0} OCR engine, where the images were reconstructed using a MESO-based CeNN, for various HD between the \textit{AppSAT}-recovered IP and the original IP.}
\label{fig:CeNN_barplot}
\end{figure}

Further, text reconstructed using the approximate CeNN IP, recovered by \textit{AppSAT} (for an HD of 20$\%$ between \textit{AppSAT}-recovered IP and the original IP), is incorrectly inferred by optical character recognition (OCR) engines like \textit{Tesseract}~\cite{smith2007overview} (Fig.~\ref{fig:CeNN_word}).\footnote{\textit{Tesseract} is the industry standard OCR used by \textit{Google} on its mobile devices and for text detection in Gmail. 
It has been trained using Google's character dataset containing millions of images, and can identify more than 100 languages.}
To substantiate the inability of an attacker to 
gain a satisfactory approximate IP of the image reconstruction hardware, we use the Long Short Term Memory (LSTM) recurrent neural network module of the \textit{Tesseract 4.0} OCR engine on all alphabets at various HDs between \textit{AppSAT}-recovered IP and original IP. As shown in Fig.~\ref{fig:CeNN_barplot}, the neural network-based OCR is unable to faithfully detect the reconstructed text at higher HDs close to 25$\%$.

Thus, we point out that, there is a clear trade-off between the accuracy of 
the original IP 
and the resilience to \textit{AppSAT} attacks, 
in terms of how closely \textit{AppSAT} 
is able to resolve the original IP. 
However, \ul{for approximate applications like image processing,
which can tolerate a certain degree of error, 
our scheme of dynamic morphing can thwart attempts to recover even an approximate version of the IP.}
Advanced IP protection mechanisms based on point-functions~\cite{xie2016mitigating,li16_camouflaging} and stripping of functionality~\cite{yasin17_CCS}
may not be suitable for protecting error-tolerant applications
such as the image-processing system considered in this section.
This is because the above mentioned techniques trade-off output corruptibility for SAT-attack resilience; the lower the output corruptibility, the stronger the resilience.
Hence, attacks working on the notion of recovering an approximate version of the protected IP (e.g., \textit{AppSAT}) are able to successfully recover 
a \textit{satisfactorily functionally similar} IP 
if protected using~\cite{li16_camouflaging,yasin17_CCS}.

We finally like to note that dynamic morphing cannot protect systems which demand highly accurate and error-free computations, e.g., cryptographic applications.
How reconfiguration at run-time might help in providing additional layer of security for these systems remains an open problem.
In general, dynamic camouflaging is suitable for applications that can tolerate a certain degree of error, including machine
learning, image processing, neuromorphic circuits etc., which also require protection against reverse engineering due to the sensitive nature of
their IP.

\section{Synthesis-Level Cost Analysis}
\label{sec:PPA_cost_analysis}

In this section, we benchmark the 
synthesis-level cost for the MESO-based
camouflaging primitive along with other spin-based devices. 
We use the ITC-99 suite for benchmarking, rather than the CeNN image-processing IP demonstrated in Section~\ref{sec:case_study}, to showcase 
the general prospects of full-chip camouflaging using spin-based devices.
We note that full-chip dynamic camouflaging may be uncalled for practical applications as in the case study above; again, this analysis here 
is for benchmarking of different devices.

\textbf{Setup:}
We compare the all-spin logic (ASL) primitive~\cite{alasad2017leveraging}, the giant spin-Hall effect (GSHE) primitive~\cite{patnaik2019spin}, and the MESO
primitive of this work in Table~\ref{tab:devices_PPA_comparison}.
The baseline designs are implemented in CMOS
and have been synthesized using \textit{Synopsys Design Compiler} with 2-input gates, in addition to inverters and buffers.

For each synthesized netlist, we replace all cells by their corresponding emerging-device model. 
Given that libraries and physical-design files for spin-based devices are not available yet, and also given that leading CAD vendors like
\textit{Synopsys} and \textit{Cadence} do not support system-level simulations of such spin-based devices yet, this setup is a practical approach.
For the MESO
primitive, we also include the peripheral MUXes shown in Fig.~\ref{fig:MESO_peripherals}, and we characterize them using \textit{Cadence
Virtuoso} for the 15-nm CMOS node using the \textit{NCSU FreePDK15} FinFET library, for a supply voltage of 0.8V.

For benchmarking, for example, the ITC-99 benchmark b17\_C 
comprises 24,228 2-input and inverter/buffer instances.
Using the GSHE primitive, along with its peripherals, each of these instances 
would consume a power of $0.2673$ $\mu$W~\cite{patnaik2019spin}.
For the MESO primitive, again with peripherals, each gate would consume $0.0615$ $\mu$W.
With simple arithmetic calculations we conclude that the GSHE-based logic would consume 6.5 mW while the MESO-based logic would consume 1.5 mW for
b17\_C. For area calculations, the same approach is taken.
For timing calculations, we keep track of the gates in the critical path, and the delay numbers are summed up.
For example, we observe 50 gates in the critical path for b17\_C. For the GSHE-based logic, each of these gates would incur a delay of
1.83 ns~\cite{patnaik2019spin}, resulting in a total delay of 91.5 ns. For MESO-based logic, with peripherals,
a delay of 0.2579 ns incurs for each instance, which totals to 12.895 ns.

\begin{table}[ht]
\centering
\scriptsize
\setlength{\tabcolsep}{1mm}
\caption{Comparison of Selected Emerging Device Primitives}
\label{tab:devices_PPA_comparison}
\input{figures/tab-comparison-GSHE}
\\[1mm]
The delay for Obfuscated MESO is the sum of the switching time for the intrinsic device (230 ps), the switching time for the interconnect (2.9 ps), and the delay induced by the peripheral MUXes.
The corresponding switching energies are 9.3 aJ for the device and 0.18 aJ for the interconnect;
these values are extracted from Table~4 of the supplementary material of~\cite{manipatruni2019scalable}.
The peripheral MUXes (shown in Fig.~\ref{fig:MESO_peripherals}) have been simulated using \textit{Cadence Virtuoso} for the
15-nm CMOS node using the NCSU FreePDK15 FinFET library, for a supply voltage of 0.8V.
The area for an intrinsic MESO device, without peripherals, is
$0.014$ $\mu \text{m}^2$~\cite{manipatruni2019scalable}.
\end{table}

\textbf{Results:}
We provide the comparison between selected emerging device primitives in Table~\ref{tab:devices_PPA_comparison}.
The results for full-chip camouflaging are presented in Table~\ref{tab:ppacomparison}.
We note that ASL-based~\cite{alasad2017leveraging} and GSHE-based~\cite{patnaik2019spin} full-chip camouflaging incurs
excessive power and timing overheads. 
On the other hand, MESO-based camouflaging 
offers substantial reductions relative to 
these spin devices and can be expected to perform even better than CMOS-based camouflaging schemes.\footnote{In general, for smaller camouflaging scales and
	hybrid designs (i.e., emerging spin devices along with CMOS),
area and power gains would scale down accordingly, whereas performance will remain similar, given that
	the emerging devices dominate the switching times.}
This is because the polymorphic MESO device consumes significantly lower switching energy, in the order of $\sim\!\!10$ atto Joules,
due to its energy-efficient electric-field-driven reversal.

\begin{table}[ht]
\centering
\scriptsize
\setlength{\tabcolsep}{0.9mm}
\renewcommand{\arraystretch}{1.2}
\caption{Comparison between Area, Power, and Performance for ASL-based~\cite{alasad2017leveraging}, GSHE-based~\cite{patnaik2019spin}, 
and MESO-based full-chip camouflaging on selected ITC-99 benchmarks. 
Absolute values are provided. 
Area is in $\mu$m$^{2}$, Power in mW, and Delay in ns. N/A indicates not available.}
\label{tab:ppacomparison}
\begin{tabular}{*{10}{c}}
\hline
\textbf{Benchmark}
& \multicolumn{3}{c}
{\textbf{ASL-based~\cite{alasad2017leveraging}}} & \multicolumn{3}{c}
{\textbf{GSHE-based~\cite{patnaik2019spin}}} 
& \multicolumn{3}{c}{\textbf{MESO-based}}\\
\cline{2-10}
& \textbf{Area} & \textbf{Power} & \textbf{Perf.} 
& \textbf{Area} & \textbf{Power} & \textbf{Perf.}
& \textbf{Area} & \textbf{Power} & \textbf{Perf.} \\
\hline
b15\_C 
& N/A & 2,702 & 54 
& 223.6 & 2.1 & 71.4 
&  183.1
&  0.5 
&  10.1  
\\ \hline
b17\_C 
& N/A & 8,494 & 71
& 702.6 & 6.5 & 91.5
&  575.4
&  1.5 
&  12.9  
\\ \hline
\textit{b18\_C}
& N/A & 21,783 & 137 
& 1,800.2 & 16.6 & 115.3
&  1,474.3
&  3.8 
&  16.3  
\\ \hline
\textit{b19\_C} 
& N/A & 42,027 & 165 
& 3,473.7 & 32.1 & 177.5
&  2,844.9
&  7.4  
&  25  
\\ \hline
\end{tabular}
\end{table}

We also compare synthesis-level PPA cost with a prior CMOS- and LUT-based scheme~\cite{baumgarten2010preventing} in Table~\ref{tab:ppacomparison_NEW}. 
As mentioned in Section~\ref{sec:toward_dynamic_camo}, 
functional polymorphism can also be implemented using CMOS-based reconfigurable units, such as FPGA LUTs.
We source the implemented scheme of~\cite{baumgarten2010preventing} from the set of benchmarks provided in~\cite{subramanyan15}.
On average, the scheme~\cite{baumgarten2010preventing} incurs area and power overheads of 193\% and 206\%, respectively, over original designs.
The MESO-based reconfiguration scheme does not incur such cost for area and power, but rather significant savings; only for delay/performance,
the MESO-based scheme incurs a higher cost than the CMOS-based scheme.
Therefore, the use of MESO devices can offer significant advantages for dynamic camouflaging, especially for circuits which are not reliant on
high performance.

\begin{table}[tb]
\centering
\scriptsize
\setlength{\tabcolsep}{0.43mm}
\renewcommand{\arraystretch}{1.2}
\caption{Comparison between Area, Power, and Performance for LUT-based obfuscation~\cite{baumgarten2010preventing} and MESO-based primitive for dynamic reconfiguration on selected ISCAS-85 benchmarks.
Absolute values are provided. 
Area is in $\mu$m$^{2}$, Power in mW, and Delay in ns.
}
\label{tab:ppacomparison_NEW}
\begin{tabular}{*{10}{c}}
\hline
\textbf{Benchmark}
& \multicolumn{3}{c}
{\textbf{Original (CMOS)}} 
& \multicolumn{3}{c}
{\textbf{LUT-based (CMOS)~\cite{baumgarten2010preventing}}} 
& \multicolumn{3}{c}{\textbf{MESO-based}}\\
\cline{2-10}
& \textbf{Area} & \textbf{Power} & \textbf{Perf.} 
& \textbf{Area} & \textbf{Power} & \textbf{Perf.}
& \textbf{Area} & \textbf{Power} & \textbf{Perf.} \\
\hline
c432
& 164.65 & 0.03 & 2.79 
& 543.71 & 0.11 & 2.96 
& 4.89 
& 0.01
& 5.67 
\\ \hline
c880
& 239.93 & 0.03 & 3.31 
& 780.98 & 0.12 & 3.48
& 6.34 
& 0.02   
& 7.48  
\\ \hline
c1908
& 250.57 & 0.05 & 3.72 
& 674.31 & 0.14 & 3.89 
& 5.79
& 0.02  
& 4.9 
\\ \hline
c2670
& 396.87 & 0.06 & 3.16 
& 1,193.0 & 0.18 & 3.24 
& 10.31   
& 0.03 
& 7.22  
\\ \hline
c3540
& 780.18 & 0.14 & 3.85 
& 2,308.08 & 0.42 & 3.91 
& 22.52  
& 0.06
& 7.74  
\\ \hline
c5315
& 1,029.95 & 0.17 & 3.63 
& 2,764.01 & 0.43 & 3.73 
& 28.76  
& 0.07  
& 6.45  
\\ \hline
c7552
& 1,138.48 & 0.23 & 3.93 
& 2,936.11 & 0.55 & 3.88 
& 28.81  
& 0.08   
& 7.99  
\\ \hline
\textbf{Average Cost} & -- & -- & -- &
\textbf{193\%} & \textbf{206\%} & \textbf{3\%} & \textbf{-97\%} & \textbf{-56\%} & \textbf{96\%} \\
\hline
\end{tabular}
\end{table}

\section{Conclusion and Future Work}
\label{sec:conclusion}

Functional polymorphism has been largely unexplored in the context of securing hardware. 
We present \textit{dynamic camouflaging} as a novel design-for-trust technique, based on the foundations of run-time polymorphism and post-fabrication reconfigurability exhibited by emerging spin-based devices.
Dynamic camouflaging serves well to secure the supply chain 
end-to-end, including the foundry, the test facility, 
and the end-user. 
We show that securing error-tolerant IPs, such as image processors, is suitable 
from the standpoint of dynamic camouflaging.
Finally, MESO-based full-chip camouflaging can offer
savings in PPA when compared to both ASL-based and GSHE-based camouflaging approaches.

As a part of future work,
we aim to explore viable techniques for securing
non-error-tolerant systems like cryptographic applications and/or mission critical systems
via dynamic camouflaging.
Besides, we will explore the design and implementation of system-level control circuitry for dynamic camouflaging.

\section*{Acknowledgments}

The work of Satwik Patnaik was supported by the Global Ph.D. Fellowship at NYU/NYU AD.
Besides, this work was carried out in part on the HPC facility at NYU AD.
This work was supported in part by the 
Semiconductor Research Corporation (SRC) and 
the National Science Foundation (NSF) 
through ECCS 1740136.

\ifCLASSOPTIONcompsoc

\ifCLASSOPTIONcaptionsoff
  \newpage
\fi

\bibliographystyle{IEEEtran}
\bibliography{main} 

\begin{IEEEbiography}[{\includegraphics[width=1in,height=1.25in,clip,keepaspectratio]{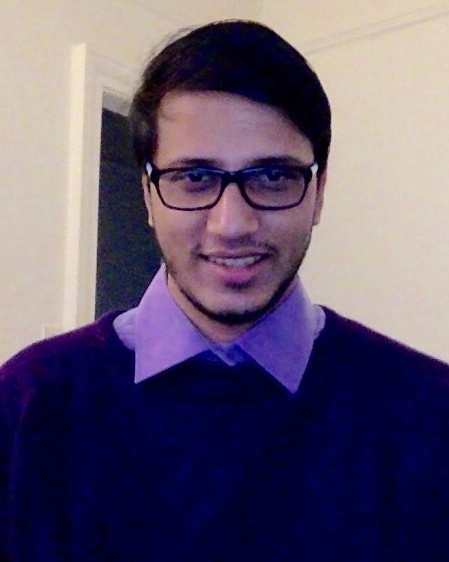}}]{Nikhil Rangarajan} (S'15--M'20)
is a Postdoctoral Associate at the Division of Engineering, New York University Abu Dhabi, UAE. 
He has Ph.D. and M.S. degrees in Electrical Engineering from New York University, NY, USA. 
His research interests include spintronics, nanoelectronics, device physics and hardware security. He is a member of IEEE. 
\end{IEEEbiography}

\begin{IEEEbiography}[{\includegraphics[width=1in,height=1.25in,clip,keepaspectratio]{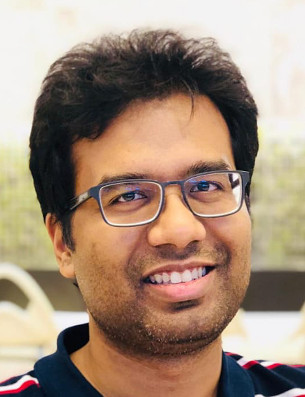}}]{Satwik Patnaik} (S'16)
received B.E.\
in Electronics and Telecommunications from the University of
Pune, India and
M.Tech.\ in Computer Science and Engineering with a specialization in VLSI Design from Indian Institute of Information Technology and
Management, Gwalior, India. 
He is a final year Ph.D.\ candidate at the Department of Electrical and Computer Engineering at the 
Tandon School of Engineering with New York University, Brooklyn, 
NY, USA.
He is
also a Global Ph.D.\ Fellow with New York University Abu Dhabi, U.A.E.
He received the Bronze Medal in the Graduate category at the ACM/SIGDA Student Research Competition (SRC) held at ICCAD 2018, and the best paper award at the Applied Research Competition (ARC) held in conjunction with Cyber Security Awareness Week (CSAW), 2017.
His current research interests 
include Hardware security, Trust and reliability issues for CMOS and emerging devices 
with particular focus on low-power VLSI Design.
He is a student member of IEEE and ACM.
\end{IEEEbiography}

\begin{IEEEbiography}[{\includegraphics[width=1in,height=1.25in,clip,keepaspectratio]{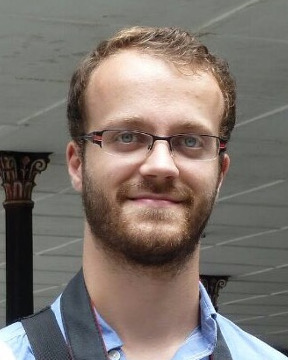}}]{Johann Knechtel}
(M'11)
received the M.Sc.\ in Information Systems Engineering (Dipl.-Ing.) in 2010 and the Ph.D.\ in Computer Engineering
(Dr.-Ing., summa cum laude) in 2014, both from TU Dresden, Germany.  
He is a Research Scientist
at the New York University, Abu Dhabi, UAE.  Dr.\ Knechtel was a Postdoctoral Researcher in 2015--16 at the Masdar Institute of Science and Technology, Abu Dhabi.  
From 2010 to 2014, he was a Ph.D.\ Scholar with the DFG Graduate School on ``Nano- and Biotechnologies for Packaging of Electronic
Systems'' hosted at the TU Dresden.  
In 2012, he was a Research Assistant with the Dept.\ of Computer Science and Engineering, Chinese University of Hong Kong, China.  
In 2010, he was a Visiting Research Student with the Dept.\ of Electrical Engineering and Computer Science, University of Michigan, USA.
His research interests cover VLSI Physical Design Automation, with particular focus on Emerging Technologies and Hardware Security.
\end{IEEEbiography}

\begin{IEEEbiography}[{\includegraphics[width=1in,height=1.25in,clip,keepaspectratio]{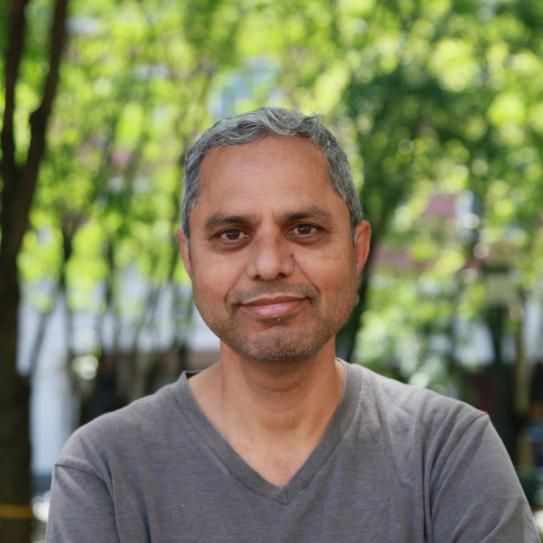}}]{Ramesh Karri} (SM'11--F'20)
obtained the B.E.\ degree in ECE from Andhra University and the Ph.D.\ degree in computer science and engineering from the University of
California at San Diego, San Diego.  He is currently a Professor of electrical and computer engineering with New York University.  He co-directs
the NYU Center for Cyber Security (http://cyber.nyu.edu).  He also leads the Cyber Security thrust of the NY State Center for Advanced
Telecommunications Technologies at NYU.  He co-founded the Trust-Hub (http://trust-hub.org).  He organizes the Embedded Systems Challenge
(https://csaw.engineering.nyu.edu/esc), the global red-team-blue-team hardware hacking event. He has published more than 200 articles in leading
journals and conference proceedings.  His research and education activities in hardware cybersecurity include trustworthy ICs; processors and
cyber-physical systems; security-aware computer-aided design, test, verification, validation, and reliability; nano meets security; hardware
security competitions, benchmarks, and metrics; biochip security; and additive manufacturing security.  
Dr. Karri's work on hardware
cybersecurity received best paper award nominations (ICCD 2015 and DFTS 2015) and awards (ITC 2014, CCS 2013, DFTS 2013 and VLSI Design 2012, ACM Student Research Competition at DAC 2012, ICCAD 2013, DAC 2014, ACM Grand Finals 2013, Kaspersky Challenge, and Embedded Security Challenge).  
He received the Humboldt Fellowship and the National Science Foundation CAREER Award.  He served as a
Program/General Chair of conferences, including the IEEE International Conference on Computer Design (ICCD), the IEEE Symposium on Hardware
Oriented Security and Trust (HOST), the IEEE Symposium on Defect and Fault Tolerant Nano VLSI Systems, NANOARCH, RFIDSEC, and WISEC.  He serves
on several program committees (HOST, ITC, VTS, ETS, ICCD, DTIS, and WIFS).  He served as an Associate Editor of the IEEE TRANSACTIONS ON
INFORMATION FORENSICS AND SECURITY from 2010 to 2014 and has been an Associate Editor of the IEEE TRANSACTIONS ON CAD since 2014, ACM Journal of
Emerging Computing Technologies since 2007, ACM Transactions on Design Automation of Electronic Systems since 2014, IEEE ACCESS since 2015, the
IEEE TRANSACTIONS ON EMERGING TECHNOLOGIES IN COMPUTING since 2015, the IEEE Design and Test since 2015, and the IEEE EMBEDDED SYSTEMS LETTERS
since 2016. He served as an IEEE Computer Society Distinguished Visitor from 2013 to 2015.  He co-founded the IEEE/ACM Symposium on Nanoscale
Architectures (NANOARCH).  He served on the Executive Committee of the IEEE/ACM Design Automation Conference leading the Security@DAC initiative
from 2014 to 2017.  He has given invited keynotes, talks, and tutorials on Hardware Security and Trust (ESRF, DAC, DATE, VTS, ITC, ICCD, NATW, LATW, CROSSING, and HIPEAC).
\end{IEEEbiography}

\begin{IEEEbiography}[{\includegraphics[width=1in,height=1.25in,clip,keepaspectratio]{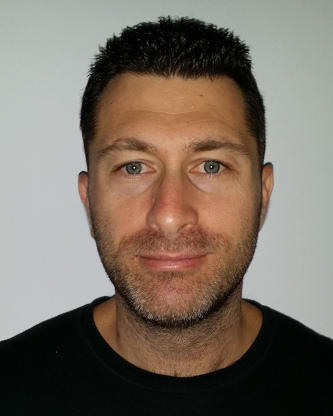}}]{Ozgur Sinanoglu} (M'11--SM'15)
is a Professor of Electrical and Computer Engineering at New York University Abu Dhabi. He earned his B.S.\ degrees, one in
Electrical and Electronics Engineering and one in Computer Engineering, both from Bogazici University, Turkey in 1999. He obtained his MS
and PhD in Computer Science and Engineering from University of California San Diego in 2001 and 2004, respectively. He has industry
experience at TI, IBM and Qualcomm, and has been with NYU Abu Dhabi since 2010. During his PhD, he won the IBM PhD fellowship award twice.
He is also the recipient of the best paper awards at IEEE VLSI Test Symposium 2011 and ACM Conference on Computer and Communication Security 2013. 

Prof.\ Sinanoglu's research interests include design-for-test, design-for-security and design-for-trust for VLSI circuits, where he has more than 180 conference and journal papers, and 20 issued and pending US Patents. 
Prof.\ Sinanoglu has given more than a dozen tutorials on hardware security and trust in leading CAD and test conferences, such as DAC, DATE, ITC, VTS, ETS, ICCD, ISQED, etc. 
He is serving as track/topic
chair or technical program committee member in about 15 conferences, and as (guest) associate editor for IEEE TIFS, IEEE TCAD, ACM JETC,
IEEE TETC, Elsevier MEJ, JETTA, and IET CDT journals. 

Prof.\ Sinanoglu is the director of the Design-for-Excellence Lab at NYU Abu Dhabi. 
His recent research in hardware security and trust is being funded by US National Science Foundation, US Department of Defense, Semiconductor Research Corporation, Intel Corp and Mubadala Technology.
\end{IEEEbiography}

\begin{IEEEbiography}[{\includegraphics[width=1in,height=1.25in,clip,keepaspectratio]{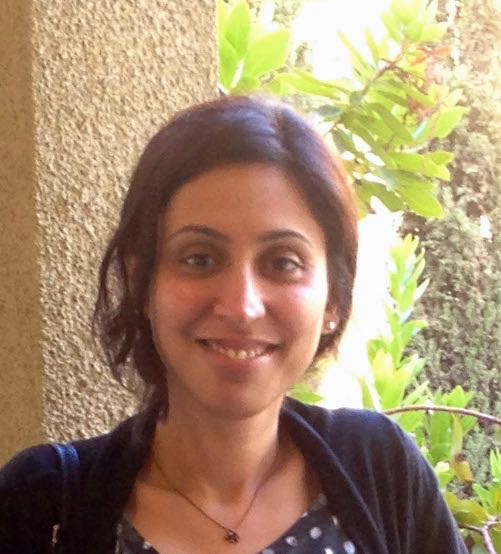}}]{Shaloo Rakheja} (M'13)
is an Assistant Professor of Electrical and Computer engineering with the Holonyak Micro and Nanotechnology Laboratory, University of Illinois at Urbana-Champaign, Urbana, IL, USA, where she works on
nanoelectronic devices and circuits. She was previously an
Assistant  Professor  of  Electrical  and  Computer  engineering with New  York  University,  Brooklyn,  NY,  USA.
Prior to joining NYU, she was a Postdoctoral Research Associate with the Microsystems Technology
Laboratories, Massachusetts Institute of Technology, Cambridge, USA. She obtained her M.S.\ and Ph.D.\ degrees in Electrical and Computer
Engineering from Georgia Institute of Technology, Atlanta, GA, USA. 
\end{IEEEbiography}

\end{document}

%% file: figures/tab-comparison-GSHE.tex
\begin{tabular}{cccc}
\hline
\textbf{Publication} 
& \textbf{Energy} 
& \textbf{Power} 
& \textbf{Delay}\\
\hline

ASL~\cite[a]{alasad2017leveraging} 
& 0.58 pJ 
& 351.52 $\mu$W 
& 1.65 ns 
\\ \hline

ASL~\cite[b]{alasad2017leveraging} 
& 1.16 pJ 
& 351.52 $\mu$W 
& 3.3 ns 
\\ \hline

ASL~\cite[c]{alasad2017leveraging} 
& 0.13 pJ 
& 342.11 $\mu$W 
& 0.38 ns 
\\ \hline

GSHE (intrinsic)~\cite{patnaik2019spin}
& 0.33 fJ 
& 0.2125 $\mu$W 
& 1.55 ns 
\\ \hline

Obfuscated GSHE~\cite{patnaik2019spin}
& 0.49 fJ 
& 0.2673 $\mu$W 
& 1.83 ns \\ 
(with MUXes) &  &  &  
\\ \hline

\hline
\textbf{MESO (intrinsic)} 
& \textbf{9.3 aJ} 
& \textbf{0.0404 $\mu$W} 
& \textbf{0.23 ns} 
\\ \hline

\textbf{Obfuscated MESO}
& \textbf{16.04 aJ}
& \textbf{0.0622 $\mu$W} 
& \textbf{0.2579 ns} \\ 
\textbf{(with MUXes and interconnects)} &  &  &  
\\ \hline

\end{tabular}